\documentclass[
preprint,    
 amsmath,amssymb,
 aps,
]{revtex4-2}

\usepackage{graphicx}
\usepackage{dcolumn}
\usepackage{bm}
\usepackage{epstopdf, epsfig}
\usepackage{subcaption}
\usepackage{float}
\usepackage{xcolor}
\usepackage{latexsym}
\usepackage{amsfonts}
\usepackage{amssymb}
\usepackage{rotating}
\usepackage{comment}
\usepackage{ulem}
\begin{document}

\title{Assessment of local isotropy, Kolmogorov constant, and modified eddy viscosity-based modeling for particle-laden turbulent channel flows}


\author{Naveen Rohilla}
\author{Partha S Goswami}%
 \email{psg@.iitb.ac.in}
\affiliation{%
 Department of Chemical Engineering, IIT Bombay,
 	Mumbai, India, 400076 \\
}%




\date{\today}

\begin{abstract}
A large number of models which address the dynamics of particle-laden turbulent flows have been developed based on the assumption of local isotropy and use the Kolmogorov constant that correlates the spectral distribution of turbulent kinetic energy with the turbulent dissipation rate. Many turbulence models (Stochastic and LES models) use the Kolmogorov constant in the formulation. Compilation of a large number of experimental data for different flow configurations has revealed that the Kolmogorov constant is independent of Reynolds number in the limit of high Reynolds number \cite{sreenivasan1995universality}. However, several numerical studies at low and intermediate Reynolds numbers which address the flow situations of practical importance consider that the Kolmogorov constant remains unchanged irrespective of whether the flow is single phase or multiphase. 
In the present work, we assess the variation of local isotropy of fluid fluctuations with the increase in particle loading in particle-laden turbulent channel flows. We also estimate the Kolmogorov constant using second-order velocity structure functions and compensated spectra in case of low Reynolds number turbulent flows. Our study reveals that the Kolmogorov constant decreases in the channel center with an increase in the particle volume fraction for the range of Reynolds number investigated here. The estimated variation of the Kolmogorov constant is used to express the Smagorinsky coefficient as a function of solid loading in particle-laden flows. Then, a new modeling technique is adopted using the large eddy simulation (LES) to predict the fluid phase statistics without solving simultaneous particle phase equations. This new methodology also helps understand the drastic decrease in turbulence intensity at critical particle volume loading.

\end{abstract}

\maketitle


\section{Introduction}

Particle-laden turbulent flows are encountered in many geophysical and industrial processes. One of the major focuses in this area is understanding the dynamics of fluid and solid phases. With the advent of high-speed computational systems, although performing direct numerical simulations for small system sizes has become possible, modeling still plays a vital role in studying large-scale systems of practical importance. Most of the modeling techniques are based on the approximation of local isotropy in the inertial and dissipation range. 
Kolmogorov's similarity hypothesis for the inertial subrange states that for every turbulent flow at a sufficiently high Reynolds number, the statistics of the motions of scale ($r$) in the range, $\eta \ll r \ll L$, have a universal form that is uniquely determined by $\epsilon$ and independent of $\nu$ \cite{kolmogorov1941energy}. Here, $L $ is the integral length scale, $\eta$ is the Kolmogorov length scale, $\nu$ is the kinematic viscosity, and $\epsilon$ is the mean viscous dissipation rate of turbulent kinetic energy. The spectral energy in the inertial subrange is expressed as  $E(k)= C \epsilon^{2/3}k^{-5/3}$,  where  $k$ is the wavenumber and $C$ is the proportionality prefactor known as the Kolmogorov constant. The Kolmogorov constant is obtained based upon the Kolmogorov hypothesis
for different flows such as boundary layers \cite{sreenivasan1995universality, saddoughi1994local} , channel flows \cite{antonia1997second, choi2004lagrangian}, homogeneous isotropic turbulence \cite{sawford2011kolmogorov, donzis2010bottleneck, yeung1997universality}, etc. The Kolmogorov constant for different experiments and simulations have been summarized by \citet{sreenivasan1995universality},  \citet{yeung1997universality}, and \citet{lien2002kolmogorov}.  The stochastic turbulence models \cite{heinz2002kolmogorov, thomson1987criteria, wilson1996review, reynolds2003application, pope1985pdf, pope2011simple, Shotorban2006, marchioli2017large} and other turbulence models which have 
been developed based on large scale turbulent structures, like Smagorinsky model \cite{smagorinsky1963general, pope2001turbulent, sagaut2006large}, and 
other eddy viscosity based models \cite{sagaut2006large} use Kolmogorov constant. Another implication of the Kolmogorov constant ($C_0$) is that the turbulent diffusivity is expressed as, $D_t = 2\sigma_u^4/C_0 \epsilon $ where $T_L = 2 \sigma^2/C_0 \epsilon$ is the integral time scale and $\sigma_u^2$ is the variance of velocity fluctuation.

In his seminal work, \citet{sreenivasan1995universality} has summarized that the Kolmogorov constant is universal and independent of the flow configuration and Reynolds number. At a sufficiently high Reynolds number where local isotropy is satisfied at dissipation scales and inertial range, the Kolmogorov constant attains a universal value \cite{yeung1997universality}. However, at moderate and low Reynolds numbers, the Kolmogorov constant may differ from the universality \cite{sreenivasan1995universality, yeung1997universality}. \citet{antonia1997second} performed experiments for channel flow and observed a lower value of the Kolmogorov constant. They analyzed the second and third-order velocity structure functions and reported that small-scale isotropy is a necessary condition for the existence of a universal inertial range. \citet{yeung1997universality} mentioned that for the presence of inertial range, isotropy should also be present along with $-5/3$ scaling. The Kolmogorov constant may attain a different value if isotropy is not satisfied in the inertial range.
\citet{heinz2002kolmogorov} discussed the variations of the Kolmogorov constant for equilibrium turbulent boundary layer and homogeneous isotropic stationary turbulence. He stated that for Stochastic modeling, the value is near two, and anisotropic velocity and acceleration fluctuations dominate the energy budget. Furthermore, the value is near six if those contributions disappear. The above studies point out the deviation of the Kolmogorov constant from universality due to anisotropy at low and moderate Reynolds numbers for unladen cases. 

In the case of particle-laden flows, the addition of particles modifies the turbulence intensity in turbulent flows. Here, we briefly discuss the capability of point-particle approximation-based numerical simulation to predict second moments of fluid phase fluctuations and drag reduction in particle-laden flows. The various parameters, such as the ratio of particle diameter to integral length scale, Stokes number, particle Reynolds number, etc., affect the turbulence of the carrier phase \cite{kulick1994particle, li2001numerical, Takeo2001, Vreman2009, vreman2015turbulence, yu_2017, yu_2021, muramulla2020disruption}. The various authors have attempted to quantify the regimes of turbulence attenuation and augmentation \cite{Gore1989, hetsroni1989particles,  crowe2000models, hosokawa2003turbulence,  righetti2004particle, Tanaka2008, noguchi2009particle, luo_2016}. Various authors have observed the increase in turbulence attenuation in different flow configurations \cite{KulickJD1994, Vreman2009,  zhao2010turbulence, bari2010aluminum, vreman2015turbulence, zade2018experimental, rohilla2022applicability, kumaran2020turbulence, muramulla2020disruption}, and has also been reviewed by Ref. \cite{balachandar2010turbulent, kuerten2016point, elghobashi2019direct, brandt2022particle}.  It is in debate whether spherical particles can cause drag reduction or not, which has been observed for point particle simulations \cite{yu_2017}. A drag reduction by point particle simulations has been reported in many numerical studies \cite{zhao2010turbulence, dubief2004coherent, vreman2007turbulence, dritselis2008numerical, muramulla2020disruption}, and in experiments by Ref.~\cite{bari2010aluminum, kartushinsky2005experimental}. However, drag reduction has not been observed in particle resolved DNS (PR-DNS) studies \cite{picano2015turbulent, fornari_2016, yu_2017, costa2020interface, costa_2021} where either particle inertia is very low or the effect of gravity has not been included.  \citet{yu_2021} did fully resolved DNS in upward vertical channel flow including gravity.
A decrease in the wall friction is observed for settling coefficient ($u_i$) less than 0.3
compared to the unladen cases. A further increase in settling coefficient ($u_i > 0.3$) results in higher wall-friction. \citet{zhu2020interface} also observed the low wall-friction for spherical particles than oblate particles for particle settling coefficient of 0.3 (Fig.2 of their paper). This happens due to the attenuation of the large vortices by spherical particles. This study was done while keeping a constant bulk flow rate for an upward channel flow including gravity.  \citet{zade2018experimental} have performed experiments for square duct with different particles sizes ($2H/d_p = 9, 16$ and 40),  Reynolds numbers ($Re_{2H} \sim 10000-27000$) and volume fractions (5\%, 10\%, and 20\%) where particles considered are almost neutrally buoyant. Here, 2H is the duct's full height. It is found that the friction factor is significantly high compared to a single phase at a low Reynolds number ($Re_{2H} =  10042$). However, an increase in Reynolds number results in a decrease in friction factor. The friction factor decreases with an increase in particle diameter for a volume fraction less than 10\%. This leads to drag reduction compared to the single-phase flow even at $Re_{2H} \sim 27000$. The decrease in drag is related to the attenuation of the turbulence. However, a non-monotonic drag modification is observed for a volume fraction of 20\% with the increase in particle diameter. At this volume loading, there is a decrease in the fluid fluctuations and Reynolds stress, but drag is increased due to an increase in particle-induced stress. 

\citet{costa_2021} performed the fully resolved simulations in a channel flow with bulk Reynolds number of 5600 and observed an increase in drag at a particle volume fraction of $3\times 10^{-5} - 3.4\times 10^{-4}$. However, they commented that a drag reduction might be observed at a higher particle volume fraction which has been observed for point-particle simulations. Also, various efforts have been put to validate the point-particle approach with fully resolved simulations and experiments (\cite{wang2019inertial}). \citet{costa2020interface} have compared the statistics of point particles and fully resolved cases and concluded that the inclusion of the Saffmann lift improves the particle statistics in the near-wall region. \citet{mehrabadi2018direct} did the comparison of point particle and particle resolved DNS for decaying isotropic turbulent flow. The authors found that for comparison of point particles and PR-DNS, correction for undisturbed velocity and finite Reynolds number are required in the point particle simulations to match the particle acceleration density function and second-order moments statistics with PR-DNS at high Stokes number. A Stokes number ($St_\eta $) of 100 is taken in their study ($St_\eta$ is based on the particle relaxation time and initial Kolmogorov time scale). For $St_\eta = 1$, there is not much difference in the simulations with and without correction of undisturbed velocity for point particles and observed a good match with the particle resolved simulations for the calculation of fluid kinetic energy and fluid dissipation rate. For $St_\eta = 100$, it is observed that the point particles with the Schiller-Naumann and undisturbed corrections show a good match with particle-resolved simulations for the fluid kinetic energy and fluid dissipation rate. A good accuracy of statistics between the point particle and fully resolved simulations has been demonstrated by \citet{frohlich2018validation}. The above discussion suggests that there is no consensus in the literature on the drag reduction in particle-laden wall-bounded flows, which depends on the parameters like fluid Reynolds number and particle Reynolds number, and also on the particle terminal velocity. More systematic numerical simulations and their validation with controlled experiments are required to conclude the drag reduction phenomenon of particle-laden flows and the limitations of point-particle and PR-DNS methodology.

An increase in turbulence attenuation also leads to a further increase in anisotropy as a significant decrease is observed in wall-normal and spanwise directions than in the streamwise direction \cite{KulickJD1994, richter2013momentum, gualtieri2013clustering, richter2015turbulence, zhao2010turbulence, rohilla2022applicability, shringarpure2012dynamics}. \citet{gualtieri2013clustering} have commented that care should be taken while applying Kolmogorov theory as anisotropy is increased for particle-laden flows. \citet{ferrante2003physical} and \citet{ahmed2000mechanisms} have also discussed the increase in anisotropy for particle-laden homogenous isotropic and homogenous shear turbulence, respectively. In a recent study, \citet{rohilla2022applicability} have demonstrated that the  LES models,  like Smagorinsky and dynamic Smagorinsky models, perform poorly in predicting the turbulence modulation and the critical volume loading at which turbulence collapses completely. The scale-similarity and mixed models are found to perform better than  Smagorinsky, and dynamic models in predicting local energy flux \cite{boivin2000prediction}. The authors mentioned that the former models could capture the backscatter and thus perform well. Even though the dynamic model allows the backscattering effect. In earlier works, the deviation between DNS and LES in predicting statistical properties was attributed to the modeling error \cite{dritselis2011large, rohilla2022applicability}. All these observations demand a rigorous analysis to check the local isotropy of small scales, which is the basis of LES formulation. In the present study, we want to explore whether particles can alter the extent of anisotropy in the gas phase, and if so, what is the effect of increased anisotropy on the Kolmogorov constant?
Direct numerical simulations (DNS) are performed for turbulent channel flow at Reynolds numbers  3300 and 5600 based on average gas velocity and channel width to answer the above questions. Kolmogorov constant has been computed following different methods using the simulation results. When the Kolmogorov constant is studied, the bottleneck and intermittency effects are worth discussing. These effects are studied by \citet{donzis2010bottleneck} where direct numerical simulations are performed for isotropic turbulence for a range of Taylor Reynolds numbers ($Re_\lambda = 38-1000$). The authors discussed the scaling of the bottleneck effect and the related problems in calculating the Kolmogorov constant. It is reported that the bottleneck effect would vanish at a Taylor Reynolds number of approximately $ 2\times 10^5$. \citet{kaneda2003energy} observed that the Kolmogorov's theory (\cite{kolmogorov1941energy}) is valid for $Re_\lambda < 700$, and the scaling is steeper than (-5/3) for $Re_\lambda > 700$. \citet{ishihara2016energy} have proposed the scaling law for higher Reynolds numbers similar to work by \citet{kolmogorov1962refinement}. In earlier studies such as Ref. \cite{donzis2010bottleneck, yeung1997universality}, it has been observed that the presence of inertial range is not possible for $Re_\lambda < 200$. However, the Kolmogorov constant for lower $Re_\lambda$ can be defined from the peak point of compensated spectra, or second-order velocity structure functions \cite{jimenez1993structure, sreenivasan1995universality, antonia1997second, choi2004lagrangian}. In the present work, the simulations are carried out at low $Re_\lambda$ where we do not expect a clear inertial range. Thus the Kolmogorov constant is defined from the peak point of compensated spectra and second-order velocity structure functions. A constant value of Kolmogorov prefactor exist for $Re_\lambda > 10^{4}$ (\cite{lien2002kolmogorov}). However, such a high Reynolds number requires high computational resources for the particle-laden turbulent flows considering particle-particle and particle-wall interactions at high volume loading, which is beyond the scope of this study. The aim of the present study is not to comment on the universality of the Kolmogorov constant which is achieved at a high Reynolds number. It highlights the increase in local anisotropy of small and large scales of turbulence for particle-laden cases and consequently affects the Kolmogorov constant.
The variation of the Kolmogorov constant as a function of particle volume loading has been used to predict the fluid phase statistics. In the proposed modeling approach, the variation of the Smagorinsky coefficient is estimated from the variation of the Kolmogorov constant. The simulations are performed to predict the dynamics of the fluid phase without solving the particle phase equations simultaneously, and thus it is computationally less expensive. The present analysis of variation of the Kolmogorov constant with particle volume loading and the new methodology will provide insight to develop advanced models for two-phase turbulent flows. This work is important for the current scenario because many studies are performed at similar Reynolds numbers ($\sim Re_b = 5600$) \cite{Armenio1999, marchioli2008some, Kuerten2005, Kuerten2006, Vreman2009, dritselis2011large, zamansky2013acceleration, duque2021influence, rohilla2022applicability} which are of practical importance.

The paper outline is as follows: In Sec.~\ref{sec:simulation_parameters}, the fluid and particle phase equations are discussed along with the flow configurations and parameters. In Sec.~\ref{sec:results}, the simulation results are presented for different Reynolds numbers and Stokes numbers. The new methodology for the subgrid-scale model to capture the fluid phase statistics in particle-laden turbulent flows without the particles is discussed in Sec.~\ref{sec:new_model}. The summary of the work is discussed in the last section.

\section{Simulation parameters}
\label{sec:simulation_parameters}

The fluid phase is considered incompressible and described by the continuity and Navier-Stokes equation as follows.

\begin{equation}
\frac{\partial u_i}{\partial x_i}=0
\end{equation}
\begin{equation}
\frac{\partial u_i}{\partial t} + \frac{\partial u_i u_j}{\partial x_j} = - \frac{1}{\rho_f} \frac{\partial p}{\partial x_i} + {\nu} \frac{\partial^2 u_i}{\partial x_j \partial x_j} + \frac{f_i({\bf x},t)}{\rho_f}
\label{DNS eqn}
\end{equation}

Where $u_i$ is the velocity, $p $ is the pressure, $ \rho_f$ is the fluid density, and $\nu$ is the kinematic viscosity. $f({\bf x},t)$ is the feedback force density due to the solid phase. The pseudo-spectral method has been used to solve the Navier-Stokes equation. A second-order Adams-Bashforth scheme for the nonlinear term and Crank-Nicholson time discretization has been used for the linear terms. The lift and drag forces are included in the feedback force term, which can be expressed as, 
\begin{equation}
f_i({\bf x},t) = - \sum_I (F_{i,I}^D + F_{i,I}^L) \delta ({\bf x}-{\bf x}_I),
\end{equation}
where ${\bf x}$ is the fluid node, ${\bf x}_I$ is the position of $I^{th}$ particle,  $F_{i,I}^L$ and $F_{i,I}^D$ are the lift and drag forces on the particle I, and $\delta({\bf x}-{\bf x}_I)$ is the Dirac delta function in three dimensions. Any fluid volume fraction term ($\phi_v^{air}\approx1$) in the fluid-phase momentum equation is not included as the particle volume fraction is much lower compared to the continuum air volume fraction~\cite{dritselis2008numerical, dritselis2011large,richter2015turbulence, vreman2015turbulence, ghosh2022dynamics, ghosh2022statistical}. 

In the present study, the point-particle approach \cite{bagchi2003effect, mehrabadi2018direct} is considered with drag and lift corrections as discussed in detailed in Ref.~\cite{muramulla2020disruption}. The point particles are tracked in the Lagrangian frame, and Newton's second law describes their motion. The particle-wall and particle-particle collisions have also been considered. The particle motion is described by Eq.~(\ref{pEqn}).
\begin{equation}
m_p \frac{dv_{i,I}}{dt} = F_{i,I}^D + F_{i,I}^L + \sum_{I\neq J}^{} F_{i,IJ} + F_{i,Iw} + m_pg,
\label{pEqn}
\end{equation}
where $m_p$ is the mass  and $v_{i,I}$  is the velocity of $I^{th}$ particle,  $F_{i,I}^D$ is the drag force, and $ F_{i,I}^L$ is the lift force exerted on the particle.  In Eqn.~\ref{pEqn}, g is the gravitational acceleration, $F_{i,IJ}$ is the interaction force between $ I^{th} $ and  $ J^{th} $ particle, $F_{i,Iw}$ is the interaction force between $I^{th }$ particle and wall. In the present study, the effect of gravity and lift are included as it is reported that the  Implementation of the lift force improves the particle statistics in near-wall region \cite{marchioli2007influence, costa2020interface}. \citet{marchioli2007influence} have explored the effect of gravity and lift on the particle distribution in wall-bounded flows.  The hard-sphere approach is taken to account the elastic collisions between the particle-particle and particle-wall. The drag force is calculated using inertia corrected drag law (Schiller-Naumann correlation \cite{naumann1935drag}) as given below. 
\begin{equation}
F_{i,I}^D = 3 \pi \mu d_p (\widetilde{u}_{i,I}(x,t) - v_{i,I}) (1+ 0.15Re_p^{0.687})
\label{drag_law}
\end{equation}
Although a point particle approximation has been used, the grid size in the wall-normal direction can be smaller than the particle size in the near-wall region. Thus, the force on the grid is calculated from the fraction of particle surface present in the cell. The fluid velocity is interpolated at the particle location to calculate the drag and lift. The detailed simulation procedure for the calculation of feedback force, near-wall corrections in lift and drag, and corrections for the undisturbed velocity field at the particle locations are discussed in our earlier work \cite{muramulla2020disruption}. The implementation of Saffmann lift and correction for undisturbed velocity are important to accurately predict the spatial particle distribution \cite{wang2019inertial, costa2020interface, mehrabadi2018direct, brandt2022particle}, which have been included in this study. The particle-to-fluid density ratios considered in the present study are $\approx$ 1000 or higher. Therefore, the buoyancy and Basset history effects are neglected in the particle's equation of motion. 

For Reynolds numbers of 3300 and 5600, 128 \& 64, and 192 \& 160 Fourier modes are used in streamwise and spanwise directions, respectively~\cite{muramulla2020disruption}. For wall-normal direction, 65 and 129 Chebyshev modes are used for $Re_b = 3300$ and 5600, respectively.  The corresponding Reynolds numbers ($Re_c$) based on centerline turbulent velocity and half-channel width are 2000 and 3360, and Reynolds numbers ($Re_\tau$) based on unladen frictional velocity and half-channel width are 115 and 180. The transitional regime for the channel flow occur in the range $1300\leq Re_b \leq 1800$,  $975 \leq Re_c \leq 1200$, and  $62.5\leq Re_\tau \leq 73.5$ \cite{patel_head_1969, carlson_widnall_peeters_1982, sano2016universal, zhang_2017}. Thus, the simulations in our work are carried out at $Re_b$ of 1.8 and 3 times the upper limit of the transitional regime, and the flow is fully turbulent. The domain lengths normalized with unladen viscous units are $2921\times 232 \times 486$ ($L_x^+ \times L_y^+ \times L_z^+$) and $4562 \times 363 \times 760$ for Reynolds numbers of 3300 and 5600, respectively. The first grid point in the wall-normal direction is such that $y^+$  is less than one. The ($^+$) symbol indicates that the quantity is normalized with viscous scales. The time step used in the simulations is ($0.0033h/\overline{u}$) where $h$ is channel width and $\overline{u}$ is fluid bulk velocity. The pressure gradient is adjusted to maintain a constant bulk flow rate. The simulations have been performed in a vertical channel with  $8\pi\delta * 2\delta* (4/3)\pi\delta$ in streamwise (x), wall-normal (y), and spanwise (z) directions, respectively. Where  $\delta$ is half channel width, no-slip boundary conditions are applied on the walls in the y-direction. The bulk Reynolds numbers ($Re_b = \rho_f \times \bar{u} \times 2\delta/\mu_f$) are fixed at 3300 and 5600 based on the channel width ($2\delta$) and average fluid velocity ($\overline{u}$), which corresponds to $Re_\tau$ of 115 and 180 respectively based on the unladen frictional velocity and half-channel width. The range of Stokes numbers is mentioned in Table~\ref{table Stokes}. 
\begin{table}[ht]
\centering
\caption{The Stokes number ($St$) is defined as $St = \tau_p/\tau_f$ where $\tau_p = \rho_p d_p^2/18\mu_f$, $\tau_f = 2\delta/\overline{u}$, $\rho_p$ is the particle density,  $d_p$ is the particle diameter, $\mu_f$ is the fluid dynamic viscosity, $\delta$ is the half channel width and $\bar{u}$ is the average fluid velocity. Here, $Re_b$ is the fluid bulk Reynolds number.}
\begin{tabular}{c c c}
& & \\ 
\hline
\hline
$ Re_b $	& $\rho_p$ 	& St \\
\hline
3300	& 1000	& 52.73 \\
	& 2000	& 105.47 \\
	& 4000	& 210.93 \\
5600 & 1200& 105.47\\	
 & 2400& 210.93\\	
\hline
\hline
\end{tabular}
\label{table Stokes}
\end{table}



\section{Results}
\label{sec:results}

\begin{figure}
	\begin{subfigure}[b]{1\textwidth}
	\minipage{0.5\textwidth}
	\includegraphics[width=\textwidth]{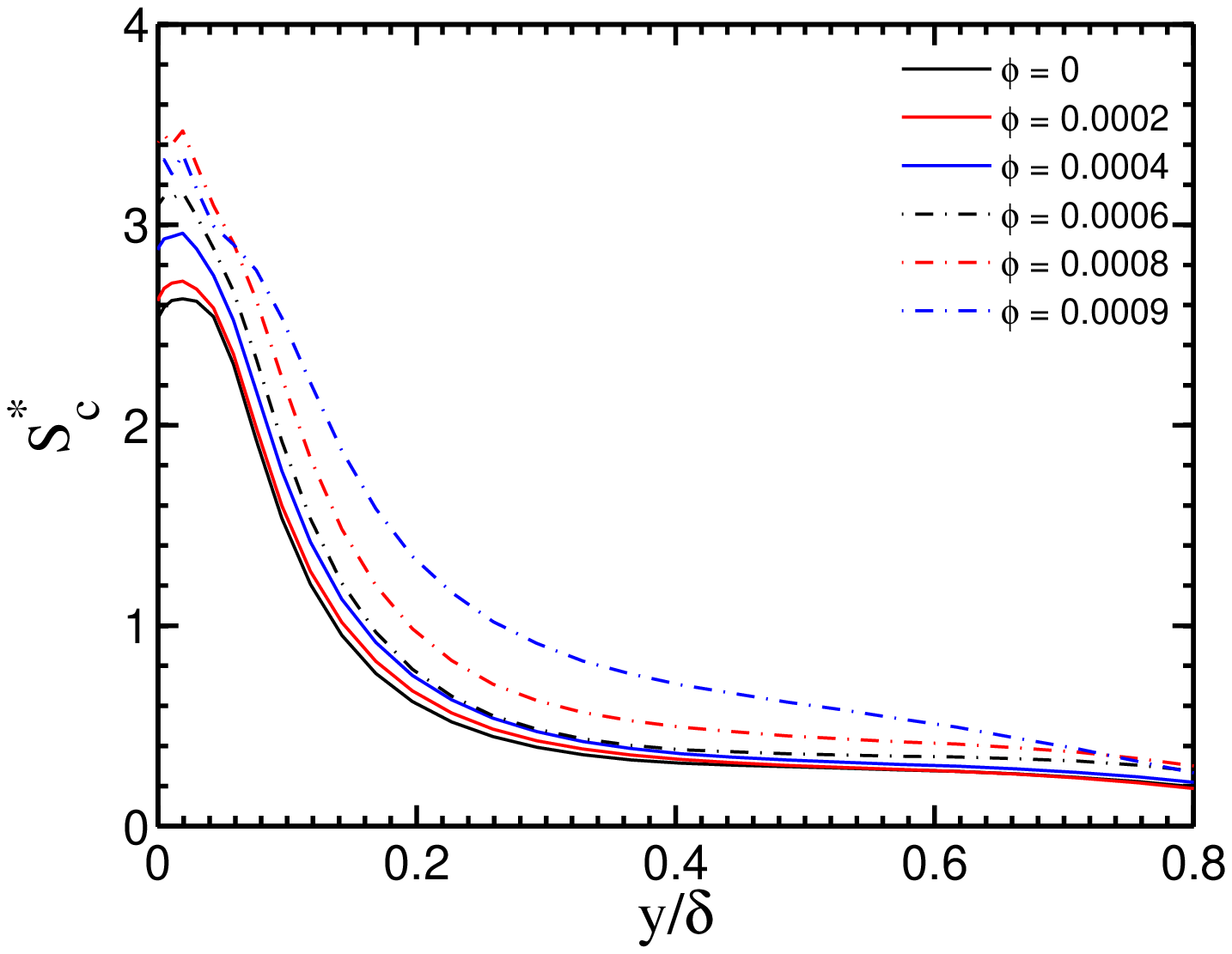}
	\caption{}
	\endminipage 
	\minipage{0.49\textwidth}
	\includegraphics[width=\textwidth]{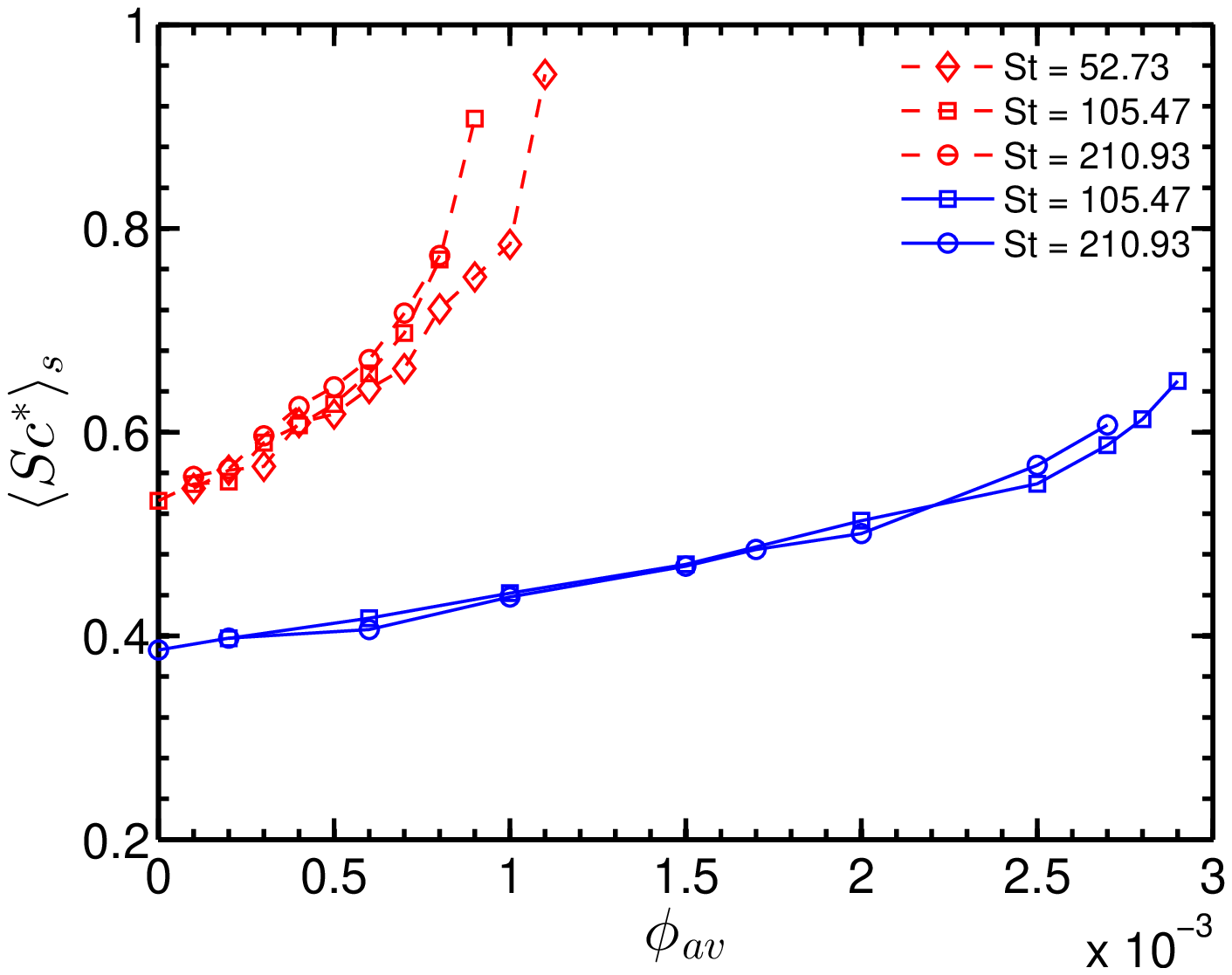}
	\caption{}
	\endminipage
	\end{subfigure} 
	\caption{The ratio of Kolmogorov time scale to mean shear time scale, ($ S_c^* = S (\nu /\epsilon)^{1/2} $), for different volume fractions. (a) $ S_c^*$ for $Re_b = 3300$ and St = 105.47. (b) The spatial averaged $ S_c^*$ across the channel width for $Re_b = 3300$ and 5600 with different Stokes numbers (St) and average volume fraction ($\phi$ or $\phi_{av}$). In Fig.~(b), the symbols with dashed lines are for $Re_b = 3300$ and symbols with solid lines are for $Re_b = 5600$ in Fig.(b). In Fig.~(a), $\delta$  is half-channel width.  }
	\label{Sc_plot}
\end{figure}

The simulations are performed for bulk Reynolds numbers of 3300 and 5600 for a range of volume fractions and different Stokes numbers using a pseudo-spectral code used in our earlier studies \cite{kumaran2020turbulence, muramulla2020disruption}. It is observed that the turbulence attenuation increases with an increase in volume loading steadily up to a certain volume fraction, and then there is a sudden collapse in the turbulence intensities\cite{kumaran2020turbulence, muramulla2020disruption, rohilla2022applicability}. The particle volume loading at which turbulence collapse happens is referred to as critical particle volume loading (CPVL). Other authors have also observed the complete turbulence collapse in different flow configurations  \cite{mito2006effect, shringarpure2012dynamics, capecelatro2018transition, wang2021effect, duque2021influence, yu_2021}. A detailed analysis of the effect of volume fraction and Stokes number on the turbulence attenuation is presented in earlier studies \cite{li2001numerical, mito2006effect, dritselis2016direct, capecelatro2018transition, kumaran2020turbulence, muramulla2020disruption}. In this work, we quantify the modification in the extent of anisotropy of turbulence fluctuations, which is assoaciated with the modulation of turbulence due to presence of dispersed phase. The local isotropy of the small scales across the channel width can be assessed from the ratio of Kolmogorov time scale to mean shear time scale \cite{saddoughi1994local, corrsin1957some, antonia1992isotropy}. The necessary condition for the small scale to be isotropic was provided by \citet{corrsin1957some}  as
\begin{equation}
\left( \frac{\nu}{\epsilon}\right)^{1/2} \ll \frac{1}{S}\hspace{0.6cm } or \hspace{0.6cm }
S_c^* \ll 1
\end{equation}
where $S = dU/dy$ is the mean shear rate, $\epsilon$ is the mean energy dissipation rate and $ S_c^* = S (\nu /\epsilon)^{1/2} $. However, \citet{antonia1992isotropy} reported that this condition is too restrictive and can be relaxed with $S_c* \leq 0.2$ for the small scales to be isotropic. \citet{antonia1992isotropy} performed DNS study for channel flow and found a value of $S_c^* = 2.5$ at the wall, and it reduces to a low value for $y^+ > 60$. The $S_c^*$ is plotted as a function of wall-normal position for particle Stokes number of $ 105.47$ and fluid phase Reynolds number of 3300 at different average volume fractions ($\phi_{av}$), which is shown in Fig.~\ref{Sc_plot}(a). It is observed that the $S_c^*$ is 2.53 at the wall for unladen flow, and a decrease of almost one order of magnitude is observed away from the wall. The value of $S_c^*$ increases across the channel width as the particle volume loading is increased. For a particle loading of $9\times 10^{-4}$, $S_c^*$ at the wall is 1.5 times higher than the unladen flow. Spatial averaged $S_c^*$ across the channel ($\langle S_c^* \rangle_s$) for the bulk Reynolds numbers of 3300 and 5600 and different Stokes numbers are shown in Fig.~\ref{Sc_plot}(b). In case of unladen flows, $\langle S_c^* \rangle_s$ is higher for $Re_b = 3300$ compared to $Re_b = 5600$ which indicates that the time scale separation is less at lower Reynolds number. $\langle S_c^* \rangle_s$  increases with an increase in particle loading for both the Reynolds numbers. With an increase in particle inertia ($St$),  $\langle S_c^* \rangle_s$ increases marginally when particle volume fraction is high. This suggests an increase in anisotropy of the small scales for particle-laden cases.

\begin{figure}
	\begin{subfigure}[b]{1\textwidth}
	\minipage{0.5\textwidth}
	\includegraphics[width=\textwidth]{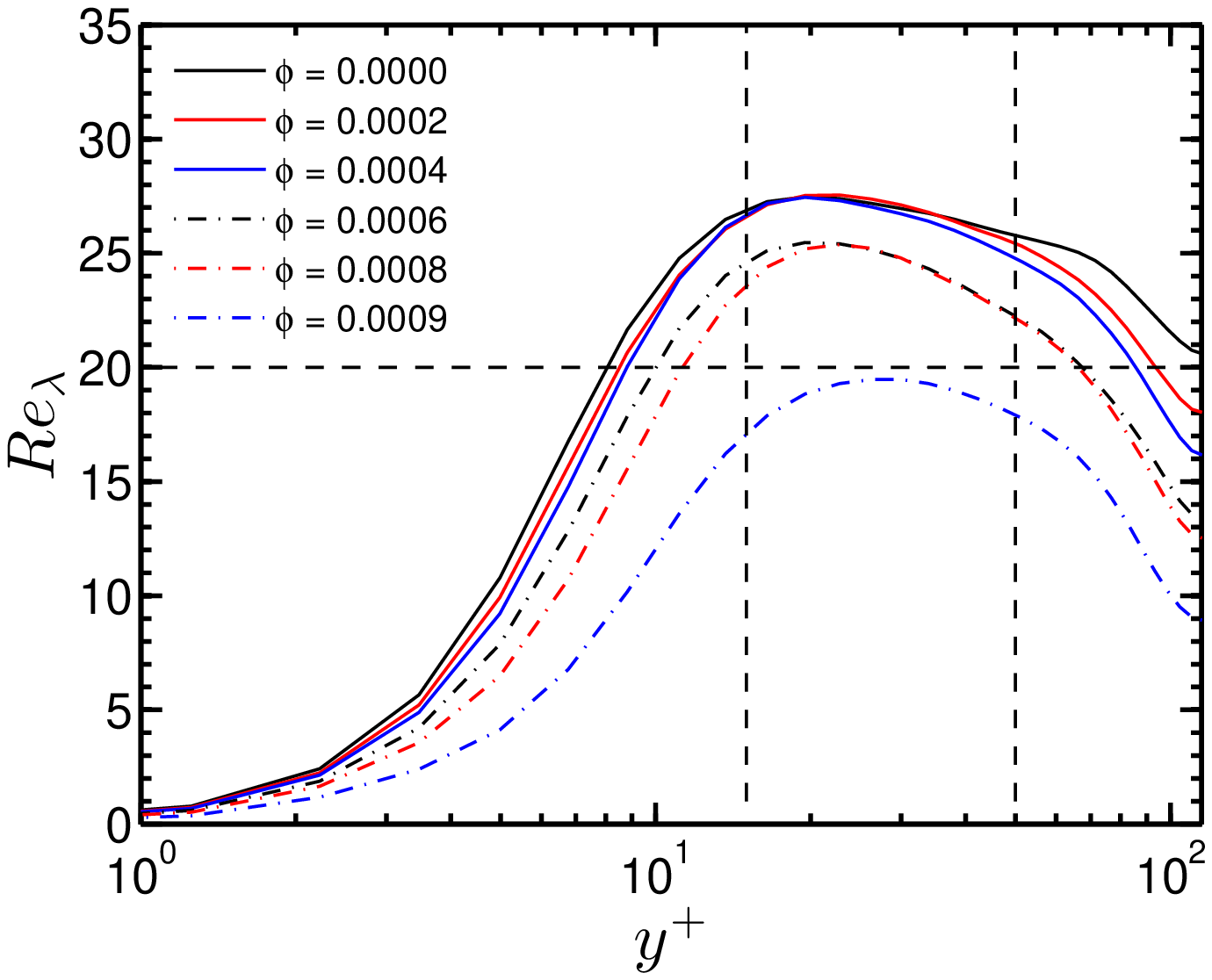}
	\caption{$Re_b = 3300$, St = 105.47}
	\endminipage 
	\minipage{0.49\textwidth}
	\includegraphics[width=\textwidth]{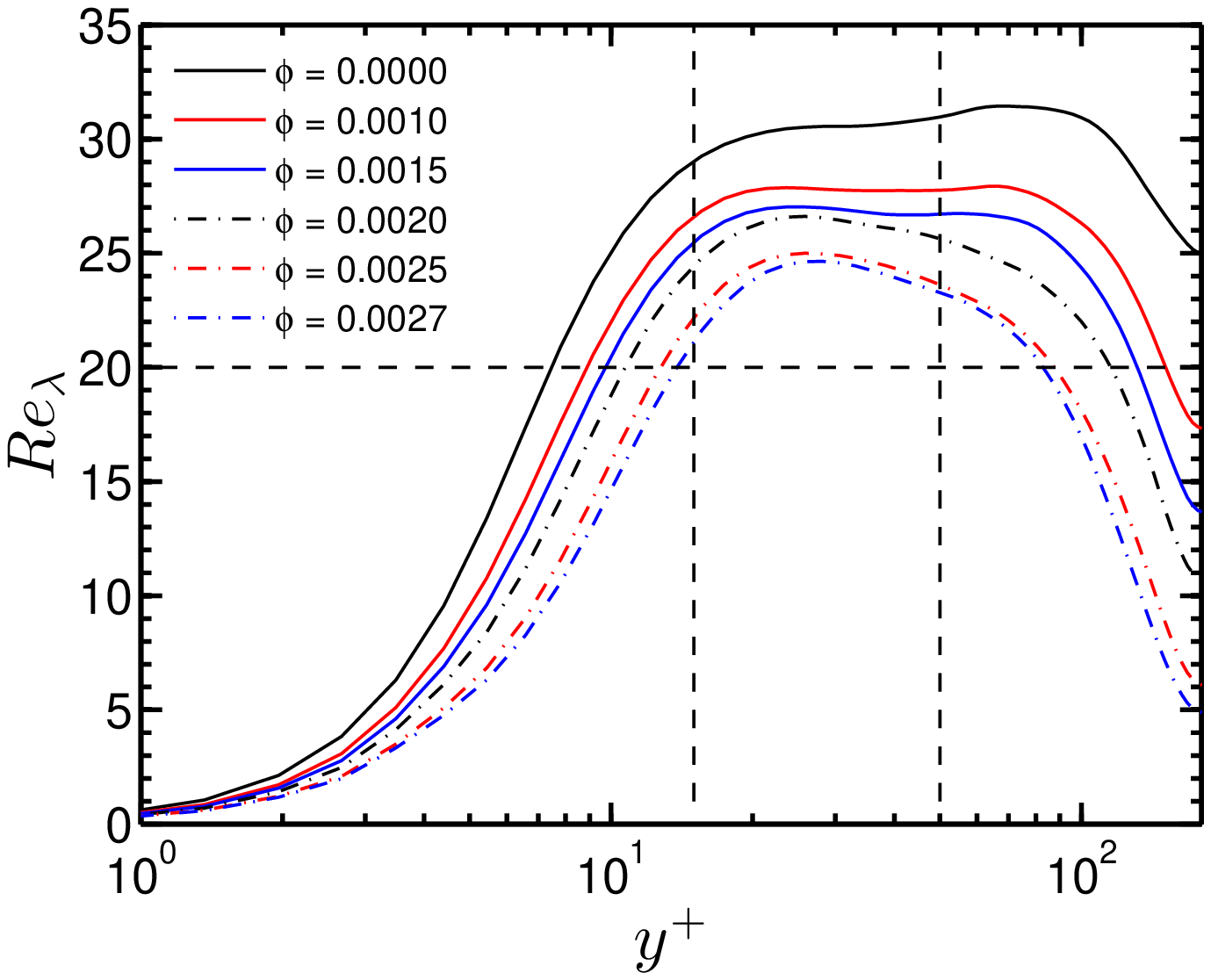}
	\caption{$Re_b = 5600$, St = 210.93}
	\endminipage \\
	\minipage{0.49\textwidth}
	\includegraphics[width=\textwidth]{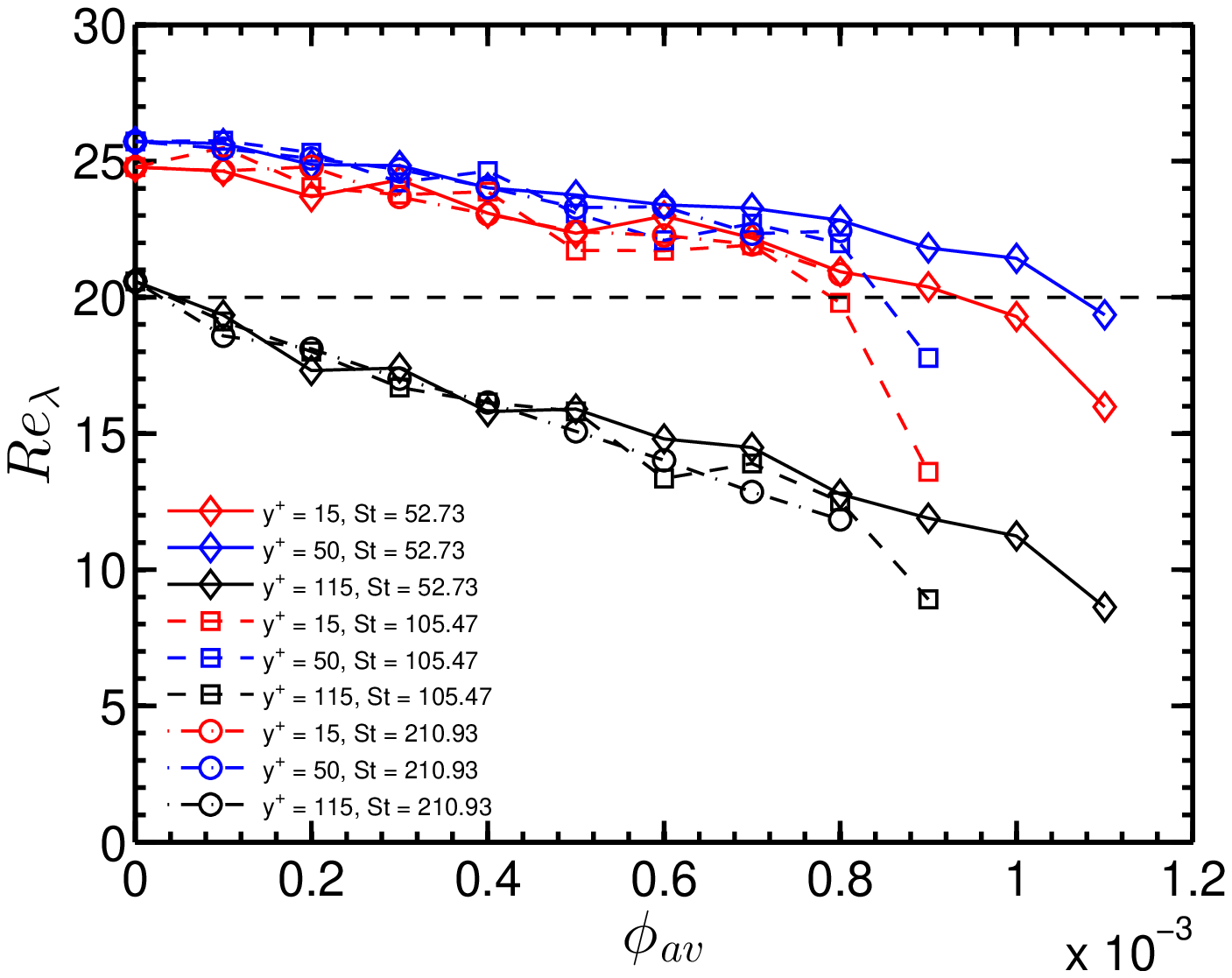}
	\caption{$Re_b = 3300$}
	\endminipage	
	\minipage{0.49\textwidth}
	\includegraphics[width=\textwidth]{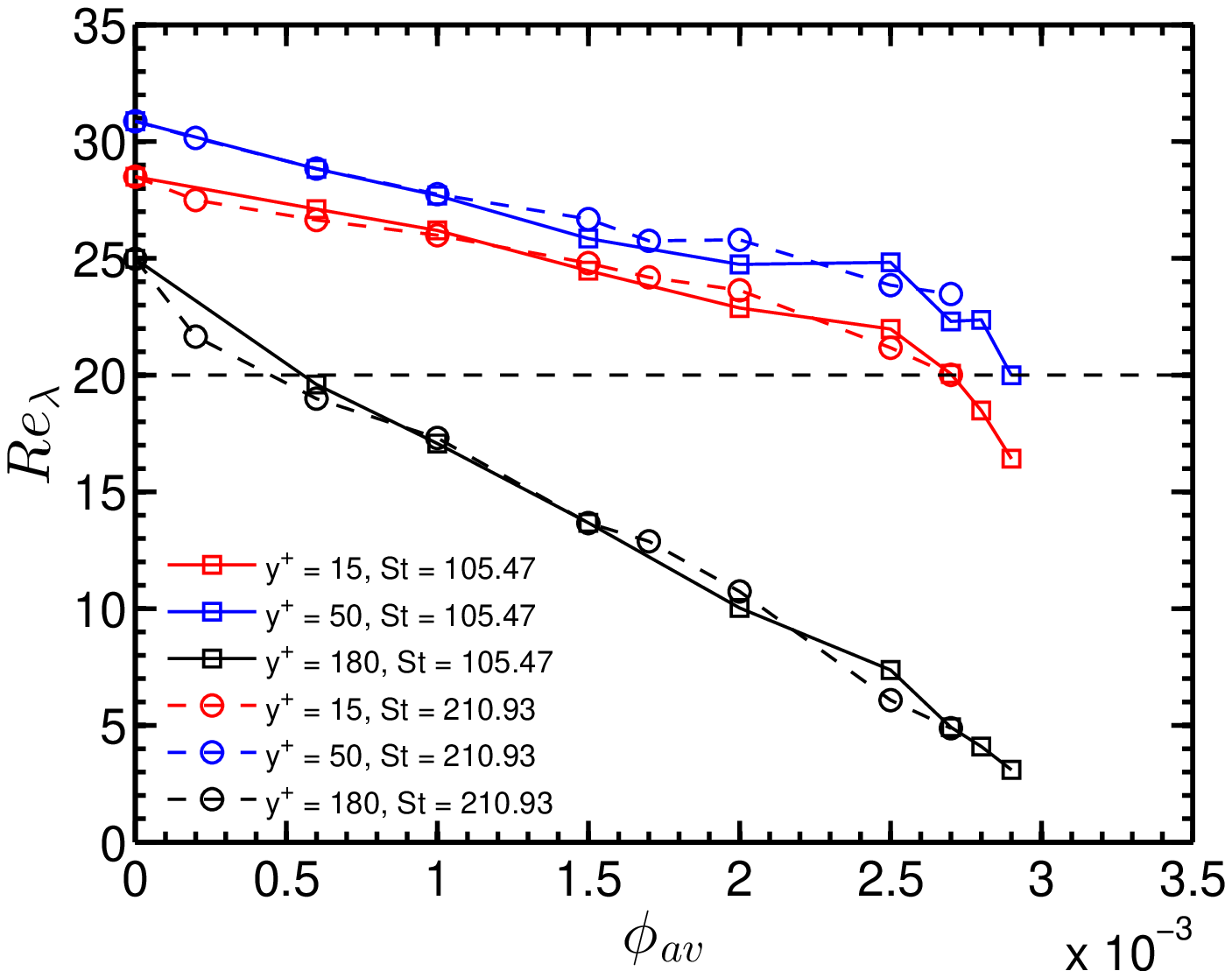}
	\caption{$Re_b = 5600$}
	\endminipage 
	\end{subfigure} 
	\caption{In Fig.~(a and b), the Taylor Reynolds number ($Re_\lambda$) is plotted in the wall-normal direction for a range of volume fractions. In Fig.~(c and d), $Re_\lambda$ at three channel locations and different Stokes numbers is plotted over a range of volume fractions. }
	\label{TaylorRe}
\end{figure}

It is observed from the Fig.~\ref{Sc_plot} that the local isotropy of small scales decreases across the channel with an increase in solid volume fraction. However, it is important to calculate the Taylor Reynolds number ($Re_\lambda$) across the channel flow, which is defined as $Re_\lambda = [(20/3)(k^{+^2}/\epsilon^+)]^{1/2}$ \cite{choi2004lagrangian, yu_2021}. Here, $k^+$ and $\epsilon^+$ are the fluctuating kinetic energy and dissipation rate normalized with viscous scales. The Taylor Reynolds number is plotted for a range of volume fraction at $Re_b = 3300$, St = 105.47 and $Re_b = 5600$, St = 210.93, Fig.~\ref{TaylorRe} (a and b). The Taylor Reynolds number is almost constant away from the wall for the unladen case, and a decrease is observed near the channel center region. The Taylor Reynolds number of 25 was observed in channel center by \citet{yu_2021} for $Re_b = 5746$  which is almost same in our case also for $Re_b = 5600$ unladen case. The $Re_\lambda$ decreases across the channel with an increase in particle volume loading. The horizontal dashed line across the graphs is plotted at $Re_\lambda = 20$, and two dashed vertical lines refer to $y^+ = 15 $ and 50. The three different channel locations are chosen for the assessment of local isotropy. These locations are $y^+ = 15, 50 $ and 115 for $Re_b = 3300$, and are $y^+ = 15, 50 $ and 180 for $Re_b = 5600$. The wall-normal distance ($y^+$) is normalized with unladen frictional velocity and kinematic viscosity. The $y^+ = 15 $ and 50 are chosen in order to have a $Re_\lambda > 20$ for all the considered cases as \citet{sreenivasan1995universality} also has collected the experimental data for $Re_\lambda > 20$ only. The channel center location is also taken for the assessment as the particle affects the turbulence intensities more in the channel center than the near-wall region \cite{kulick1994particle}. The $Re_\lambda$ at these locations for both the Reynolds numbers is plotted in Fig.~\ref{TaylorRe}(c and d) for the considered Stokes numbers. For laden cases, a large decrease in $Re_\lambda$ is observed in the channel center region as the volume fraction is increased. The decrease in $Re_\lambda$ is less and remain above 20 at $y^+ = 15 $ and 50 with an increase in volume loading, except just before CPVL. The Stokes number effect on the $Re_\lambda$ with an increase in particle volume loading is negligible except near the CPVL. 

\begin{figure}
	\begin{subfigure}[b]{1\textwidth}
	\minipage{0.49\textwidth}
	\includegraphics[width=\textwidth]{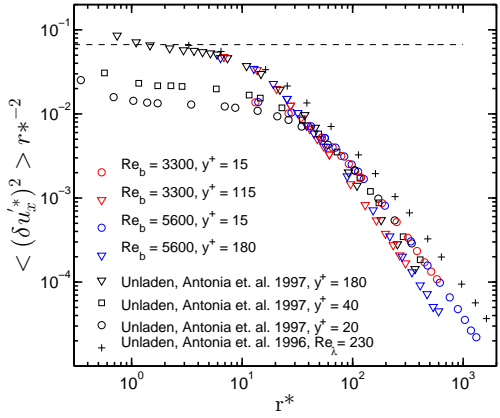}
	\caption{Unladen cases}
	\endminipage 
	\minipage{0.49\textwidth}
\includegraphics[width=\textwidth]{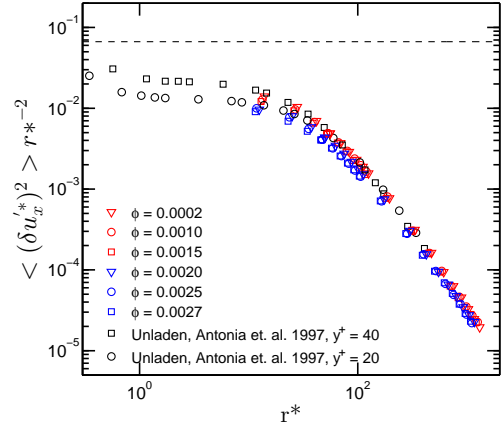}
	\caption{$y^+ = 15$,  $Re_b = 5600$ and St = 210.93}
	\endminipage \\
	\minipage{0.49\textwidth}
	\includegraphics[width=\textwidth]{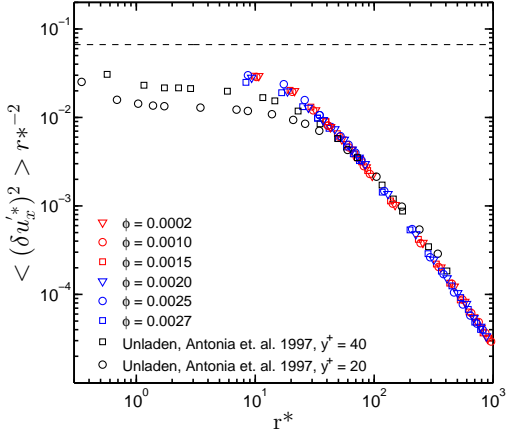}
	\caption{$y^+ = 50$,  $Re_b = 5600$ and St = 210.93}
	\endminipage 
	\minipage{0.49\textwidth}
	\includegraphics[width=\textwidth]{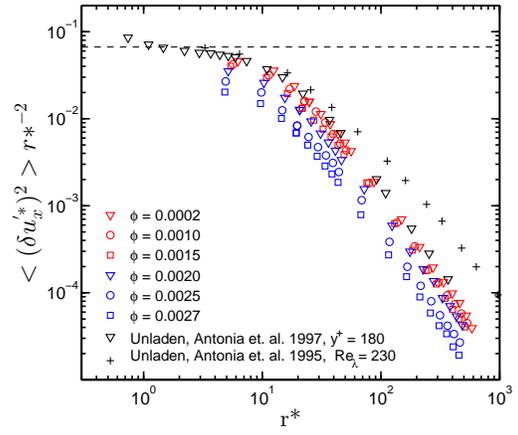}
	\caption{$y^+ = 180$,  $Re_b = 5600$ and St = 210.93}
	\endminipage 
	\end{subfigure} 
	\caption{ The second-order velocity structure functions multiplied with $r^{*-2}$ is plotted for (a) unladen cases, and for a range of volume fractions ($\phi$) at (b) $y^+ = 15$, (c) $y^+ = 50$, and (d) $y^+ = 180$ for  $Re_b = 5600$ and St = 210.93. The dashed black line is ($1/15$) ordinate. } 
	\label{u2r2}
\end{figure}

The local isotropy at the small scales can also be checked using the expressions of the second-order velocity structure function and the mean energy dissipation rate for the homogeneous isotropic turbulence. In the limit of small $r$, the moment of longitudinal velocity fluctuation is defined with the following expression \cite{kolmogorov1941energy}
\begin{equation}
\langle (\delta u'_x)^2\rangle r^{-2} =  \left(\frac{\partial u'_x}{\partial x}\right)^2,
\end{equation}
where `$ r $' is the distance between the two points and $\delta u'_x = u'_x(x+r) - u'_x(x)$ with $u'_x$ being the longitudnal fluctuation. The mean energy dissipation($\epsilon$) for homogeneous isotropic turbulence \cite{pope2001turbulent} is expressed as 
\begin{equation}
\epsilon = 15 \nu \left(\frac{\partial u'_x}{\partial x}\right)^2.
\end{equation}
 Using the above two equations for local isotropy of dissipation range, the following relation holds
\begin{equation}
\langle (\delta u^{'*}_x)^{2}\rangle (r^*)^{-2} =  \frac{1}{15}
\label{u2r2_dissipation}.
\end{equation}
Here, `(*)' denote the non-dimensionalized quantities. $\delta u'_x$ is normalized using Kolmogorov velocity scale, $u_k = (\nu \epsilon)^{1/4}$, and $r$ is normalized with Kolmogorov length scale, $\eta = \nu^{3/4}/\epsilon^{1/4}$. Using DNS, we have computed $\langle (\delta u^{'*}_x)^{2}\rangle (r^*)^{-2}$ and compared the unladen cases with Eqn.~\ref{u2r2_dissipation} and also with the experimental data reported by \citet{antonia1997second} for both the Reynolds numbers as shown in Fig.~\ref{u2r2}(a). A good agreement is observed for different wall-normal locations and Reynolds numbers. Fig.~\ref{u2r2} (a) shows that for both Reynolds numbers, the channel center location is nearly isotropic for unladen flows. However, the near-wall region  deviates from the isotropic condition, which has been reported in experiments~\cite{antonia1997second}. In Figs.~\ref{u2r2}~(b-d), we present the effect of particle volume loading on local isotropy at three different channel locations for St = 210.93 and $Re_b = 5600$. It is observed that the deviation from the local isotropy increases with an increase in particle volume loading at $y^+ = 15$ and 180. There is negligible change in the profiles at $y^+ = 50$ (Fig. \ref{u2r2} (c)) in spite a very similar decrease in $Re_\lambda$ at both $y^+ = 15$ and 50, shown in Fig.~\ref{TaylorRe} (c and d). Similar observations are followed for Reynolds number of 3300, which are not shown here for brevity. The isotropy of the inertial range is unlikely to be achieved if there is a deviation from isotropy at the small scales~\cite{antonia1997second}. Thus, it is expected that the decrease in the local isotropy of small scales with an increase in particle volume loading will affect the local isotropy at inertial range and the Kolmogorov constant. Here, we examine that effect via second-order velocity structure-function and compensated spectra. The reduction in isotropy for particle-laden flows is related to the extent of attenuation of the different components of fluid velocity fluctuations. It is observed that the decrease in the transverse velocity component is more than the streamwise component~\cite{KulickJD1994, richter2013momentum, gualtieri2013clustering, richter2015turbulence, zhao2010turbulence, rohilla2022applicability} as the particle volume fraction is increased. \citet{richter2013momentum} mentioned that inertial particles reduce the ability of the carrier phase to transfer the momentum flux in a wall-normal direction. This increased anisotropy may affect the Kolmogorov constant for particle-laden cases, which is analyzed hereafter. 

\begin{figure}[htb]
	\begin{subfigure}[b]{1\textwidth}
	\minipage{0.52\textwidth}
	\includegraphics[width=\textwidth]{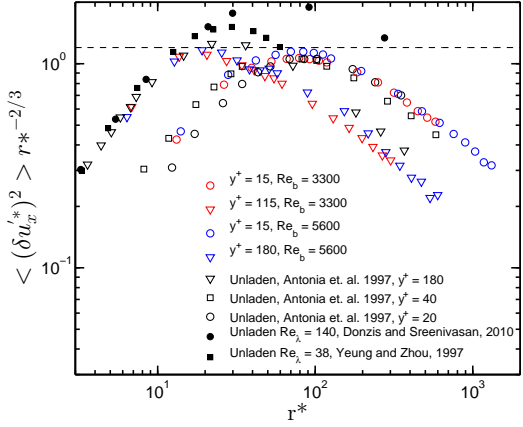}
	\caption{Unladen cases}
	\endminipage 
	\minipage{0.51\textwidth}
	\includegraphics[width=\textwidth]{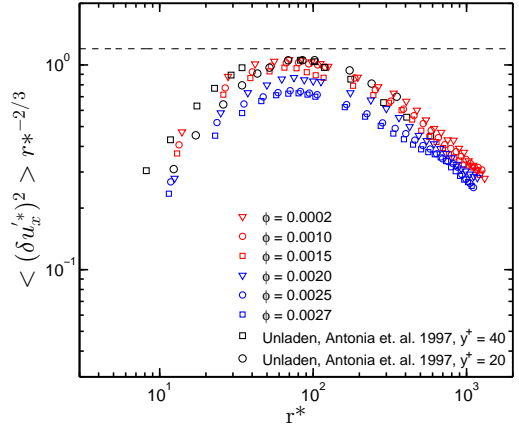}
	\caption{$y^+ = 15$, $Re_b = 5600$ and St = 210.93}
	\endminipage \\
	\minipage{0.52\textwidth}
	\includegraphics[width=\textwidth]{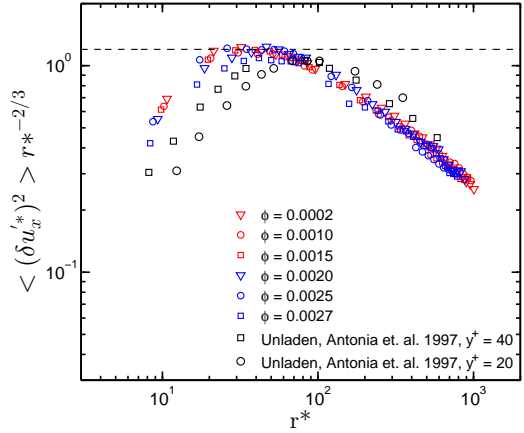}
	\caption{$y^+ = 50$, $Re_b = 5600$ and St = 210.93}
	\endminipage 
	\minipage{0.53\textwidth}
	\includegraphics[width=\textwidth]{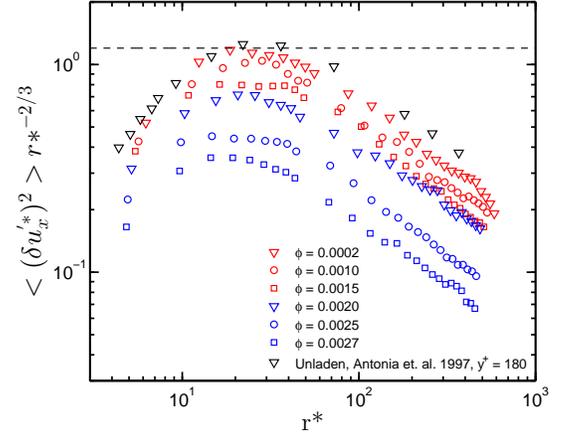}
	\caption{$y^+ = 180$, $Re_b = 5600$ and St = 210.93}
	\endminipage \\
	\end{subfigure} 
	\caption{ The second-order velocity structure functions multiplied with $r^{*-2/3}$ is plotted for (a) unladen cases, and for a range of volume fractions ($\phi$) at (b) $y^+ = 15$, (c) $y^+ = 50$, and (d) $y^+ = 180$ for  $Re_b = 5600$ and St = 210.93.  The dashed black line is at 1.2 ordinate. }
	\label{peak_alpha}
\end{figure}

The second-order velocity structure function in the inertial range is defined as~\cite{kolmogorov1941energy}, 
\begin{equation}
\langle (\delta u'_x)^2 \rangle = C_2 (\epsilon r)^{2/3},
\label{str_fun}
\end{equation}
where $\epsilon$ is the mean viscous dissipation rate and, $\delta u'_x = u'_x(x+r) - u'_x(x)$, with $u'_x$ being the longitudinal fluctuations. The $r$ is described as $\eta \ll r \ll L$, with $L $ as the integral length scale. In the above expression, $C_2$ is the Kolmogorov constant, and angular brackets denote the time averaging. An equivalent relation in terms of scaled variables can be written as $\langle (\delta u^{'*}_x)^2 \rangle = C_2 (r^*)^{2/3}$. The second-order velocity structure-function for unladen case is plotted in Fig.~\ref{peak_alpha} using the scaled form of Eqn.~\ref{str_fun}. The ($^*$) denotes the scaled variables. To validate our results, the profiles are plotted for both the Reynolds numbers at two locations, one in the near-wall region ($y^+ = 15$) and the other at the channel center (at $y^+ = 180$ for $Re_b = 5600$ and at $y^+ = 115$ for $Re_b = 3300$). Fig.~\ref{peak_alpha} (a) shows the validation for unladen flows against the experimental data for channel flow reported by \citet{antonia1997second}, and simulation data for isotropic turbulence reported by Ref.~\cite{yeung1997universality, donzis2010bottleneck}. There is a good agreement between the experimental data and the present DNS results for both the channel locations and Reynolds numbers. The simulation data of isotropic turbulence \cite{yeung1997universality, donzis2010bottleneck} predict a higher value of $C_2$ due to higher $Re_\lambda$ used in those studies. In Fig.~\ref{peak_alpha}, the peak value of $C_2$  (from plateau in the inertial range where statistical properties are only dependent on mean energy dissipation rate) is considered as Kolmogorov constant~\cite{sreenivasan1995universality,donzis2010bottleneck,sreenivasan1997phenomenology}. The majority values of $C_2$ (peak value of the second-order velocity structure-function) reported in the literature are 2 or more  \cite{choi2004lagrangian, sawford2011kolmogorov, saddoughi1994local, antonia1996comparison, yeung1997universality, donzis2010bottleneck}. However, a lower value of $C_2$ occurs due to lower Reynolds number \cite{sreenivasan1995universality} or if there is a deviation of isotropy in dissipation \cite{antonia1997second} and inertial-range \cite{yeung1997universality}.

\begin{figure}
	\begin{subfigure}[b]{1\textwidth}
	\minipage{0.49\textwidth}
	\includegraphics[width=\textwidth]{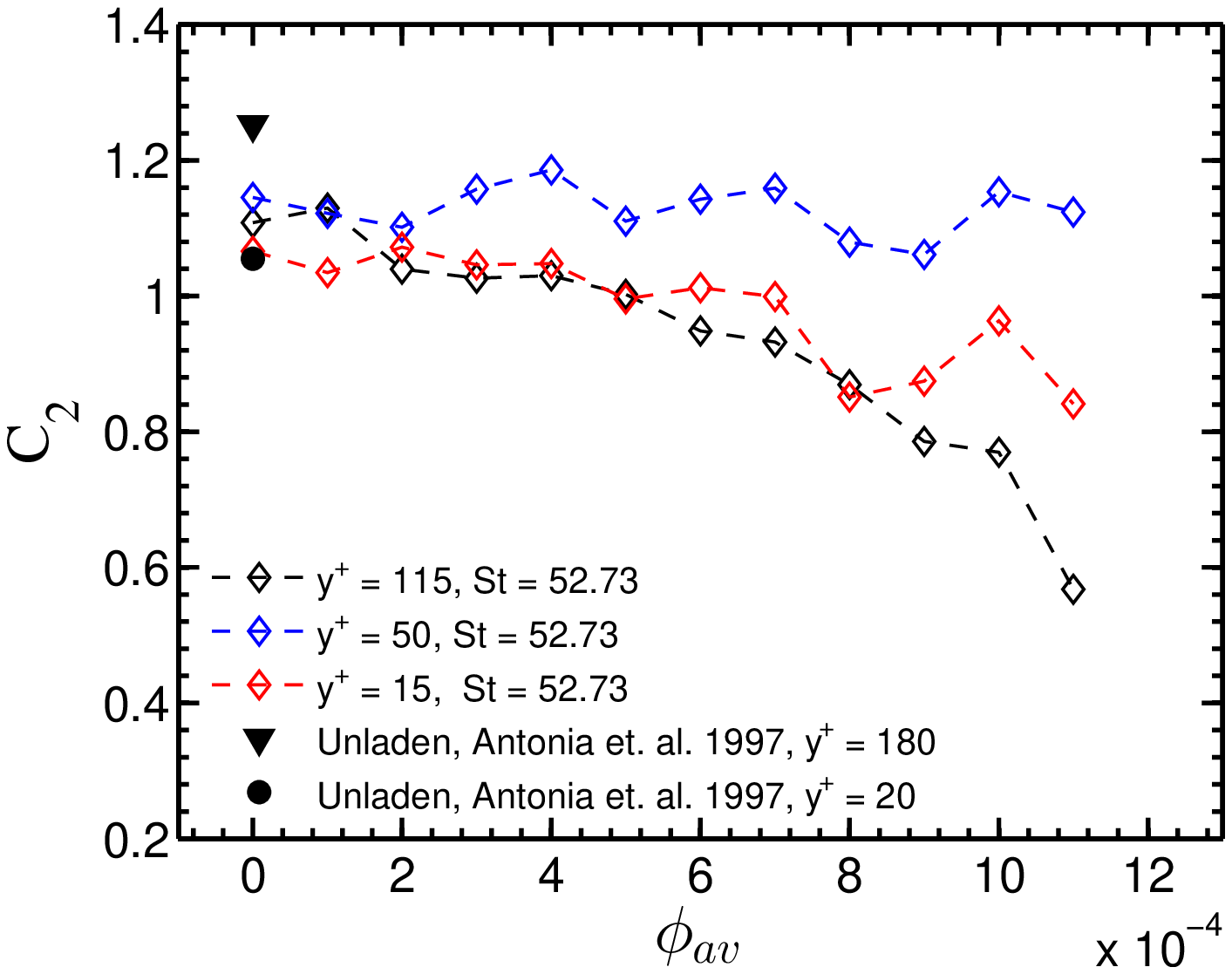}
	\caption{$Re_b = 3300$}
	\endminipage 
	\minipage{0.49\textwidth}
	\includegraphics[width=\textwidth]{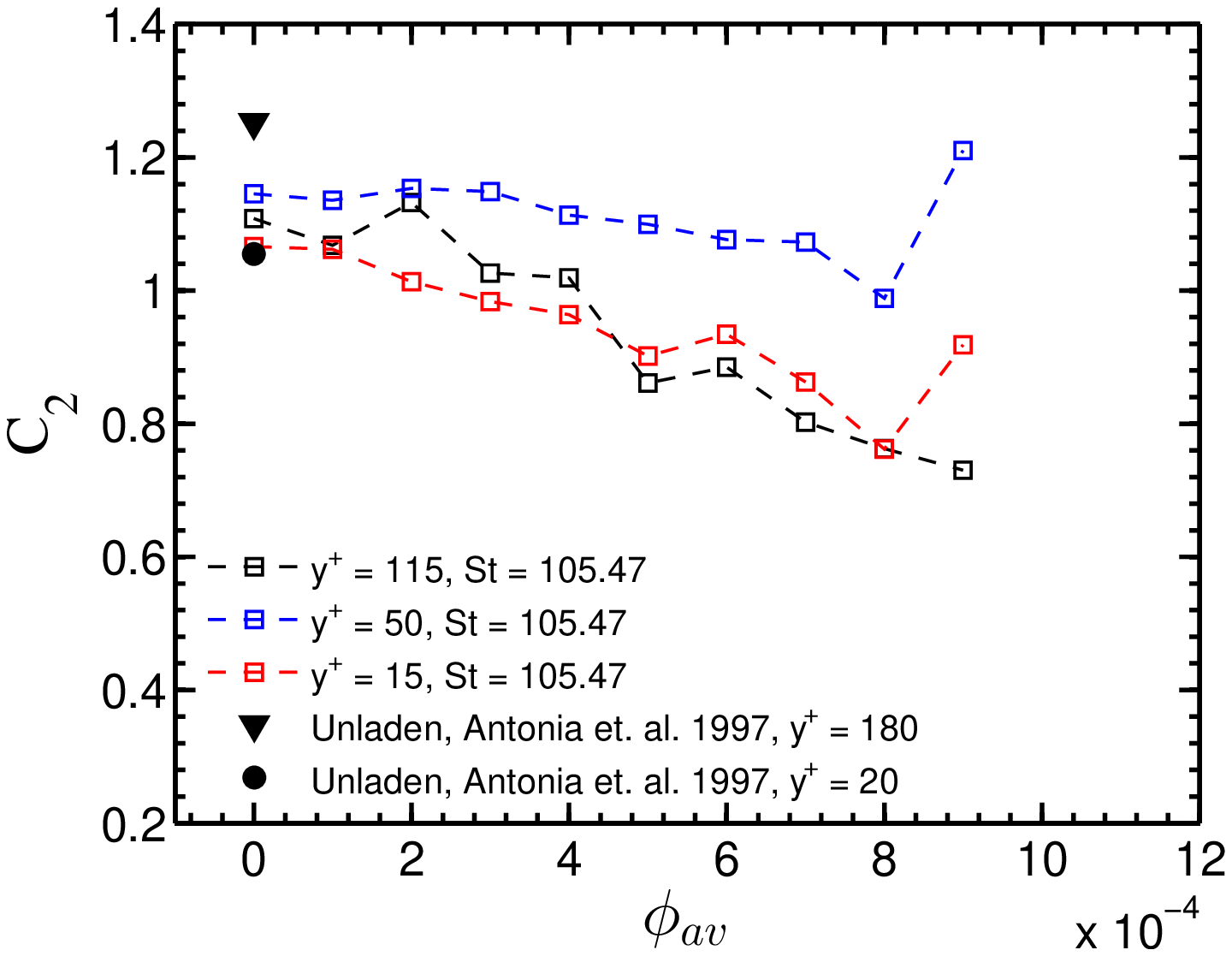}
	\caption{$Re_b = 3300$}
	\endminipage \\
	\minipage{0.49\textwidth}
	\includegraphics[width=\textwidth]{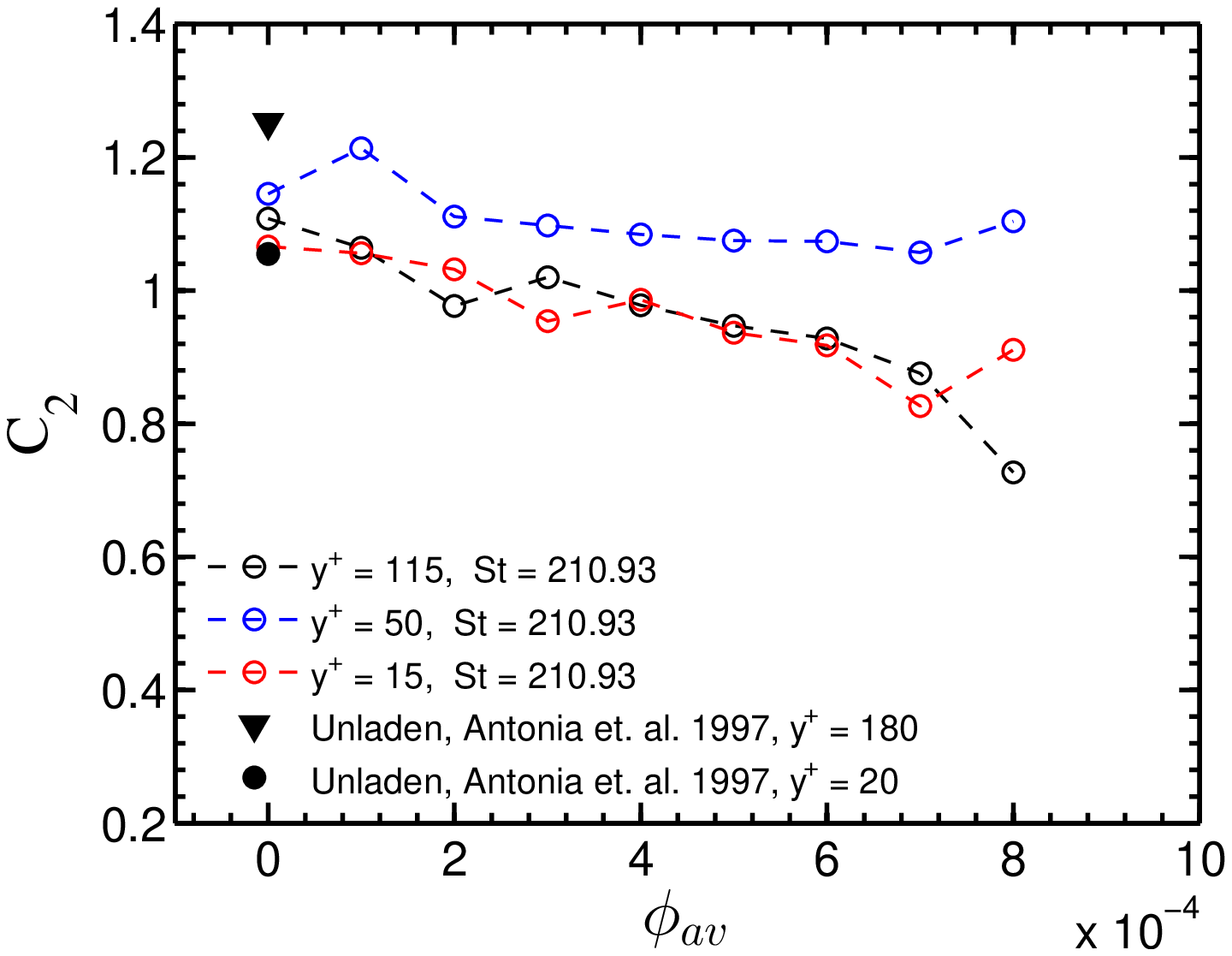}
	\caption{$Re_b = 3300$}
	\endminipage 
	\minipage{0.49\textwidth}
	\includegraphics[width=\textwidth]{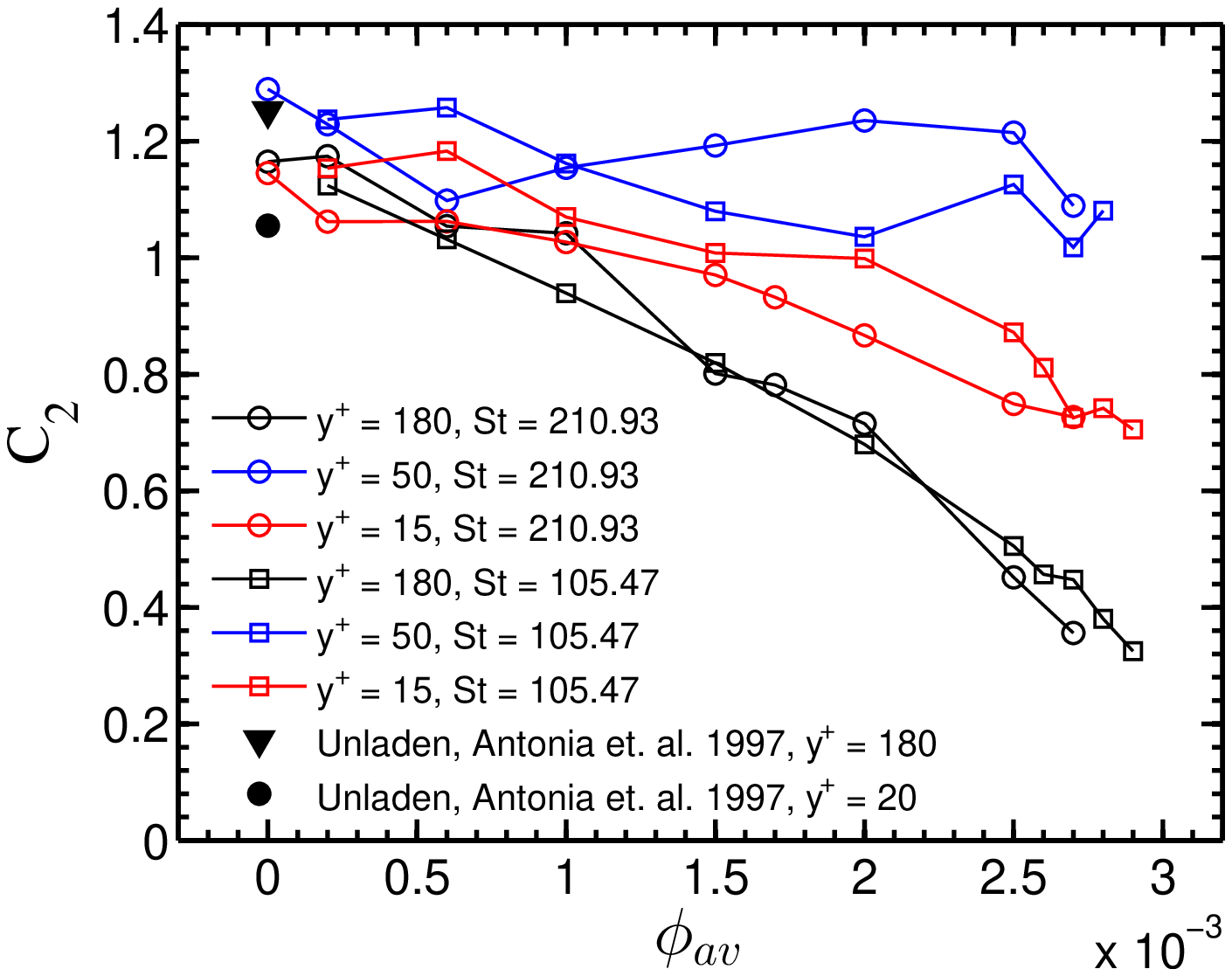}
	\caption{$Re_b = 5600$}
	\endminipage 
	\end{subfigure} 
	\caption{The peak of second-order velocity structure-function (from Fig.~\ref{peak_alpha}) is plotted at different channel locations as a function of solid volume fractions for (a) $Re_b = 3300$ and $St = 52.73$, (b)  $Re_b = 3300$ and $St = 105.47$,  (c)  $Re_b = 3300$ and $St = 210.93$, (d) $Re_b = 5600$ with $St = 105.47 $ and 210.93.}
	\label{alpha_3300}
\end{figure}

\begin{figure}[htb]
	\begin{subfigure}[b]{1\textwidth}
	\minipage{0.49\textwidth}
	\includegraphics[width=\textwidth]{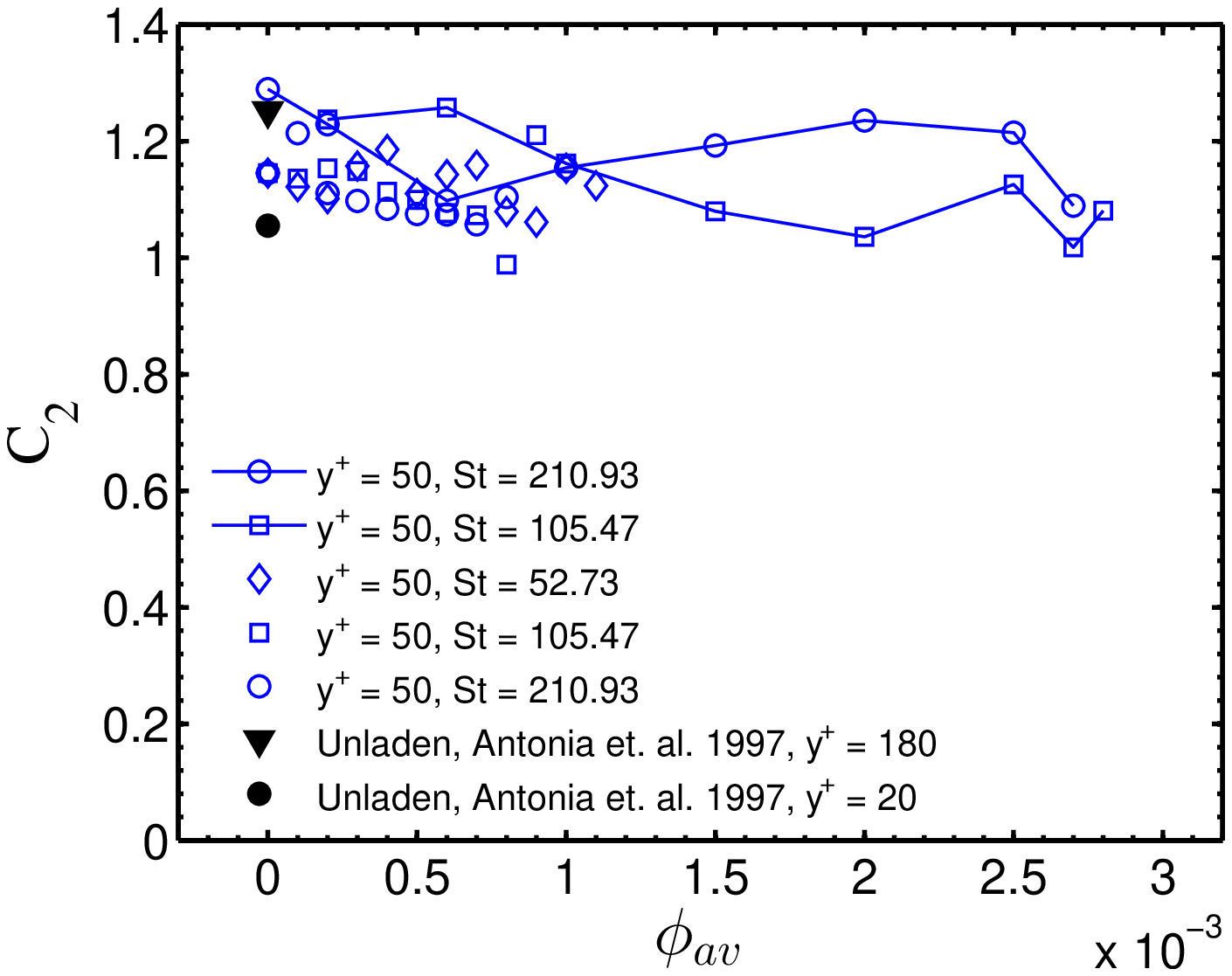}
	\caption{}
	\endminipage 
	\minipage{0.49\textwidth}
	\includegraphics[width=\textwidth]{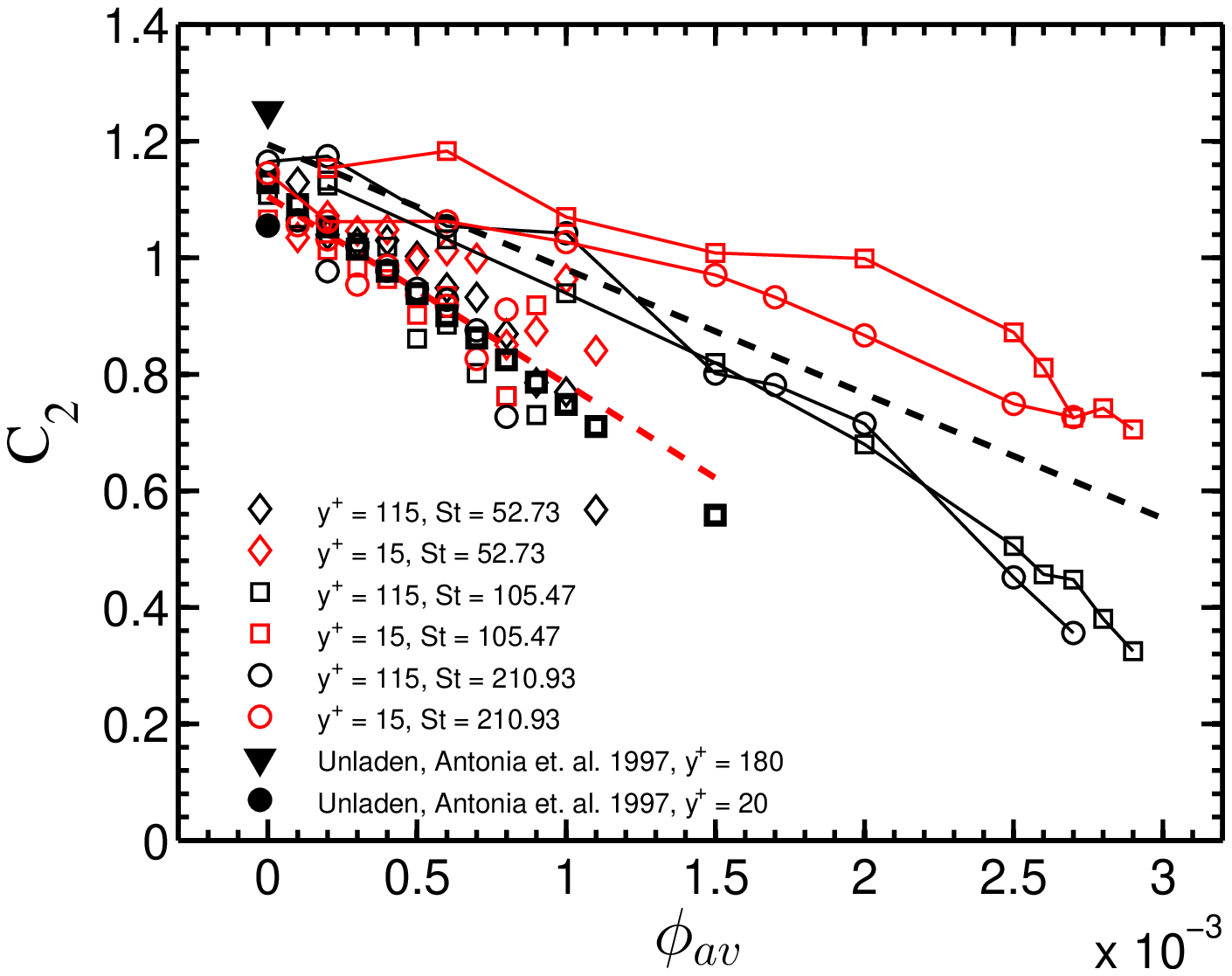}
	\caption{}
	\endminipage 
	\end{subfigure} 
	\caption{The peak of second-order velocity structure-function (from Fig.~\ref{peak_alpha}) is plotted at $Re_b = 3300$ and 5600 for different volume fraction cases at (a) $y^+ = 50 $, (b) $y^+ = 15 $ and channel center. The symbols with lines are for $Re_b = 5600$ and symbols without lines are for $Re_b = 3300$. The symbols for $Re_b = 5600$ in Fig.~(b) are same as in Fig.~\ref{alpha_3300}(d). The black and red dashed lines in Fig.~(b) are the fitting curves for $Re_b = 5600$ and $Re_b = 3300$, respectively. }
	\label{alpha_variation}
\end{figure}

\begin{figure}[htb]
	\begin{subfigure}[b]{1\textwidth}
	\minipage{0.49\textwidth}
	\includegraphics[width=\textwidth]{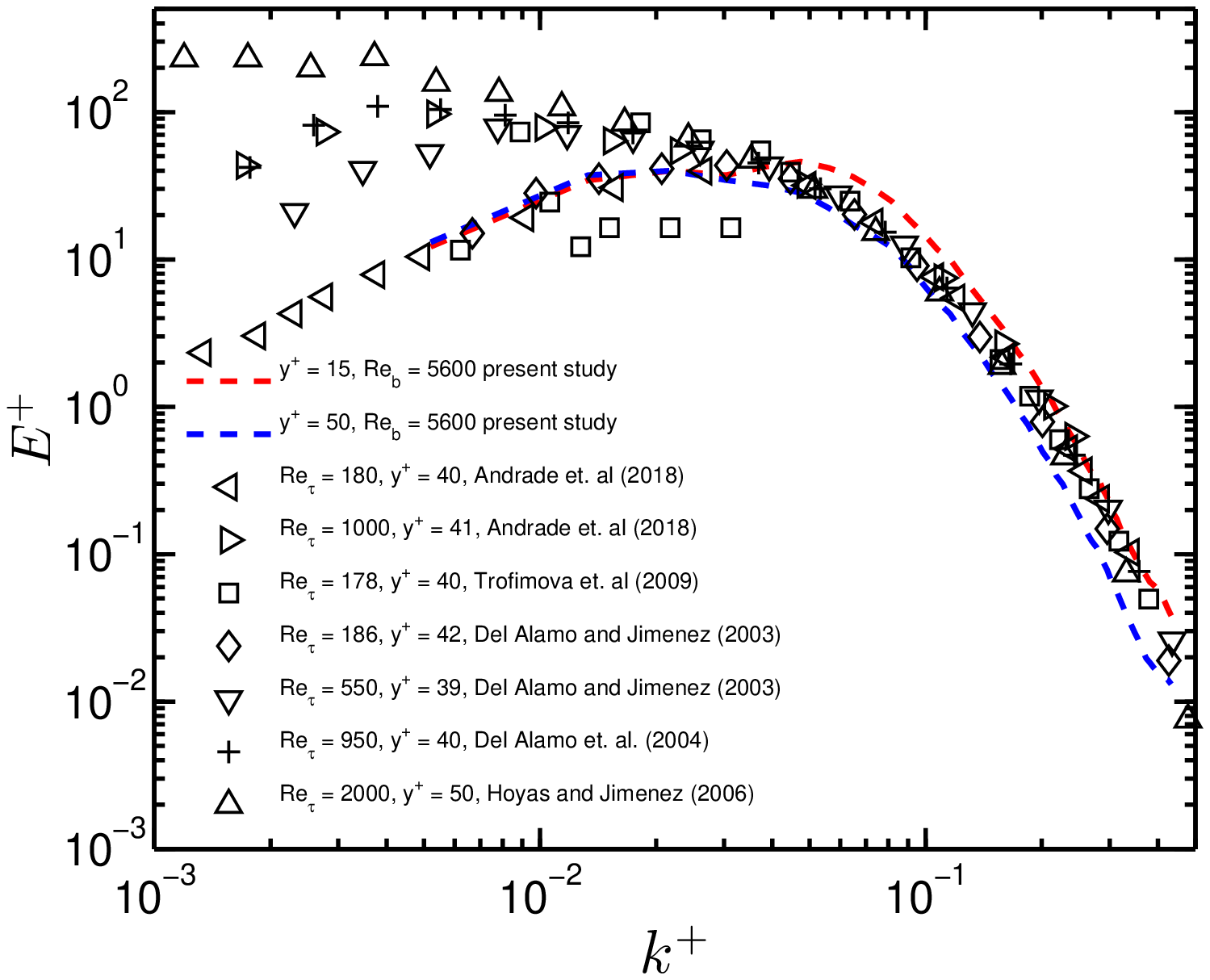}
	\caption{}
	\endminipage 
	\minipage{0.49\textwidth}
	\includegraphics[width=\textwidth]{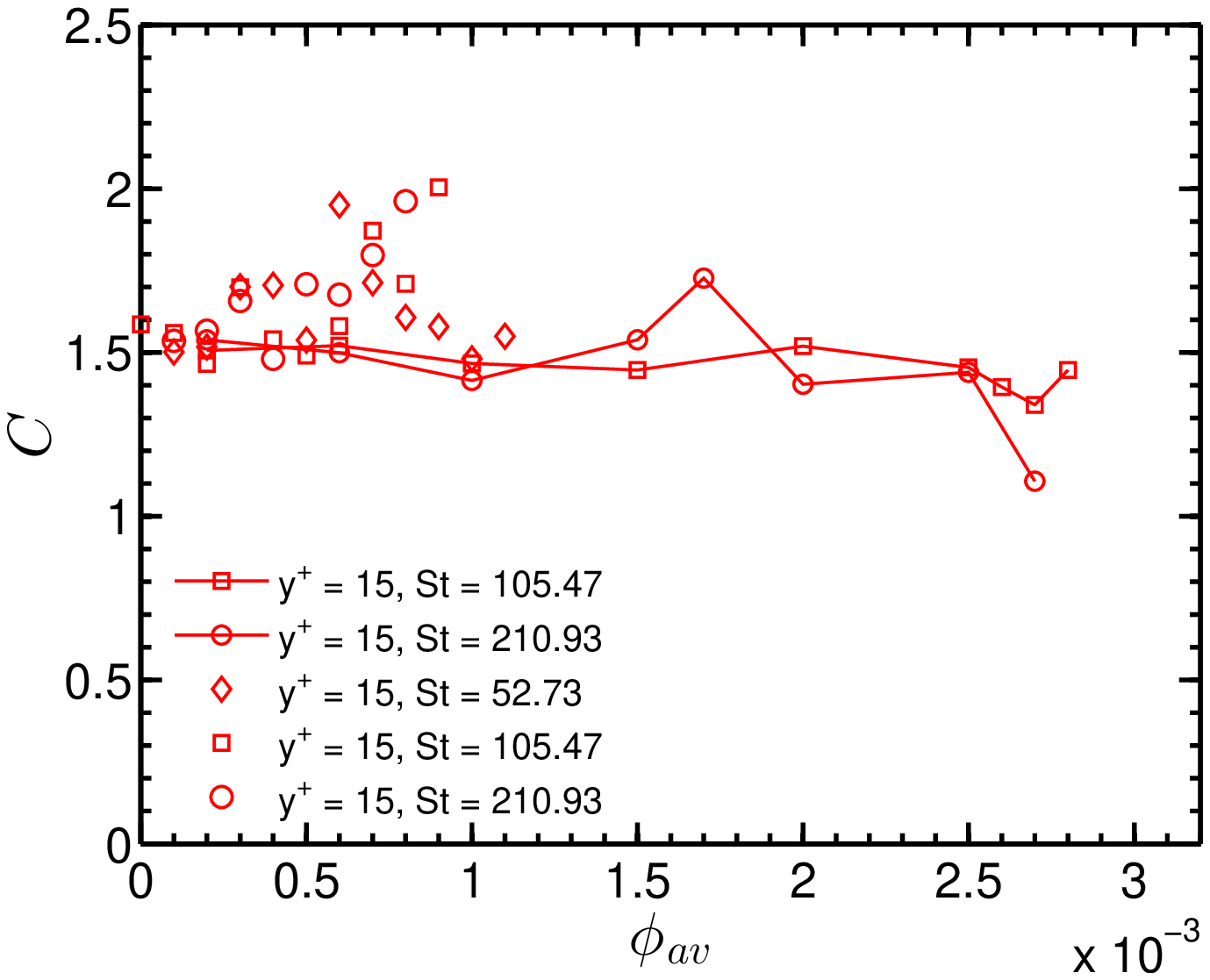}
	\caption{}
	\endminipage \\
	\minipage{0.49\textwidth}
	\includegraphics[width=\textwidth]{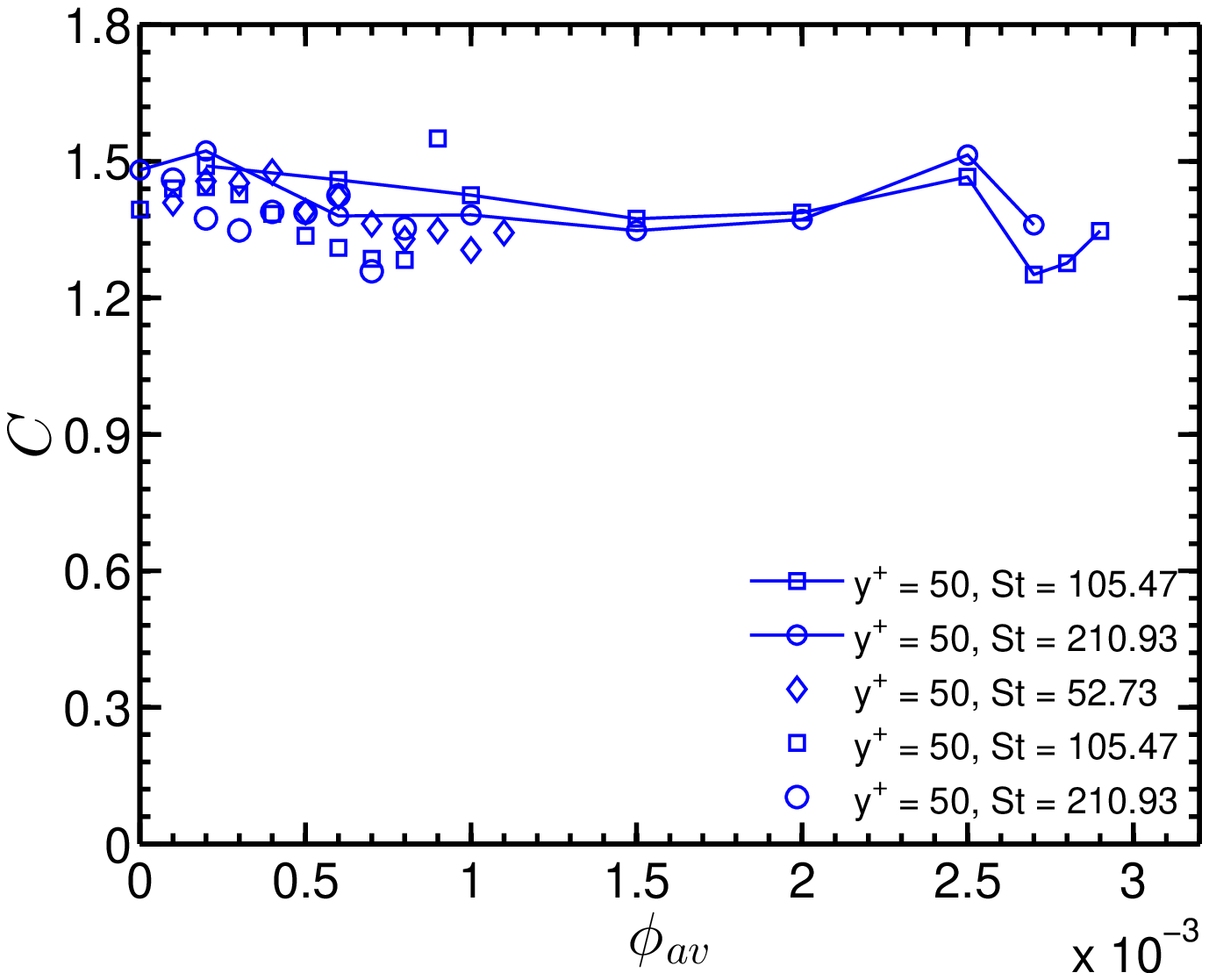}
	\caption{}
	\endminipage 
	\minipage{0.49\textwidth}
	\includegraphics[width=\textwidth]{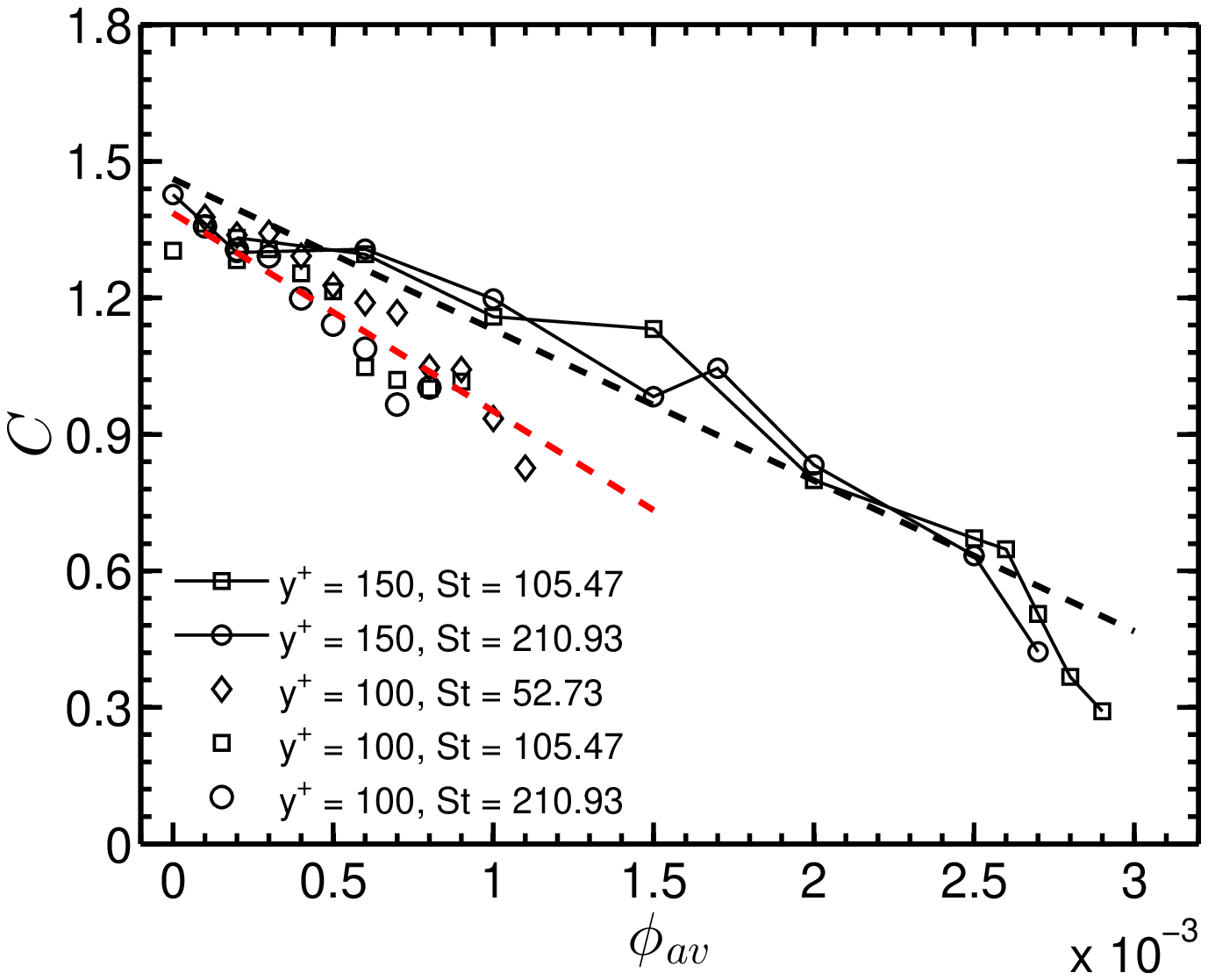}
	\caption{}
	\endminipage 
	\end{subfigure} 
	\caption{(a) Energy spectrum density at $y^+ = 15$ and 50 for $Re_b = 5600$ compared with different DNS results for the unladen cases. The Kolmogorov constant ($C$) plotted from compensated spectra for $Re_b = 3300$ and $5600$ at channel locaitions of (b) $y^+ = 15$, (c) $y^+ = 50$, and (d) at $y^+ = 100$ for $Re_b = 3300$ and  at $y^+ = 150$ for $Re_b = 5600$. In Fig.~(b - d), the symbols with lines are for $Re_b = 5600$, the symbols without the lines are for $Re_b = 3300$.  The black and red dashed lines in Fig.~(d) are the fitting curves for $Re_b = 5600$ and $Re_b = 3300$, respectively. }
	\label{Compensated_spectra}
\end{figure}

\begin{figure}[htb!]
	\begin{subfigure}[b]{1\textwidth}
	\minipage{0.49\textwidth}
	\includegraphics[width=\textwidth]{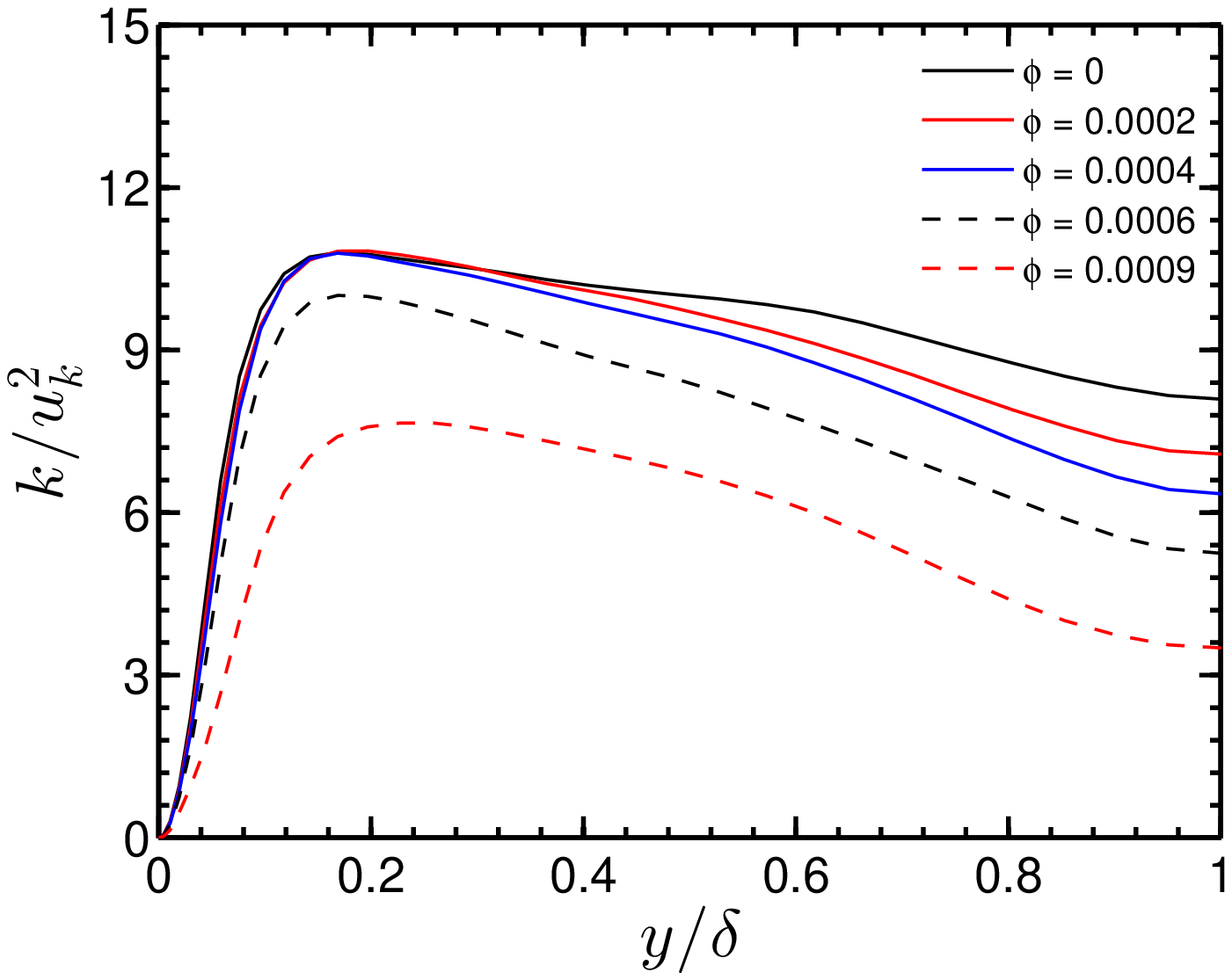}
	\caption{$Re_b = 3300$, St = 105.47  }
	\endminipage 
	\minipage{0.49\textwidth}
	\includegraphics[width=\textwidth]{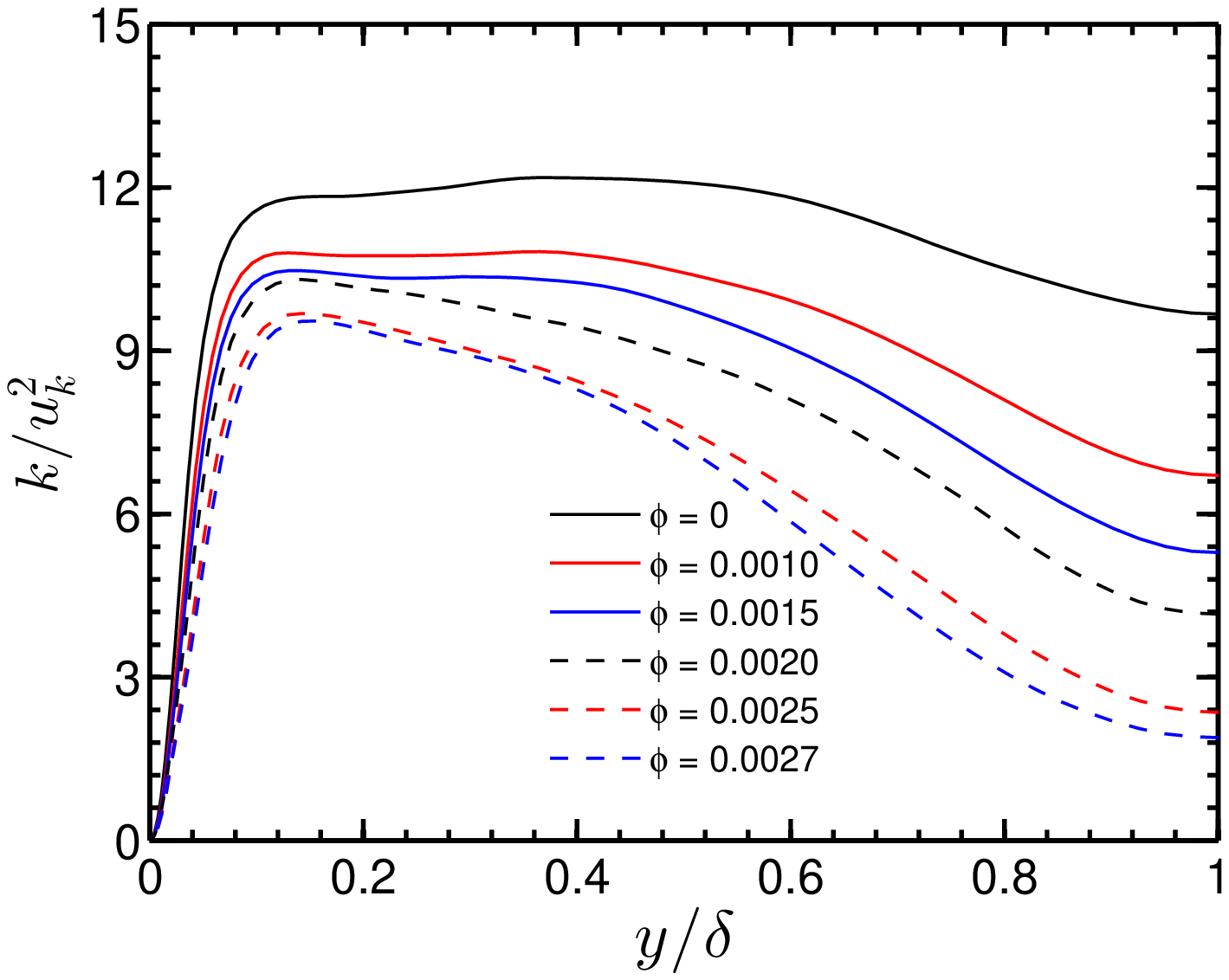}
	\caption{$Re_b = 5600$, St = 210.93 }
	\endminipage 	
	\end{subfigure} 
	\caption{The ratio of fluctuating turbulent kinetic energy ($k$) and square of Kolmogorov velocity scale ($u_k$) plotted across the channel width for different volume fractions.}
	\label{ux_uk}
\end{figure}

In Fig.~\ref{peak_alpha}~(b - d), the second-order velocity structure function profiles are shown as a function of different particle volume loadings at three-channel locations and Reynolds number of 5600 with particles St = 210.93. It is observed that at $y^+ = 15$  and 180 \cite{rohillakolmogorov}, there is a decrease in value of the scaled second-order velocity structure-function ($\langle (\delta u^{'*}_x)^2 \rangle r^{*-2/3}$) at all the $r^*$ locations for an increase in volume fraction. The peak value decreases nearly to 0.4 for $\phi = 0.0027$ from 1.2 for the unladen flow, which indicates that the two-point correlation of fluid velocity fluctuation becomes weaker with an increase in solid volume loading, shown in Fig.~\ref{peak_alpha}(d). At $y^+ = 15 $, with an increase in solid volume fraction a maximum of 25\% decrease in Taylor Reynolds number is observed, Fig.~\ref{TaylorRe}. Therefore, the decrease in the peak value of the second-order structure function ($\langle (\delta u^{'*}_x)^2 \rangle r^{*-2/3}$) may be due to the increase in anisotropy across the channel width. However, no significant change is observed in the second-order structure-function at $y^+ = 50 $ location, Fig.~\ref{peak_alpha}(b).

In Fig.~\ref{alpha_3300}, the peak values ($C_2$) of the second-order velocity structure-functions are plotted for three-channel locations, $y^+ = 15$, 50 and at the channel center (at $y^+ = 180$ for $Re_b = 5600$ and at $y^+ = 115$ for $Re_b = 3300$), for a range of solid volume fraction ($\phi_{av}$). In Fig.~\ref{alpha_3300}(a - d), the $C_2$ is plotted for both the Reynolds numbers and Stokes numbers given in Table~\ref{table Stokes}. It is observed that the Kolmogorov constant ($C_2$) decreases significantly with an increase in volume loading at the near-wall ($y^+ = 15$) and the channel center for both the Reynolds numbers and Stokes numbers considered here. 
Also, It is observed that the effect of particle loading to reduce  $C_2$ is more significant in the channel center location compared to the near-wall region ($y^+ = 15$) of the channel. Interestingly, a crossover occurs as the volume loading is increased, and the $C_2$ value becomes lower at the channel center than the near-wall region for all the Stokes numbers. The larger decrease of $C_2$ at the channel center is due to the larger decrease in $Re_\lambda$ at this position (Fig.~\ref{TaylorRe} (c and d)) and an associated increase in the anisotropy. The change in $C_2$ for the $y^+ = 50$ is non-monotonic and remain almost constant. 

Figure~\ref{alpha_variation} presents a consolidated picture of $C_2$ at different channel locations for all the Stokes, Reynolds numbers, and volume fractions. For $y^+ = 50$, shown in Fig.~\ref{alpha_variation}(a), the variation of $C_2$ with an increase in volume loading is non-monotonic with an insignificant variation over the range of volume fraction. A linear decrease in $C_2$ with volume fractions is observed at $y^+ = 15$ and at the channel center as shown in Fig.~\ref{alpha_variation}(b). The observations suggest that the Kolmogorov constant is a function of particle volume fraction ($\phi$), and wall-normal location for particle-laden turbulent flows at low Reynolds number. The effect of Stokes number on the variation of the Kolmogorov constant seems insignificant for the range reported here. Thus, from the analysis of the second-order structure function, it is observed that the Kolmogorov constant remains unaffected in the initial part of the log-law regime, and a decrease is observed close to the wall and in the channel center with an increase in particle volume loading. It is interesting to note that the value of the $Re_\lambda$, and variation in $Re_\lambda$  with an increase in particle loading is almost similar at the $y^+ = 15$ and 50 for both the Reynolds numbers (Fig.~\ref{TaylorRe} (c and d)), there is almost no change in the $C_2$ at $y^+ = 50$ as shown in Fig.~\ref{alpha_variation}(a). However, a significant decrease in $C_2$ is observed at $y^+ = 15$. It is worth noting that for unladen flow, the $Re_\lambda$ is 25 in the channel center for $Re_b = 5600$, and a similar $Re_\lambda$ occur for near-wall region for $\phi = 1.5\times 10^{-3}$, Fig.~\ref{TaylorRe} (d). However, $C_2$ near the wall (for $\phi = 1.5\times 10^{-3}$) is 20\% lower than the $C_2$ at the center for unladen flow. Similar observations are followed for other Reynolds numbers and locations as well. All the above observations suggest that $C_2$ is not only a function of channel location and $Re_\lambda$, but it is also a function of particle volume loading.


The spectral representation of turbulent kinetic energy is given as \cite{pope2001turbulent},
\begin{equation}
\hat{E}(k) = \frac{1}{2} \langle \hat{u}'_i(k) \hat{u}^{'*}_i(k) \rangle.
\label{Ek}
\end{equation}
Where $ \hat{u}'_i(k)$ is the Fourier transform of $ u'_i(x,t)$ over the homogenous directions, and 
$\hat{u}^{'*}_i(k)$ is the complex conjugate of $ \hat{u}'_i(k)$. Here, $\langle . \rangle $ denotes the ensemble averaging.
The energy spectrum density, E(k), is calculated using Eqn.~\ref{Ek}. In Fig.~\ref{Compensated_spectra}(a), the normalized energy spectrum density ($E^+ = E(k)/(u_\tau \nu)$) is compared with earlier studies \cite{andrade2018analyzing, trofimova2009direct, del2003spectra, del2004scaling, hoyas2006scaling} for verification at two-channel locations of $y^+ = 15$ and 50 which shows a good agreement. The Kolmogorov constant ($C$) is also plotted using the compensated spectra, $C = E(k)k^{5/3}\epsilon^{-2/3}$ at $y^+ = 15$, 50, and at center of the channel (for $y^+ = 150 $ for $Re_b = 5600$ and $y^+ = 100 $ for $Re_b = 3300$) for both the Reynolds numbers and different Stokes numbers, shown in  Fig.~\ref{Compensated_spectra} (b-d).  Figures.~\ref{Compensated_spectra} (b) and (c) show that there is almost no variation in C at  $y^+ = 15$ and 50 for $Re_b = 5600$ and at $y^+ = 50$ for $Re_b = 3300$  for all the Stokes numbers. However, a non-monotonic variation is observed in $C$ for $Re_b = 3300$ at $y^+ = 15$. This is in contrast to the behavior of Kolmogorov constant ($C_2$) obtained via second-order velocity structure function where a monotonic decrease in $C_2$ is observed with an increase in particle volume loading.
Fig.~\ref{Compensated_spectra}(d) shows that the value of $C$ decreases almost linearly from 1.3 to 0.4 for a change in $\phi$ from $2\times 10^{-4}$ to $2.7\times 10^{-3}$ for $Reb = 5600$ and approximately to 0.8 at $\phi = 0.0011$ for $Re_b = 3300$. The observations from the second-order structure function and compensated spectra are consistent away from the wall (at $y^+ = 50$) and the channel center region for both the Reynolds numbers and all the Stokes numbers reported here.


The larger decrease in the Kolmogorov constant at the channel center compared to the near-wall region with an increase in particle loading is associated with a higher decrease of $Re_\lambda$, Fig.~\ref{TaylorRe} (c and d), and the ratio of the fluctuating velocity to the Kolmogorov velocity at the channel center. The ratio of kinetic energy to the square of the Kolmogorov velocity scale is plotted as a function of wall-normal direction in Fig.~\ref{ux_uk} (a and b) for different particle volume loadings. It is observed that the ratio decreases faster in the channel center than in the near-wall region with an increase in volume loading. It is worth noting that the maximum turbulence production happens near the wall, and the dissipation due to the particle is maximum at the channel center \cite{muramulla2020disruption}. A decrease in the ratio of velocities signifies the reduction in scale separation of the large and the small scales; consequently, it will increase the small-scale anisotropy.


\section{New modeling approach based on modified energetics}
\label{sec:new_model}

From the above analysis of compensated spectra and second-order velocity structure function, it is observed that there is a non-monotonic decrease in the Kolmogorov constant across the channel in the presence of dispersed particles. However, a significant variation of the Kolmogorov constant, especially in the channel center, happens constantly with an increase in particle volume loading. This observation can be used to model the effect of particles on the fluid phase without actually adding the discrete solid phase. The use of the Kolmogorov constant appears in the classical Smagorinsky model \cite{smagorinsky1963general}. In the Smagorinsky model (LES approach), the eddy viscosity term model the subgrid-scale dissipation. This eddy viscosity is represented in terms of Smagorinsky coefficient which is calculated from the Kolmogorov constant \cite{sagaut2006large}. The filtered continuity and momentum equations for LES are written as,
\begin{equation}
\frac{\partial \widetilde{u}_i}{\partial x_i}=0,
\end{equation}
\begin{equation}
\frac{\partial \widetilde{u}_i}{\partial t} + \frac{\partial \widetilde{u}_i\widetilde{u}_j}{\partial x_j} = - \frac{1}{\rho_f} \frac{\partial \widetilde{p}}{\partial x_i} + {\nu} \frac{\partial^2 \widetilde{u}_i}{\partial x_j \partial x_j} +  \frac{\partial (\widetilde{u}_i\widetilde{u}_j - \widetilde{u_i u_j})}{\partial x_j}
\label{LES eqn}.
\end{equation} 
Here, $\widetilde{p} $ is the filtered pressure, $\widetilde{u}_i$ is the filtered velocity, $\nu$ is the kinematic viscosity, and $ \rho_f$ is the fluid density. The subgrid scale (SGS) stress term, ($  \widetilde{u}_i\widetilde{u}_j - \widetilde{u_i u_j}$), in the Smagorinksy model is expressed as,
\begin{equation}
-\tau_{ij} =  \widetilde{u}_i\widetilde{u}_j - \widetilde{u_i u_j} = 2 \nu_t \widetilde{S}_{ij},
\end{equation}
\begin{equation}
 \widetilde{S}_{ij}  = \frac{1}{2} \left[ \frac{\partial  \widetilde{u}_i }{ \partial x_j} + \frac{\partial  \widetilde{u}_j }{ \partial x_i} \right],
\end{equation}
where $\nu_t$ is eddy viscosity and $\widetilde{S}_{ij}$ is the filtered strain rate tensor. The eddy viscosity is written as,
\begin{equation}
\nu_t = (C_s \widetilde{\triangle})^2 |\widetilde{S}| ,
\label{nu_t}
\end{equation}
\begin{equation}
|\widetilde{S}| = \sqrt{2 \widetilde{S}_{ij} \widetilde{S}_{ij}}.
\end{equation}

In Eqn.~\ref{nu_t},   $C_s$ is the Smagorinsky coefficient, $|\widetilde{S}|$ is the magnitude of the strain rate, and $\widetilde{\triangle} = (\widetilde{\triangle}_1 \widetilde{\triangle}_2 \widetilde{\triangle}_3 )^{1/3}$ is the cube-root volume of grid size. $\widetilde{\triangle}_1 $, $\widetilde{\triangle}_2 $ and $\widetilde{\triangle}_3 $ are the grid spacing in the x, y and z directions respectively. The relation between the Kolmogorov constant and the Smagorinsky coefficient is given as \cite{sagaut2006large},
\begin{equation}
C_s = \frac{0.23}{C^{3/4}}.
\label{C_s formula}
\end{equation}

Therefore, the variation of the Kolmogorov constant (C) causes a modification in the Smagorinsky coefficient $ C_s $. The Kolmogorov constant decreases with an increase in particle volume loading, which results in an increased $ C_s $. Thus, the simulations are performed for single-phase flow with higher $ C_s $ values for the Smagorinsky model without adding the particles.  In this context, It is worth noting the earlier work of \citet{yeo2010modulation} who performed the fully resolved simulation for bubbles, neutrally buoyant and inertial particles in homogenous turbulence and commented on the possibility of representing the particle feedback effect with an additional effective viscosity. 

The present exercise aims to capture the effect of particles using a single-phase simulation by varying the Smagorinsky coefficient without adding the particles to the system. Although, we have observed that the variation of the Kolmogorov constant is a function of wall-normal location. As a first approximation, we consider a single Kolmogorov constant approximation across the channel, a function of particle concentration. It should be noted that the particles increase the anisotropy across the fluid fluctuations, and an anisotropic inhomogeneous modeling approach should be taken to model the fluid phase accurately, which is left as a future scope. However, the present method of modeling the dynamics of fluid phase in two-phase flow using a single-phase simulation is computationally less expensive. It provides insights into the mechanism of turbulence modulation. Simulations are performed with the Smagorinksy model which has been used in Ref.~\cite{rohilla2022applicability}. The number of grids used is $128\times65\times64$ and $ 64\times65\times32 $  in the streamwise (x), wall-normal (y), and spanwise (z) directions for bulk Reynolds numbers of 5600 and 3300, respectively.  $C_s = 0.125$ is used in the unladen flow simulations and the Van-Driest damping is implemented to avoid high dissipation in the near-wall region~\cite{rohilla2022applicability}.

\begin{figure}[]
	\begin{subfigure}[b]{1\textwidth}
		\centering
	\minipage{0.5\textwidth}
	\includegraphics[width=\textwidth]{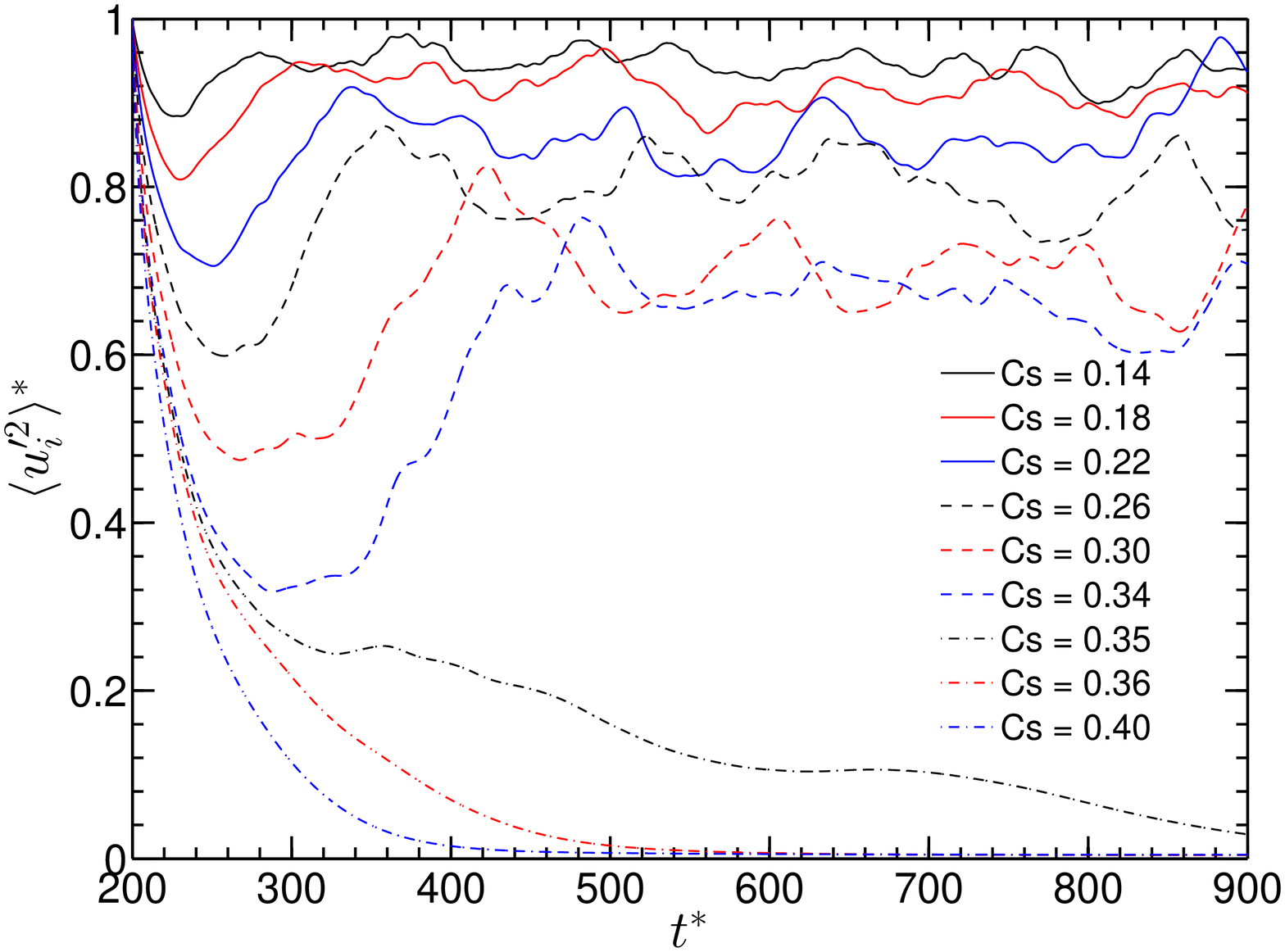}
	\caption{}
	\endminipage
	\end{subfigure} 
	\caption{The average of fluid fluctuations (normalized with the $t^* = 200$ value) across the channel evolved over the time for a range of $C_s$. The $t^*$ is normalized with channel width and fluid bulk velocity. }
	\label{L2norm}
\end{figure}

\begin{figure}
	\begin{subfigure}[b]{1\textwidth}
	\minipage{0.45\textwidth}
	\includegraphics[width=\textwidth]{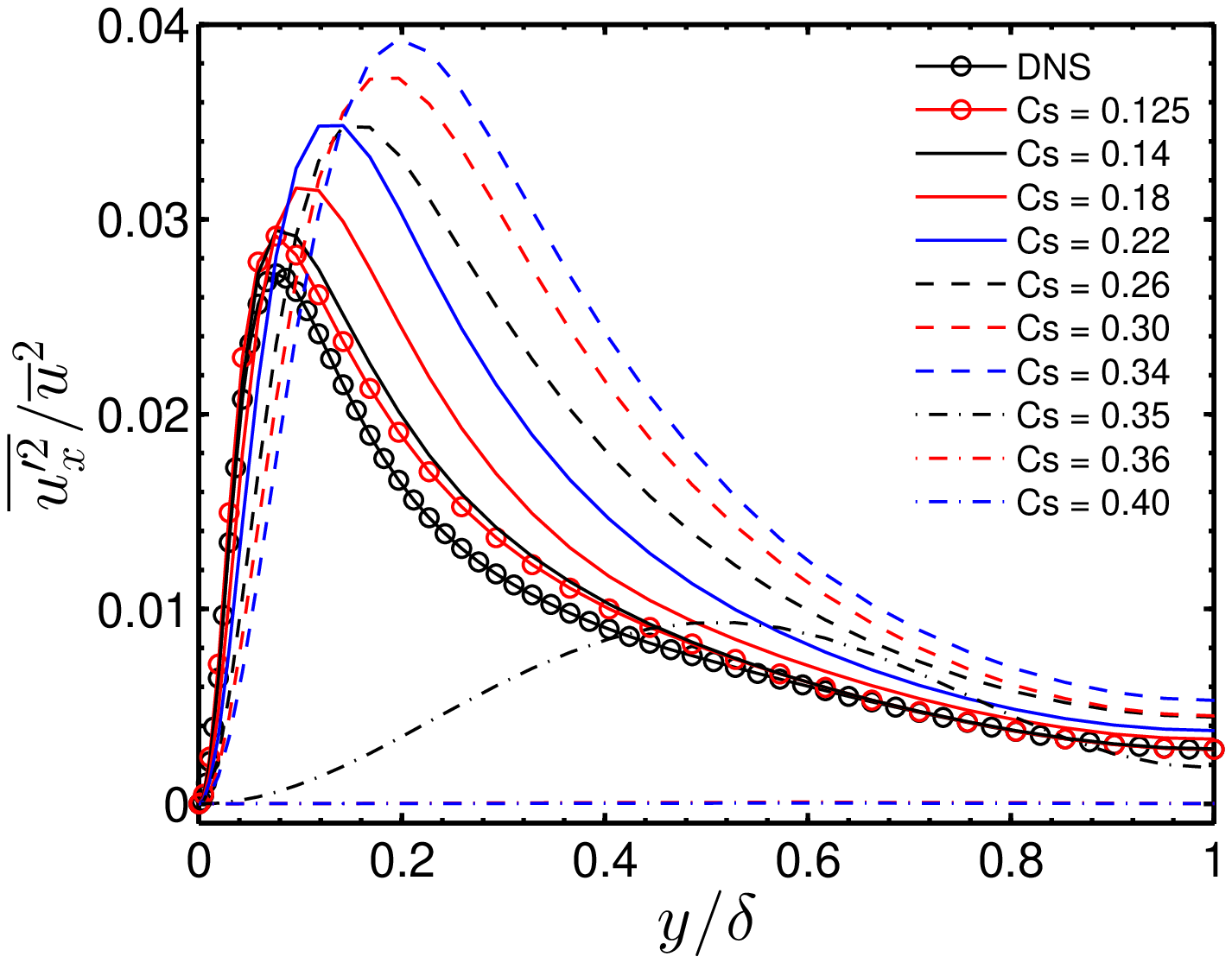}
	\caption{Streamwise fluctuations}
	\endminipage 
	\minipage{0.45\textwidth}
	\includegraphics[width=\textwidth]{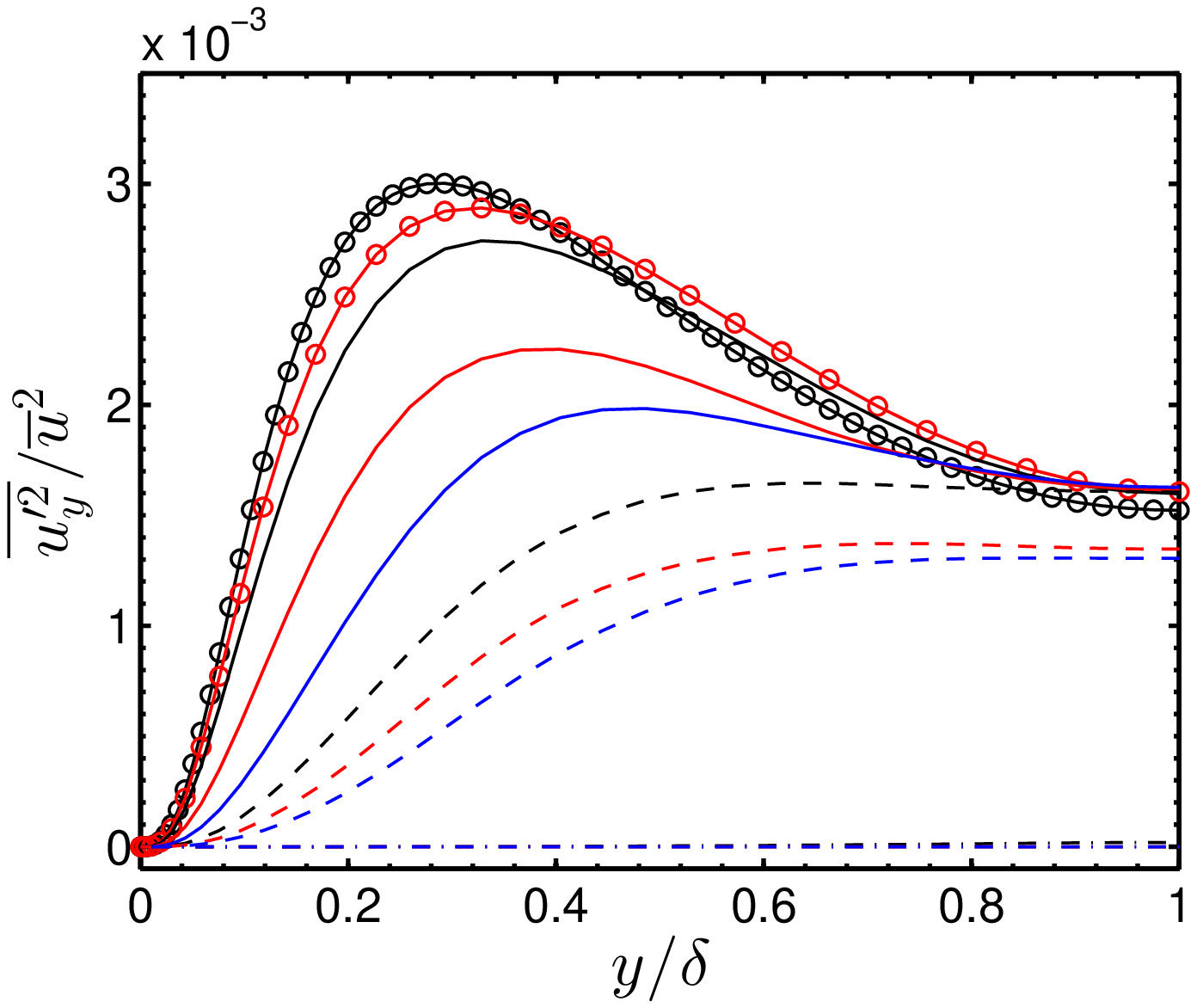}
	\caption{Wall-normal fluctuations }
	\endminipage \\
	\minipage{0.45\textwidth}
	\includegraphics[width=\textwidth]{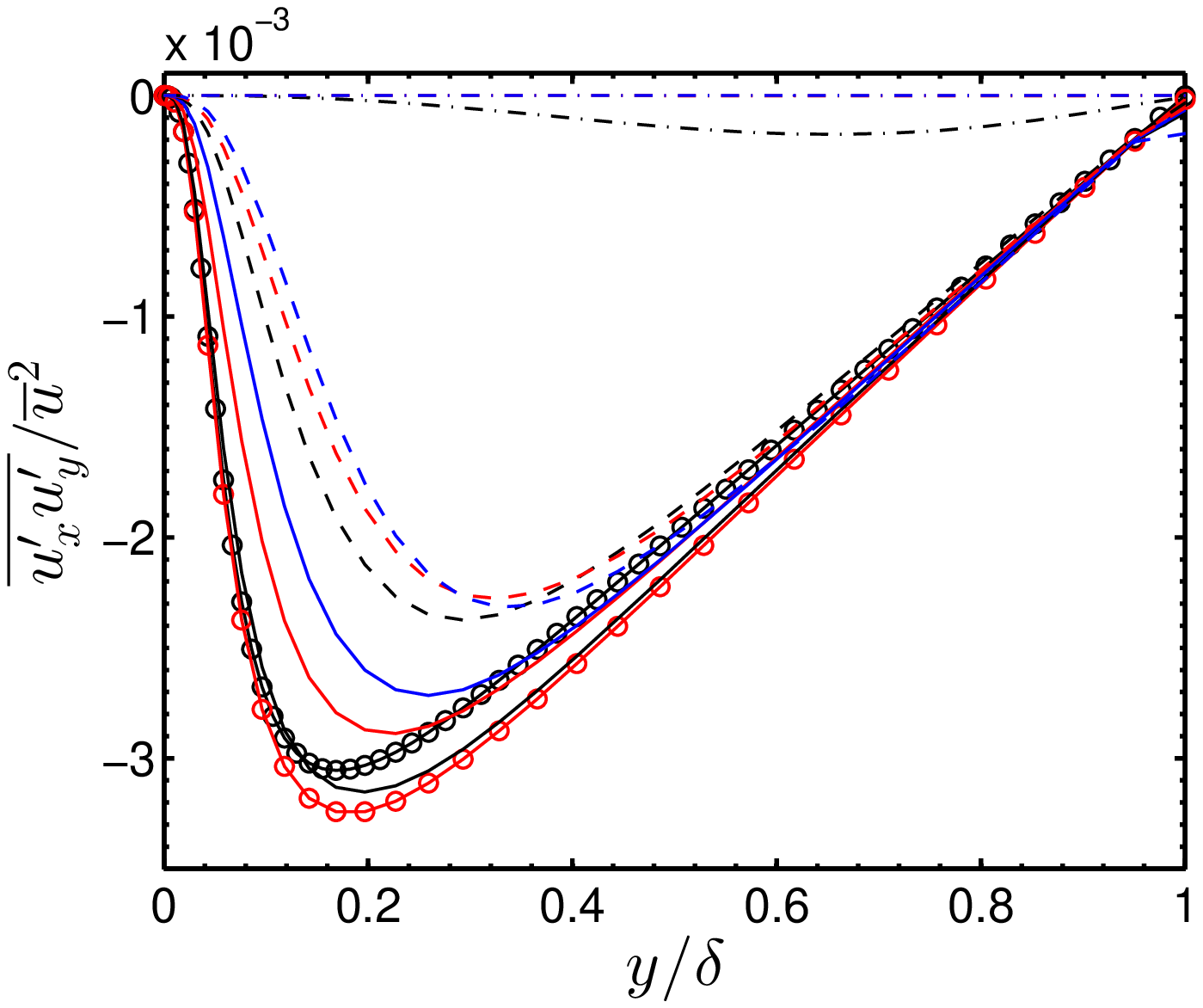}
	\caption{Reynolds stress}
	\endminipage 
	\minipage{0.45\textwidth}
	\includegraphics[width=\textwidth]{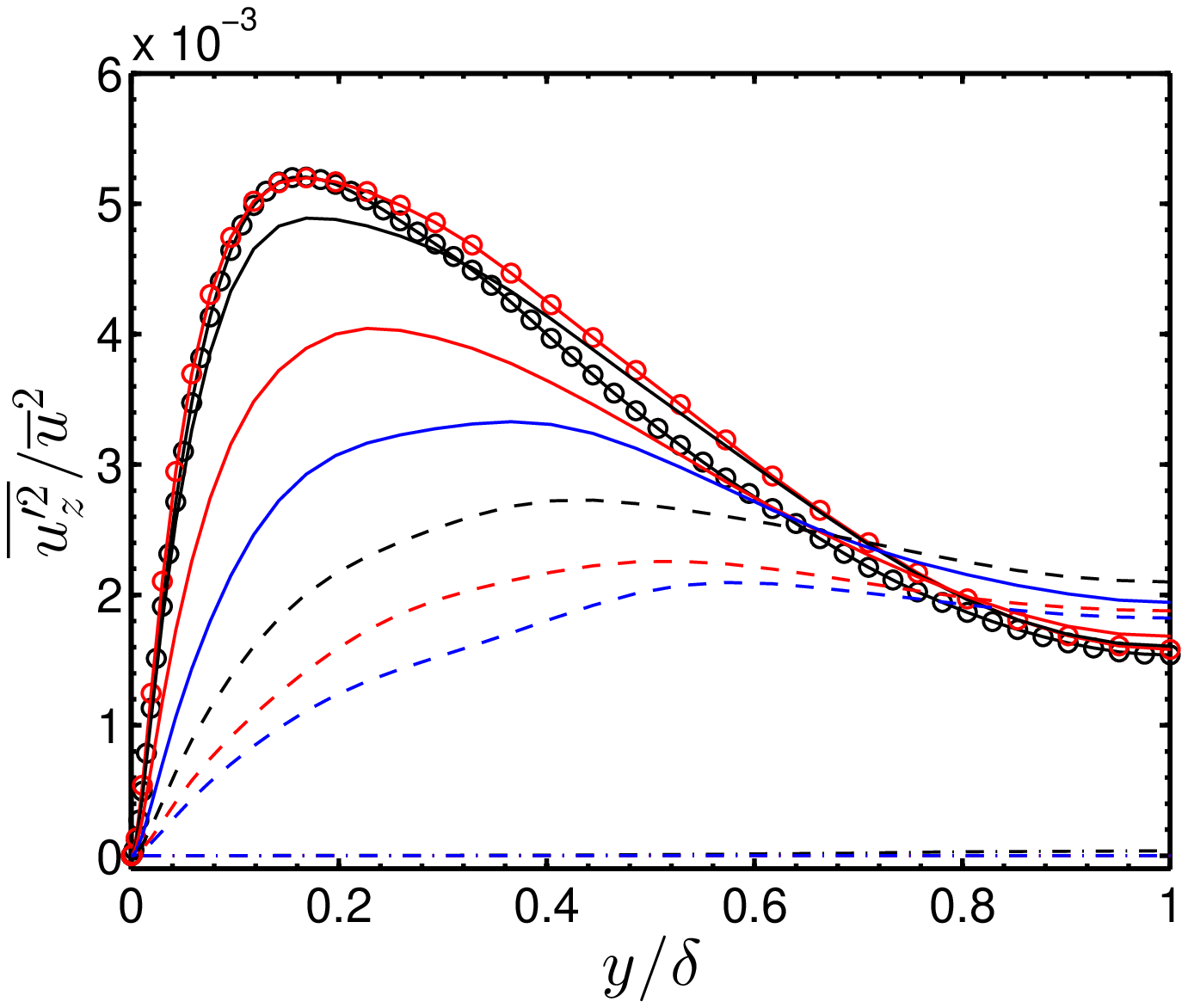}
	\caption{Spanwise fluctuations }
	\endminipage
	\end{subfigure} 
	\caption{The profiles of fluid fluctuations in the wall-normal direction. The simulations are performed with Smagorinksy model with increasing Smagorinsky coefficient ($C_s$) for $Re_b = 5600$. The symbols in Figs.~(b, c, and d) are the same as in Fig.(a). }
	\label{velocity_profile}
\end{figure}

First, we have increased the value of $C_s$ in LES to determine the effect of $C_s$ on the fluid phase fluctuations. Then, we have presented the variation of second moments of fluctuation simultaneously as a function of $C_s$ and particle volume loading ($\phi_{av}$). In Fig.~\ref{L2norm}, the temporal evolution of the normalized sum of the fluid fluctuations is plotted for $Re_b = 5600$ with different $C_s$ values. The simulation with a lower $C_s$ value reaches a stationary state quickly. For high $C_s$ such as 0.30, the sum decreases initially and reaches the stationary state after a long time. For $C_s = 0.35$ and higher values, the sum of fluid fluctuations continuously decreases, representing a decay of the intensity of fluid turbulence. In Fig.~\ref{velocity_profile}, the profiles of fluid fluctuations are plotted along the wall-normal direction for a range of $C_s$. It is observed that the streamwise fluid fluctuations initially increase with an increase in $C_s$, and a sudden decrease happens at $C_s = 0.35$. However, a continuous decrease is observed for Reynolds stress, wall-normal, and spanwise fluid velocity fluctuations, and a complete collapse of turbulence is observed at $C_s = 0.35$. As the transverse fluctuations are decreased, there is a decrease in momentum flux. The non-monotonic behavior in the streamwise fluid fluctuations has also been reported by \citet{zhou2020non} where authors have performed the DNS of particle-laden channel flow. An increase in the streamwise fluctuations was observed at a solid volume fraction of $1.8\times 10^{-4}$, and a decrease was observed with a further increase in volume fraction. The authors mentioned two competitive phenomena. First, the near-wall vortices become weaker, leading to the larger spacing between the streaks. This effect causes a decrease in streamwise fluctuations. However, in the second case, the streaks become more organized and aligned in the streamwise direction. This phenomenon increases the streamwise fluctuations. A continuous decrease in the wall-normal and spanwise fluctuations was observed by~\citet{zhou2020non}. In the present study, an increase in $C_s$ leads to an initial increase in the streamwise fluid fluctuations. Then, a sudden turbulence collapse is observed at a higher $C_s = 0.35$. It is to be noted that we have used a constant value of $C_s$ across the channel. A further analysis using inhomogeneous $C_s$  is expected to provide more quantitative modulation of fluid turbulence. The present simulation shows a monotonic decrease in the Reynolds stress, wall-normal, and spanwise fluid fluctuations (Fig.~\ref{velocity_profile} (b - d)) as reported by DNS studies~\cite{li2001numerical, vreman2015turbulence, muramulla2020disruption, zhou2020non}.

\begin{figure}[htb]
	\begin{subfigure}[b]{1\textwidth}
	\minipage{0.45\textwidth}
	\includegraphics[width=\textwidth]{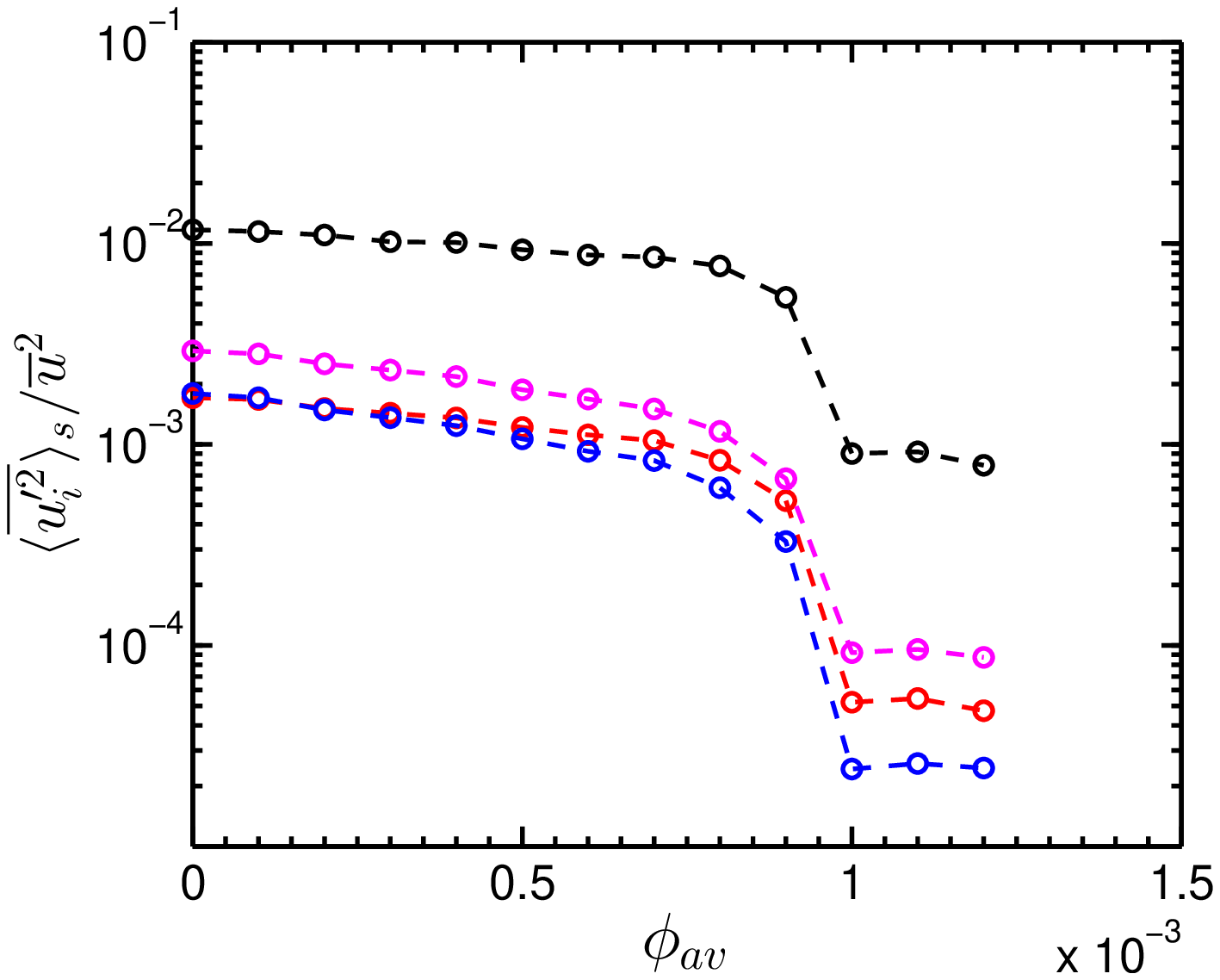}
	\caption{ $Re_b = 3300$}
	\endminipage 
	\minipage{0.45\textwidth}
	\includegraphics[width=\textwidth]{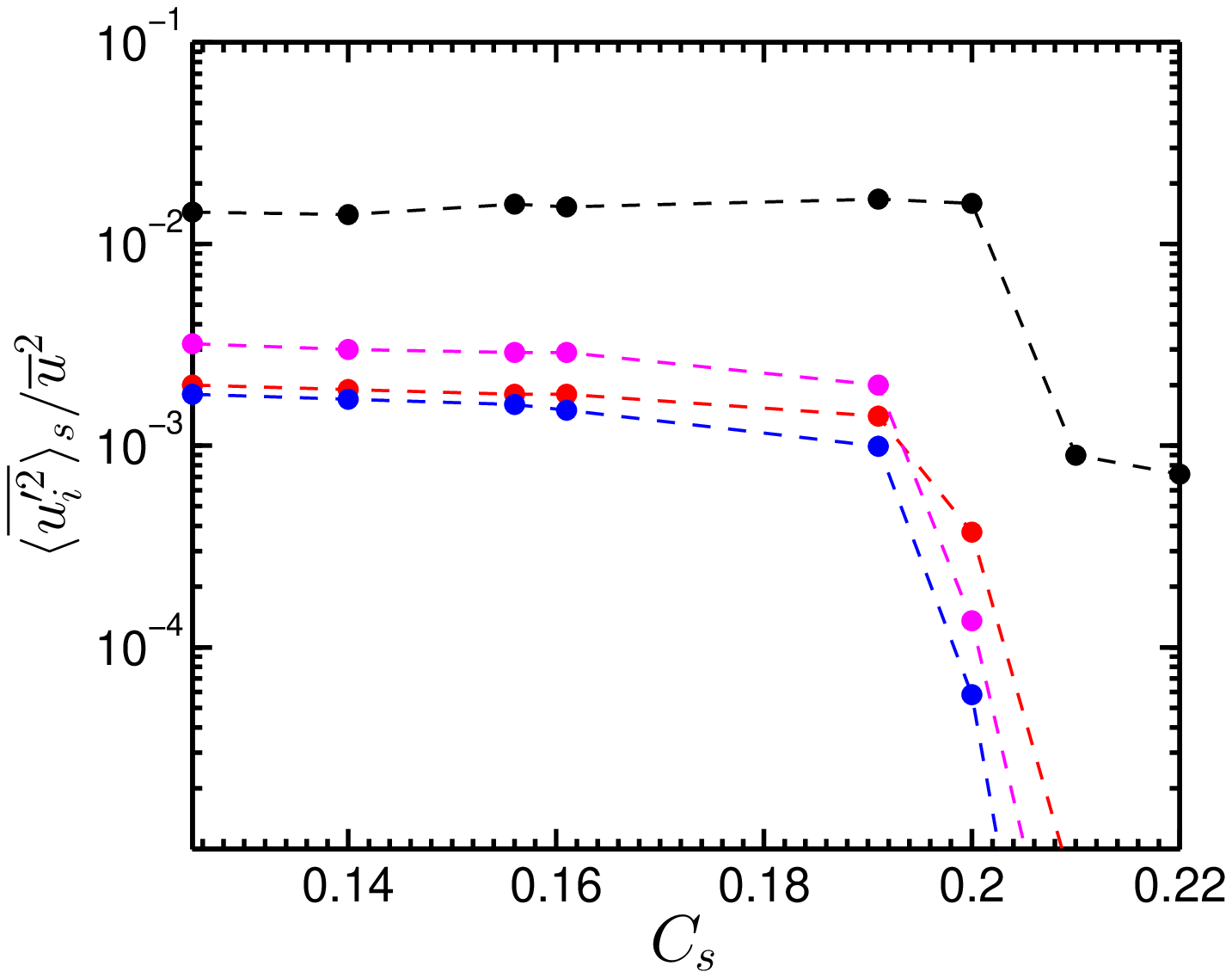}
	\caption{ $Re_b = 3300$}
	\endminipage \\
	\minipage{0.45\textwidth}
	\includegraphics[width=\textwidth]{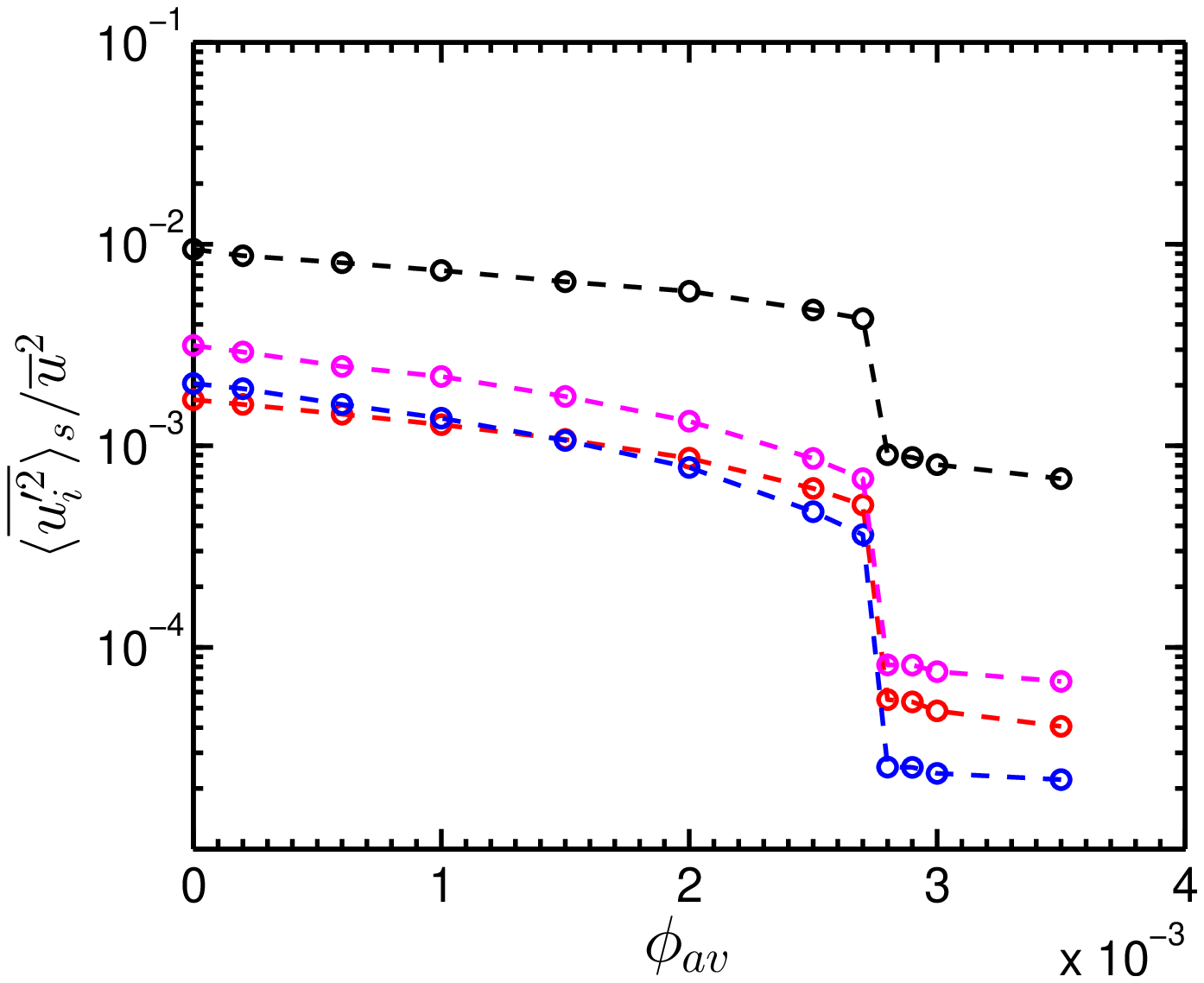}
	\caption{$Re_b = 5600$}
	\endminipage 
	\minipage{0.45\textwidth}
	\includegraphics[width=\textwidth]{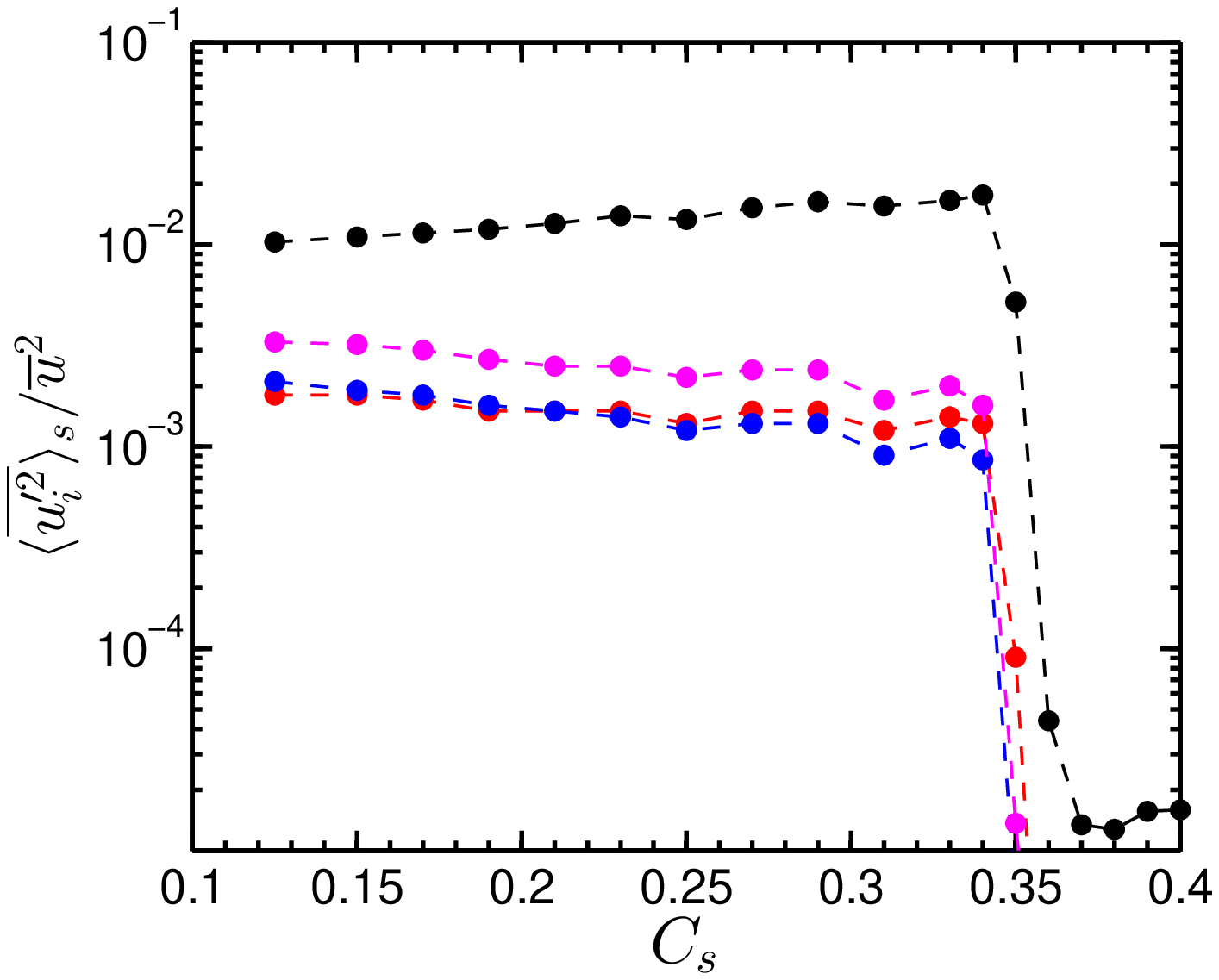}
	\caption{$Re_b = 5600$}
	\endminipage
	\end{subfigure} 
	\caption{The average fluid fluctuations normalized with fluid bulk velocity ($\bar{u}$) for $Re_b = 3300$ and 5600. Figs.~(a and c): the average fluid fluctuations are shown for particle-laden DNS \cite{muramulla2020disruption}. Figs.~(b and d): the average fluid fluctuations are shown for the Smagorinksy model over a range of Smagorinsky coefficient ($C_s$). The dashed lines with closed symbols are for the Smagorinksy model, and the dashed lines with open symbols are for the DNS. The legends are as follows, black line: Streamwise, red line: Reynolds stress, blue line: Wall-normal, and magenta line: Spanwise fluctuations. }
	\label{avg_rms}
\end{figure}
\begin{figure}[htb]
	\begin{subfigure}[b]{1\textwidth}
	\minipage{0.45\textwidth}
	\includegraphics[width=\textwidth]{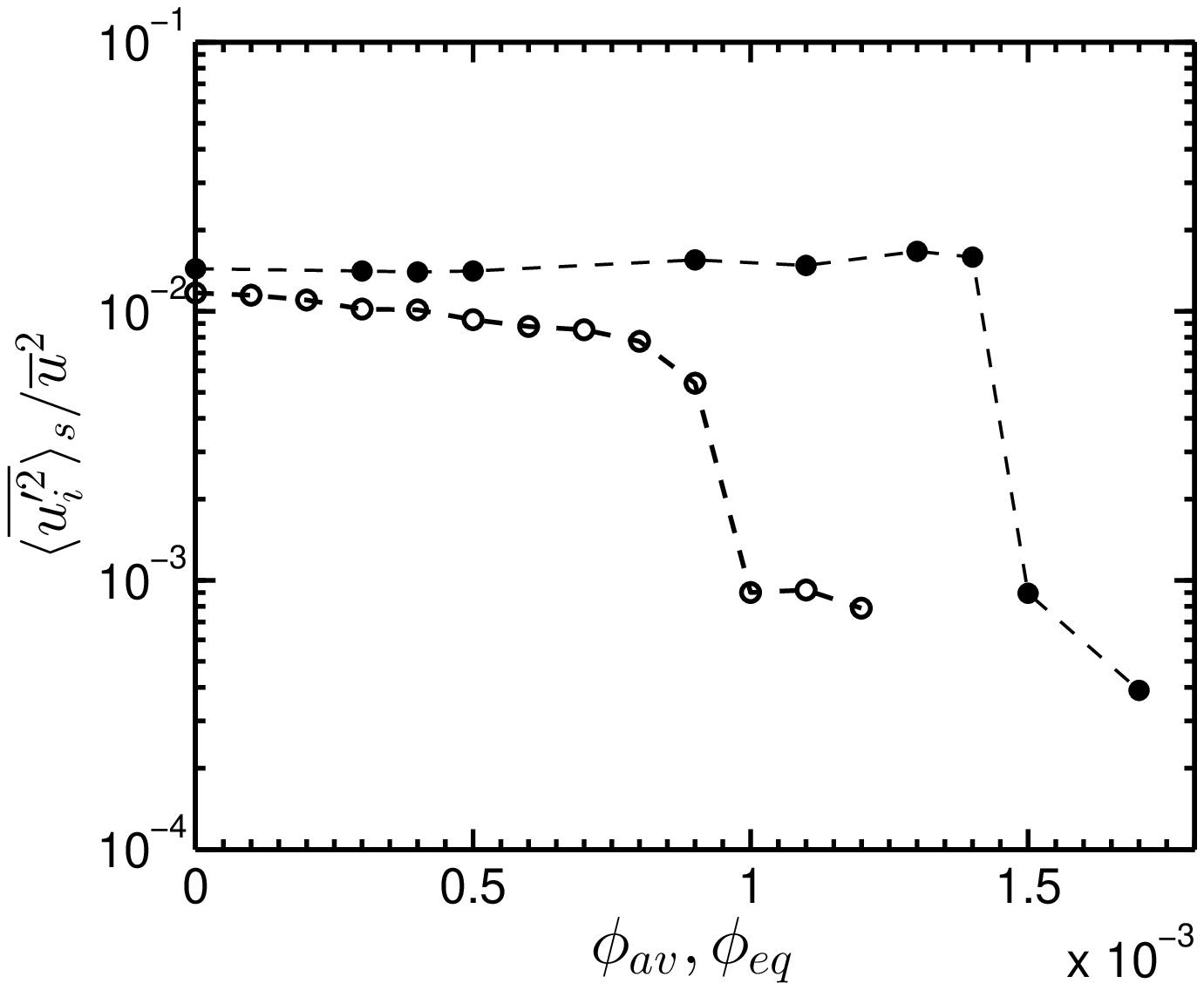}
	\caption{ $Re_b = 3300$}
	\endminipage 
	\minipage{0.45\textwidth}
	\includegraphics[width=\textwidth]{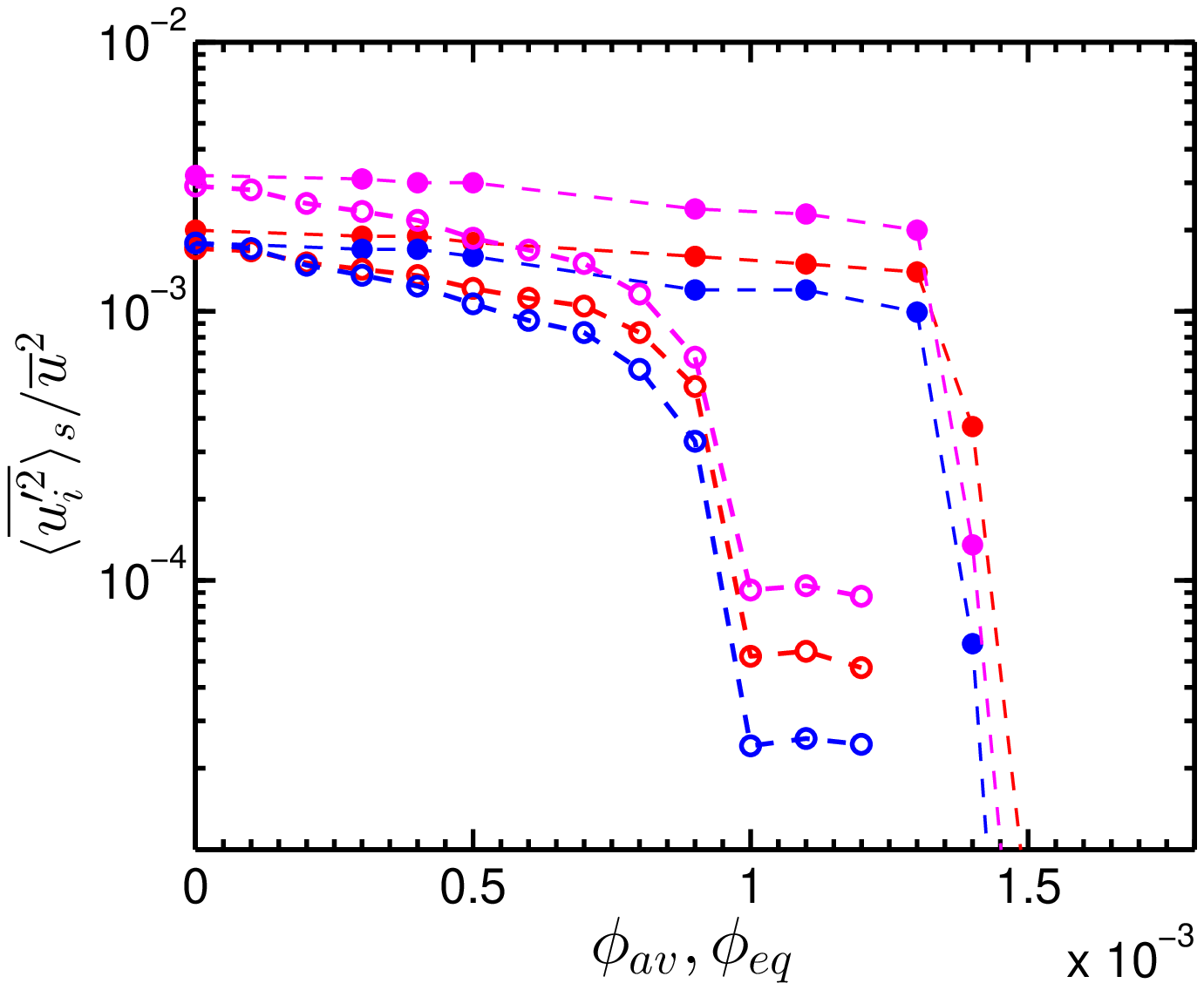}
	\caption{ $Re_b = 3300$}
	\endminipage \\
	\minipage{0.45\textwidth}
	\includegraphics[width=\textwidth]{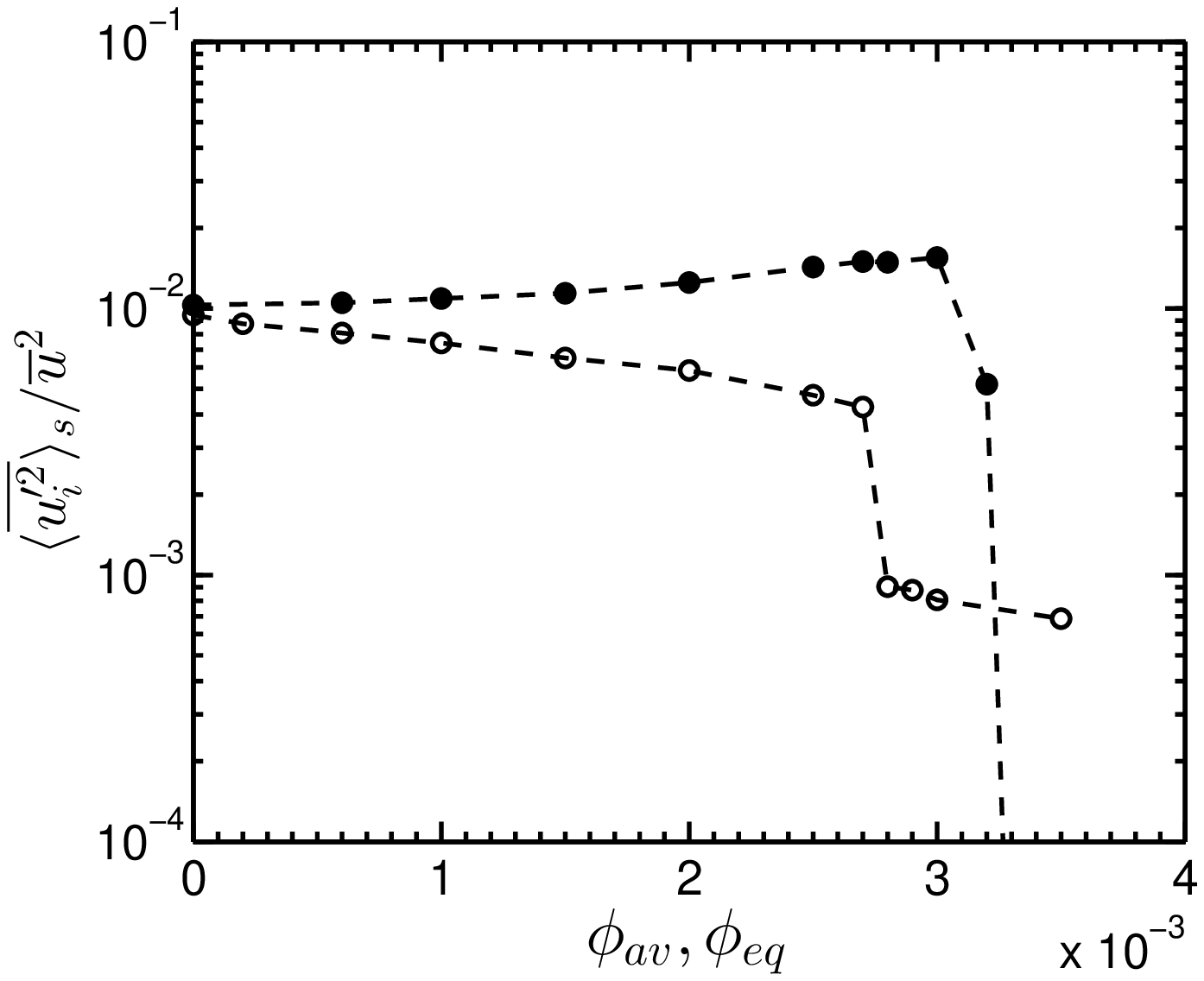}
	\caption{$Re_b = 5600$}
	\endminipage 
	\minipage{0.45\textwidth}
	\includegraphics[width=\textwidth]{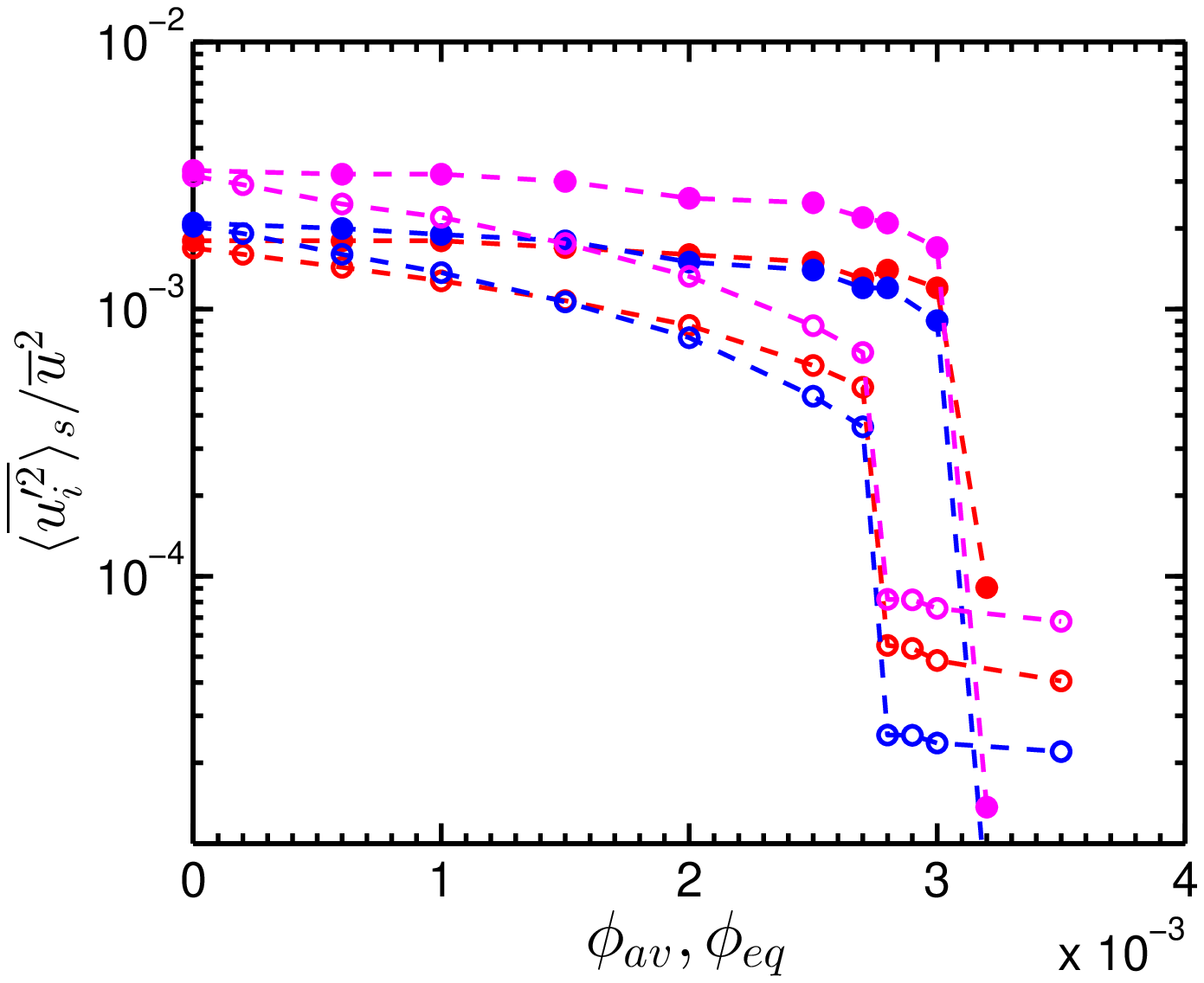}
	\caption{$Re_b = 5600$}
	\endminipage
	\end{subfigure} 
	\caption{The average fluid fluctuations (shown in Fig.~\ref{avg_rms}) for the Smagorinsky model and DNS are compared. Here, $C_s$ is represented in terms of equivalent volume fraction ($\phi_{eq}$). Fig.~(a and c): Streamwise fluid fluctuations, and Fig.~(b and d): Cross-stream, wall-normal, and spanwise fluid fluctuations. The dashed lines with closed symbols are for the Smagorinksy model, and the dashed lines with open symbols are for the DNS. The legends are as follows, black line: Streamwise, red line: Reynolds stress, blue line: Wall-normal, and magenta line: Spanwise fluctuations. }
	\label{avg_rms_S_DNS}
\end{figure}

Fig.~\ref{velocity_profile} shows that the Reynolds stress and other components of fluctuations are a strong function of wall-normal distance. Therefore, we define the channel averaged second-moments of velocity fluctuations (Eqn.~\ref{avg_fluc}) and present the effect of variation of $C_s$. 
\begin{equation}
\langle \star \rangle_s = \frac{1}{\delta} \int_{0}^{\delta} dy \langle \star \rangle
\label{avg_fluc}
\end{equation}
Here, $\langle \star \rangle_s$ is the averaged quantity over half-channel width ($\delta$). The average fluid fluctuations across the channel as a function of $C_s$ are plotted in Fig.~\ref{avg_rms}(b and d) for both the Reynolds numbers. For DNS, the average fluid fluctuations as a function of $\phi_{av}$ for both the Reynolds numbers are shown in Fig.~\ref{avg_rms}(a and c). The Stokes numbers considered are 105.47 and 210.93 for $Re_b = 3300$ and 5600 respectively for DNS. The average fluid fluctuations decreases with an increase in particle volume fraction, and a complete turbulence collapse is observed at $\phi_{av} = 10^{-3}$ and $2.8\times10^{-3}$ for $Re_b = 3300$ and 5600, respectively, shown in Fig.~\ref{avg_rms}(a and c). It is observed that the fluid fluctuations, except the streamwise fluctuations, decrease with an increase in $C_s$ value, and a drastic collapse of more than one order of magnitude is observed at $C_s = 0.20$ for $Re_b = 3300$, and at $C_s = 0.35$ for $Re_b = 5600$, shown in Fig.~\ref{avg_rms}(b and d). The decrease in Reynolds stress, wall-normal and spanwise fluctuations (Fig.~\ref{avg_rms}(b and d)) with an increase in $C_s$ is similar to the decrease observed for DNS (Fig.~\ref{avg_rms}(a and c)). The complete turbulence collapse observed in fluid fluctuations is similar to the collapse observed for particle-laden cases by Ref.~\cite{mito2006effect, shringarpure2012dynamics, capecelatro2018transition, kumaran2020turbulence, duque2021influence, muramulla2020disruption, rohilla2022applicability}. However, once turbulence is collapsed, the fluid fluctuations do not remain constant or increase as happens in the case of particle-laden cases due to particle-induced fluctuations \cite{capecelatro2018transition, kumaran2020turbulence, yu_2021}. The volume fractions ($\phi_{eq}$) which corresponds to different $C_s$ values are plotted in Fig.~\ref{avg_rms_S_DNS} where fluid fluctuations (shown in Fig.~\ref{avg_rms}) predicted by Smagorinksy model are compared with DNS. The Kolmogorov constant, computed using the second-order velocity structure function and compensated spectra, found to decrease linearly in the channel center location, shown in Fig.~\ref{alpha_variation} (b) and \ref{Compensated_spectra} (d). The equivalence between the $C_s$ (used as the Smagorinsky coefficient) and the particle volume loading has been derived as follows. From the value of $C_s$ (0.125) used for the unladen flow, $C$ is calculated using Eqn.~\ref{C_s formula}. It is assumed that the variation of $C$ as a function of $\phi_{av}$ follows the same functional form as that of $C_2$ versus $\phi_{av}$, shown in Fig.~\ref{alpha_variation}~(b). Therefore, we can estimate $C$ and $C_2$ for a range of particle volume loading ($\phi_{av}$). The equivalent volume fraction ($\phi_{eq}$) which corresponds to the $C_s$ are plotted in Fig.~\ref{avg_rms_S_DNS} along with average volume fraction ($\phi_{av}$).
For $Re_b = 3300$ and 5600, the turbulence collspase was observed at $\phi_{av} = 10^{-3}$ and $2.8\times10^{-3}$ , respectively, for particle-laden DNS \cite{muramulla2020disruption}. However, the turbulence collapse for present simulations is observed at $\phi_{eq} = 1.4\times10^{-3}$ and $3.2\times10^{-3}$ for $Re_b = 3300$ and 5600, which is very close the critical loading predicted by DNS. 

In the present study, an attempt is made to capture the effect of change in particle volume loading in a two-phase flow by a fluid phase only phase simulation with a modified Smagorinsky coefficient. Therefore, it is expected that the total dissipation caused by the viscous term and modified eddy viscosity term in the present simulation should be similar to the total dissipation by the mean viscous term and dissipation due to feedback force exerted by the particles. Therefore, we have compared the total dissipation predicted in these two cases. The kinetic energy balance equation of the mean fluid flow in the stationary state for the particle-laden flow can be described as,

\begin{multline}
- {U}_x \frac{1}{\rho } \frac{\partial {P}}{\partial x} - \frac{\partial (\overline{{u_x^{'}} {u_y^{'}}} {U}_x)}{\partial y} + \overline{{u_x^{'}} {u_y^{'}}} \frac{\partial {U}_x}{\partial y} \\
+\nu  \frac{\partial}{\partial y}({U}_x \frac{\partial {U_x}}{\partial y}) - \nu \frac{\partial {U}_x}{\partial y} \frac{\partial {U}_x}{\partial y}  - {U}_x \overline{ \frac{\rho_p f \phi_c}{ \rho_f \tau_p} ({u}_x  - {v}_x) }    = 0.
\label{energy}
\end{multline}
Here,  $\overline{{p}} = {P} $; ${p}$ is the instantaneous pressure and ${P}$ is the mean pressure. Similarly, ${u_i}$ is the instantaneous velocity, ${U}_x$ is the mean velocity, and ${u'_i}$ is the fluctuating velocity. $\rho_f \overline{{u^{'}_i} {u^{'}_j}}$ is the Reynold stress,  $\tau_p$ is the particle relaxation time, $\phi_c$ is the volume fraction in respective grid, $f$ is the drag coefficient, $\rho_p$ is the material density of the particle, and $\rho_f$ is the fluid density. In equation (\ref{energy}), the first term is the energy due to pressure work; the second and fourth terms are the divergence of energy fluxes due to Reynolds stress and fluid viscous stress, respectively; the third term is the energy utilized for the turbulence production, the fifth term is the viscous dissipation ($\epsilon_m$) due to mean shear and the sixth term is the loss of energy due to the particle drag ($F_p$). The filtered mean kinetic energy equation for unladen fluid flow is written as,
\begin{multline}
- \widetilde{U}_x \frac{1}{\rho } \frac{\partial \widetilde{P}}{\partial x} - \frac{\partial (\overline{\widetilde{u_x^{'}} \widetilde{u_y^{'}}} \widetilde{U}_x)}{\partial y} + \overline{\widetilde{u_x^{'}} \widetilde{u_y^{'}}} \frac{\partial ( \widetilde{U}_x)}{\partial y} 
+(\nu +\nu_t) \frac{\partial}{\partial y}(\widetilde{U}_x \frac{\partial \widetilde{U_x}}{\partial y}) - (\nu +\nu_t) \frac{\partial \widetilde{U}_x}{\partial y} \frac{\partial \widetilde{U}_x}{\partial y}     = 0.
\label{filtered_energy}
\end{multline}
Here,  $\overline{\widetilde{p}} = \widetilde{P} $; $\widetilde{p}$ is the instantaneous filtered pressure and $\widetilde{P}$ is the mean filtered pressure. Similarly, $\widetilde{u}_i$ is the instantaneous filtered velocity, $\widetilde{U}_x$ is the mean filtered velocity, $\widetilde{u'_i}$ is the filtered fluctuating velocity and $\rho_f \overline{\widetilde{u^{'}_i} \widetilde{u^{'}_j}}$ is filtered Reynold stress. The terms in Eqn.~\ref{filtered_energy} are similar to Eqn.~\ref{energy} with filtered quantities. The dissipation in the mean kinetic energy equation (Eqn.~\ref{energy}) is caused by the last two terms, mean viscous dissipation and particle-induced dissipation. However, the dissipation in Eqn.~\ref{filtered_energy} is due to the last term which is due to molecular ($\epsilon_m$) and eddy viscosity ($\epsilon_{eddy}$). Thus, these two dissipation terms should be comparable to predict fluid phase dynamics accurately when a single-phase simulation supplants the two-phase simulation.

\begin{figure}
	\begin{subfigure}[b]{1\textwidth}
	\minipage{0.49\textwidth}
	\includegraphics[width=\textwidth]{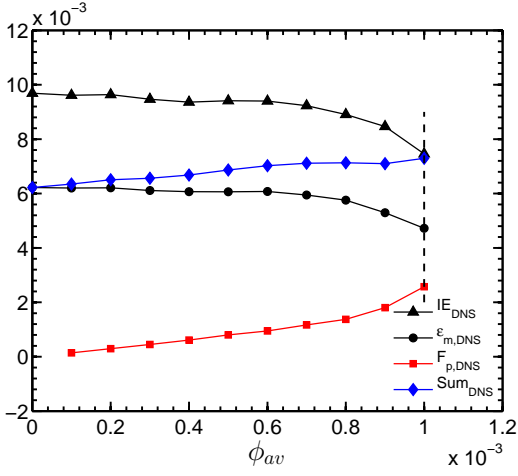}
	\caption{ $Re_b = 3300$, St = 105.47.}
	\endminipage 
	\minipage{0.49\textwidth}
	\includegraphics[width=\textwidth]{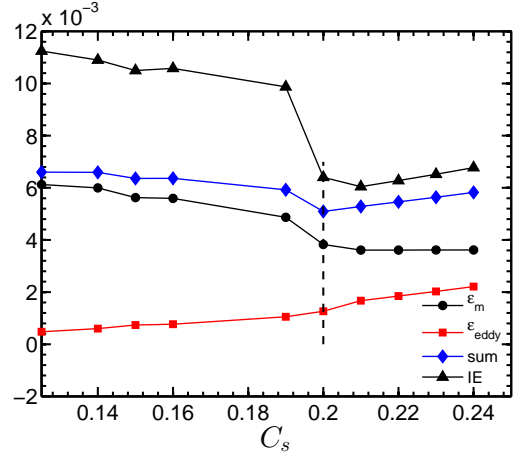}
	\caption{ $Re_b = 3300$}
	\endminipage \\
	\minipage{0.49\textwidth}
	\includegraphics[width=\textwidth]{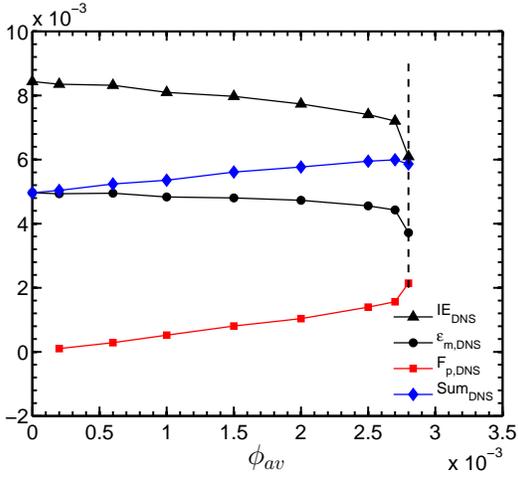}
	\caption{ $Re_b = 5600$, St = 210.93}
	\endminipage 
	\minipage{0.49\textwidth}
	\includegraphics[width=\textwidth]{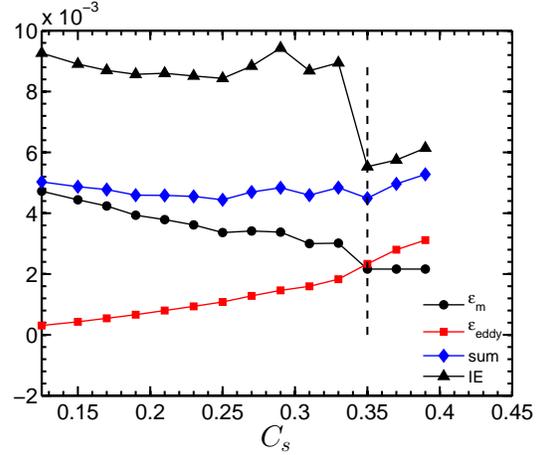}
	\caption{ $Re_b = 5600$}
	\endminipage 
	\end{subfigure} 
	\caption{The terms from the mean kinetic energy equation are presented. Fig.(a and c) are particle-laden DNS cases where $\epsilon_m$ is the mean viscous dissipation, `$F_{p,DNS}$' is the dissipation due to particles, and `sum' is the addition of feedback and mean viscous dissipation. Fig.(b and d) are single-phase simulations where the Smagorinsky coefficient is varied, $\epsilon_m$ is the viscous dissipation at mean flow, $\epsilon_{eddy}$ is the dissipation due to eddy viscosity, and `sum' is the addition of eddy and viscous dissipation. `IE' is the input energy from pressure work.}
	\label{total_dissipation}
\end{figure}

\begin{figure}
	\begin{subfigure}[b]{1\textwidth}
	\minipage{0.49\textwidth}
	\includegraphics[width=\textwidth]{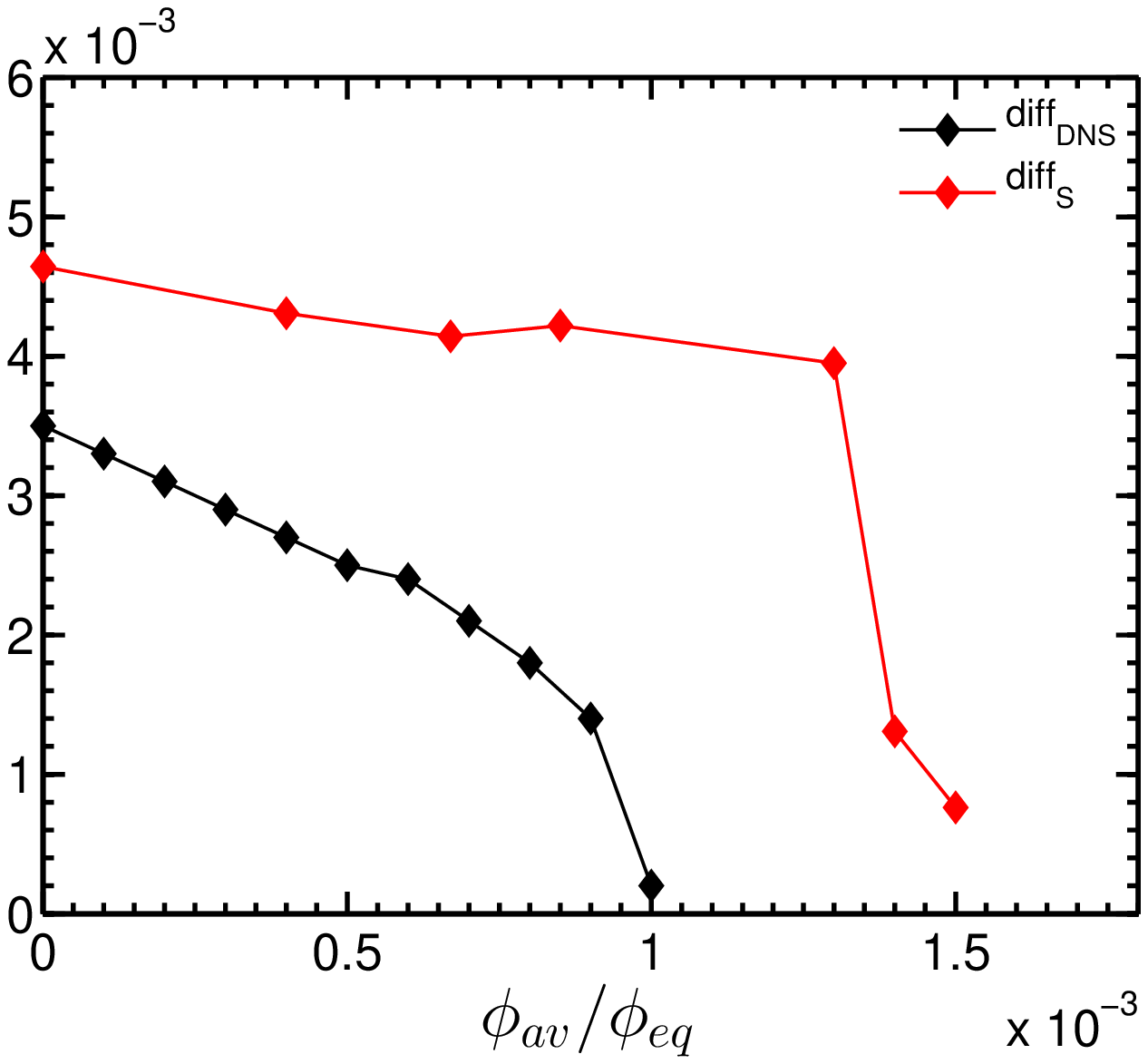}
	\caption{ $Re_b = 3300$}
	\endminipage 
	\minipage{0.49\textwidth}
	\includegraphics[width=\textwidth]{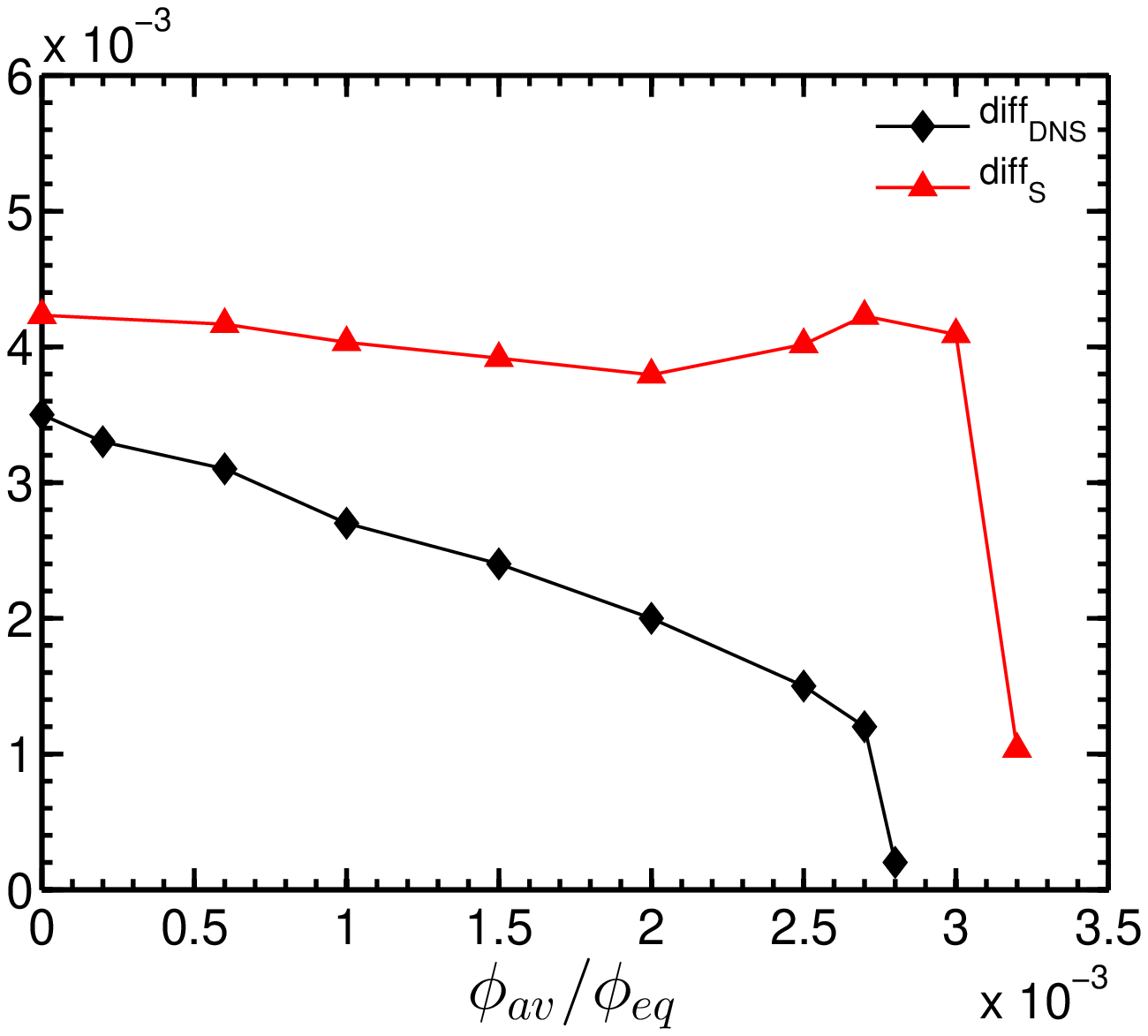}
	\caption{ $Re_b = 5600$}
	\endminipage 
	\end{subfigure} 
	\caption{ `diff = IE - sum' is the difference of input energy (IE) and total dissipation (`sum') which are plotted in Fig.~\ref{total_dissipation}. The subscripts are as S: Smagorinsky model and DNS: direct numerical simulation.}
	\label{net_turbulence}
\end{figure}

Fig.~\ref{total_dissipation}(a and c) show the terms due to mean viscous dissipation, dissipation due to particle drag, and input energy due to pressure work for particle-laden DNS cases \cite{rohilla2022applicability}. In Fig.~\ref{total_dissipation} (b and d), the Smagorinsky coefficient ($ C_s $) is varied, and input energy due to pressure work, the dissipation due to mean viscous and eddy viscosity are plotted using the Smagorinsky model for the unladen flow. It is observed that with an increase in $ C_s $, the mean viscous dissipation decreases. However, the dissipation due to eddy viscosity increases. The dissipation due to particle feedback in Fig.~(a and c) is comparable and shows a similar trend with the dissipation by modified eddy viscous term in Fig.~(b and d). Also, the total dissipation in the present unladen simulations is comparable to the total dissipation caused by the particle-laden DNS. It is to be noted that the total dissipation (mean viscous and particle induced dissipation) is almost constant in all the cases, which has also been reported by earlier work~\cite{rohilla2022applicability}. In the case of homogenous isotropic turbulence, it was observed by \citet{squires1990particle} that the total dissipation was constant for all the cases. For the mass loading of one, the decrease in the viscous dissipation was nearly $50\%$, and another $50\%$ dissipation was caused by the particles. A decrease in the input energy is observed for the particle-laden DNS, Fig.~\ref{total_dissipation}(a and c), and in the case of single phase Smagorinsky simulations, Fig.~\ref{total_dissipation}(b). In Fig.~\ref{total_dissipation}(d), a decrease in the input energy is observed at a low value of $C_s$, and a non-monotonic variation is observed before the turbulence collapse. The difference between input energy and the total dissipation, which is used in the turbulence production, is plotted in Fig.~\ref{net_turbulence}, and is compared with DNS \cite{rohilla2022applicability}. Here,  $C_s$ is represented in terms of equivalent volume fraction ($\phi_{eq}$). A decrease in the difference (`diff') is observed with an increase in $\phi_{eq}$ for the low Reynolds number ($Re_b = 3300 $).  But, it is not monotonic at moderate Reynolds number ($Re_b = 5600 $). However, a sudden collapse occur at $\phi_{eq} = 0.0014$ and 0.0032 for $Re_b = 3300$ and 5600, respectively. The turbulence collapse happens at $\phi = 0.001$ and 0.0028 for DNS at $Re_b = 3300$ and 5600 respectively. The critical volume fraction predicted by the Smagorinsky model closely matches with DNS at both Reynolds numbers.

\begin{figure}
	\begin{subfigure}[b]{1\textwidth}
	\minipage{0.49\textwidth}
	\includegraphics[width=\textwidth]{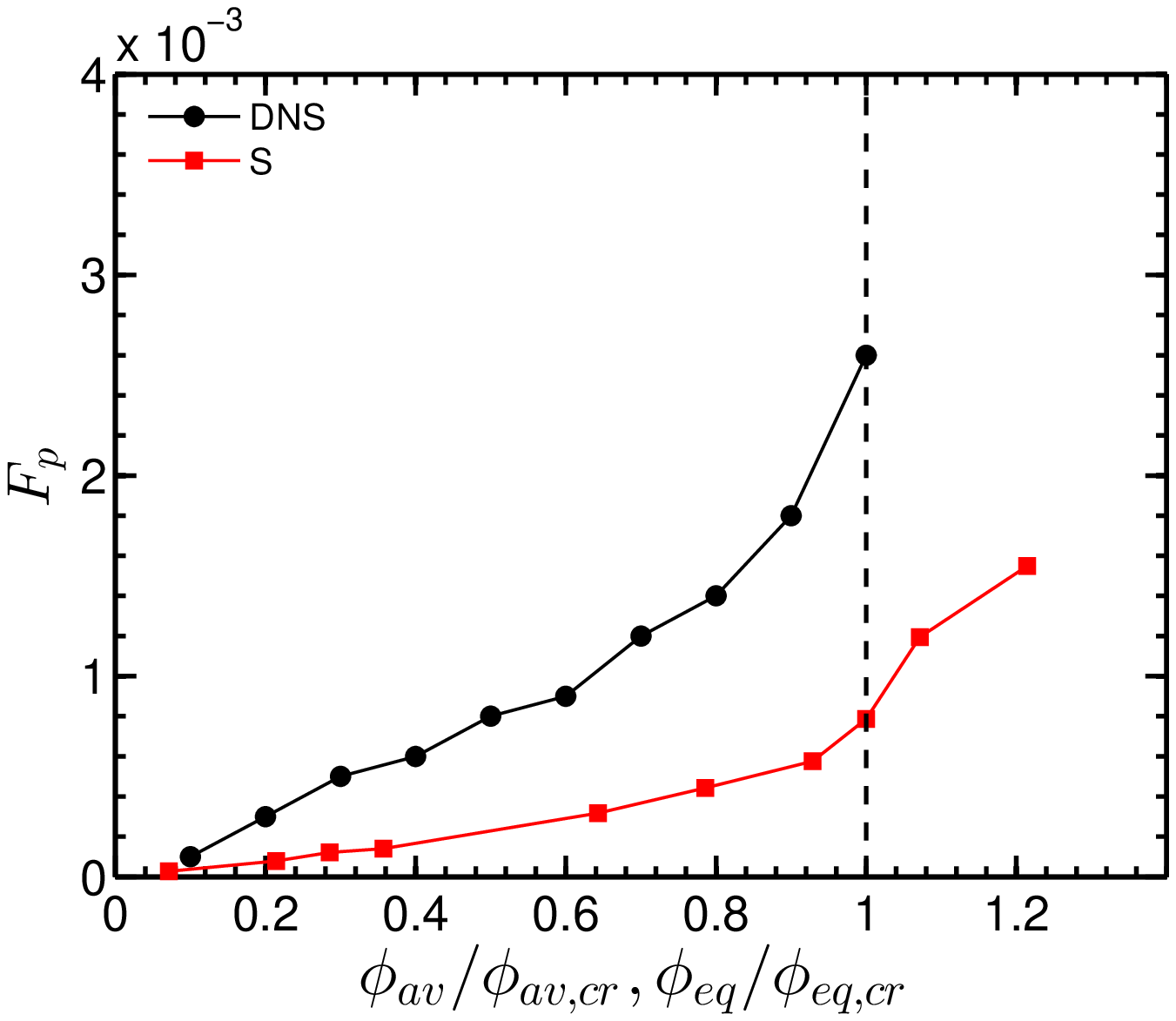}
	\caption{ $Re_b = 3300$}
	\endminipage 
	\minipage{0.49\textwidth}
	\includegraphics[width=\textwidth]{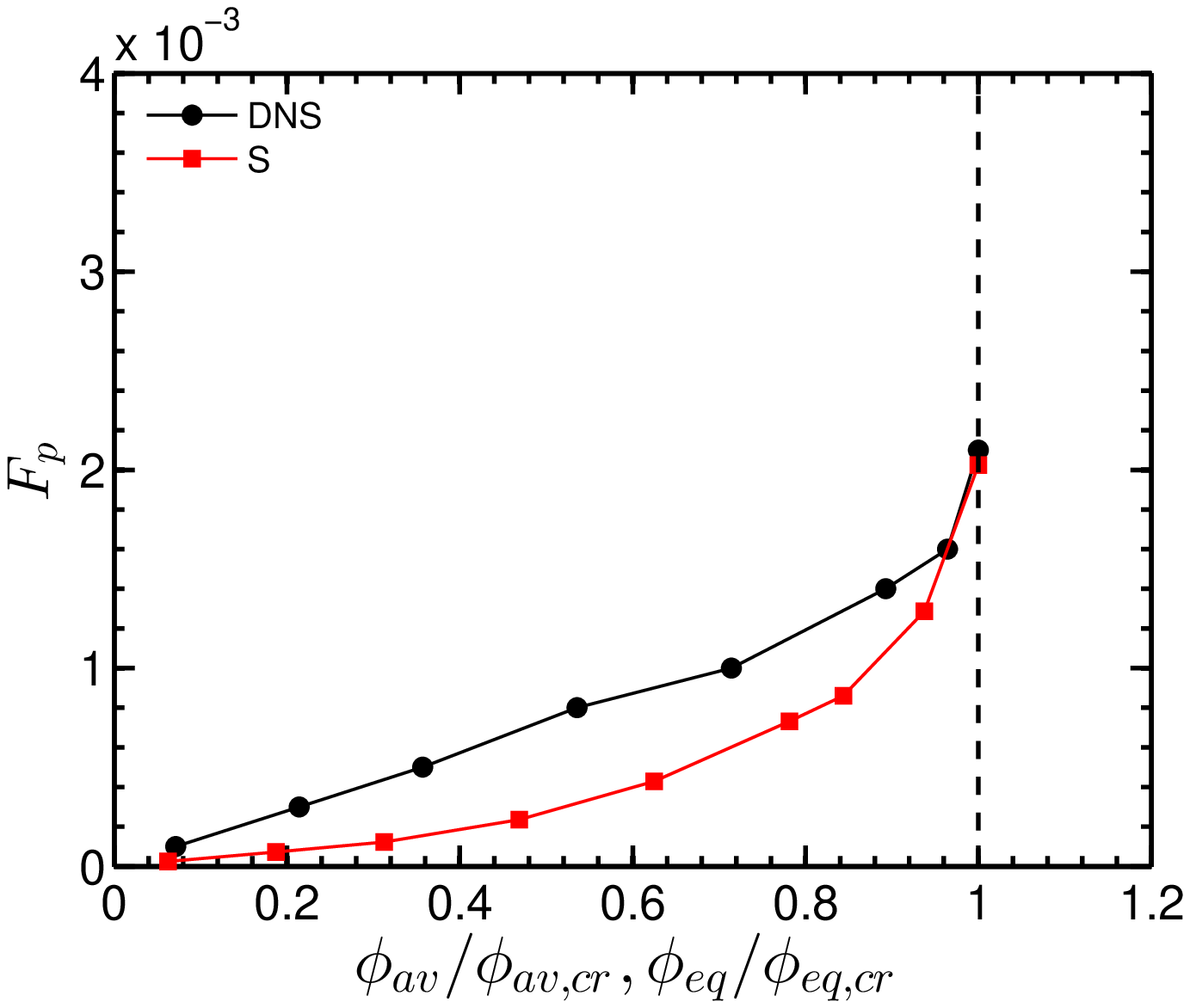}
	\caption{ $Re_b = 5600$}
		\endminipage 	
	\end{subfigure} 
	\caption{The dissipation term in DNS due to particles and an equivalent dissipation in the Smagorinksy model due to modified $C_s$ are compared. Here, $C_s$ is represented in terms of equivalent volume fraction ($\phi_{eq}$). The volume fraction ($\phi_{av}$ or $\phi_{eq}$) is normalized with a critical volume fraction at which turbulence collapse is observed in individual cases. }
	\label{Fp_DNS_LES}
\end{figure}

\begin{figure}
	\begin{subfigure}[b]{1\textwidth}
	\minipage{0.35\textwidth}
	\includegraphics[width=\textwidth]{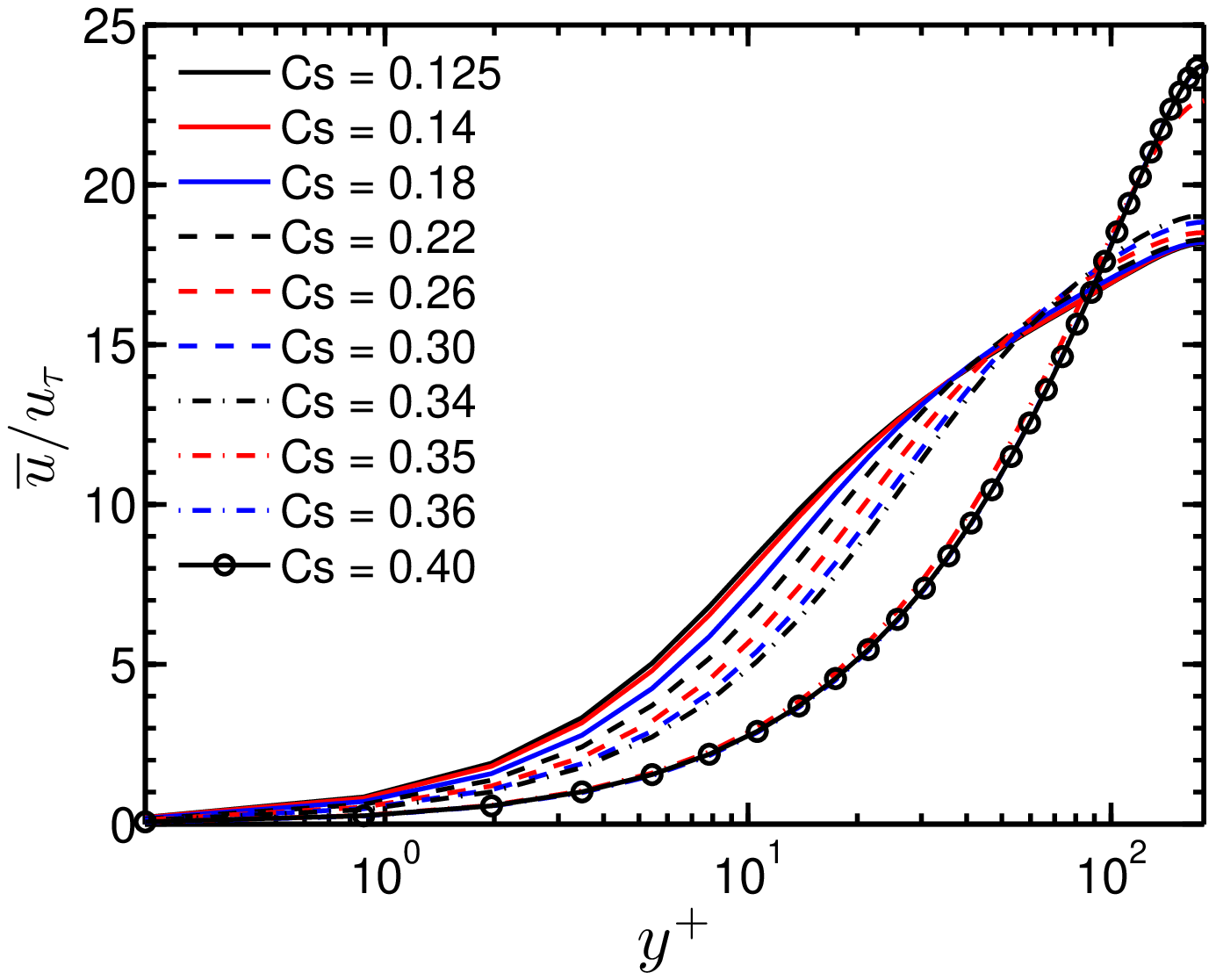}
	\caption{ $Re_b = 5600$}
	\endminipage 
	\minipage{0.35\textwidth}
	\includegraphics[width=\textwidth]{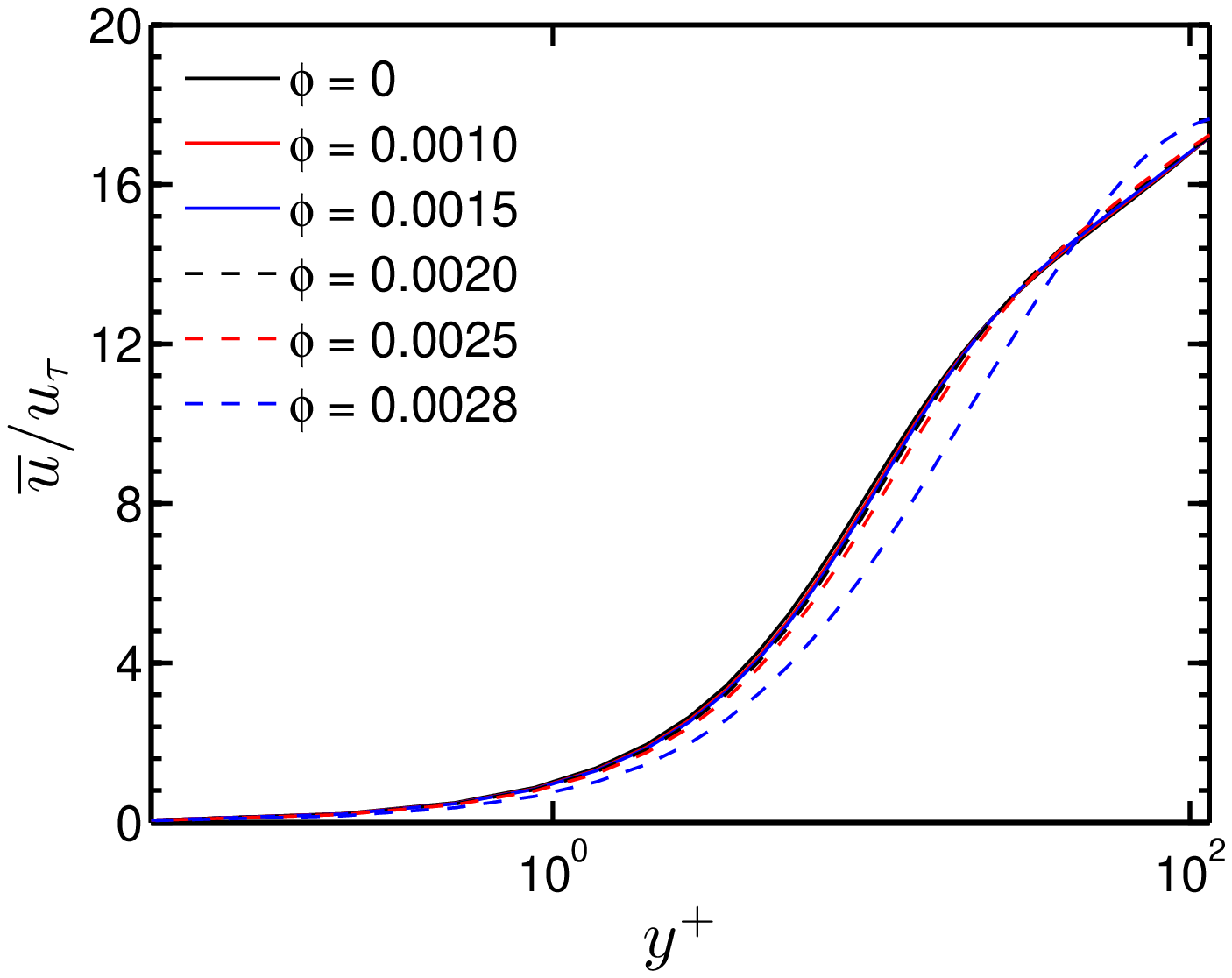}
	\caption{  $Re_b = 5600$, St = 210.93}
	\endminipage \\
	\minipage{0.35\textwidth}
	\includegraphics[width=\textwidth]{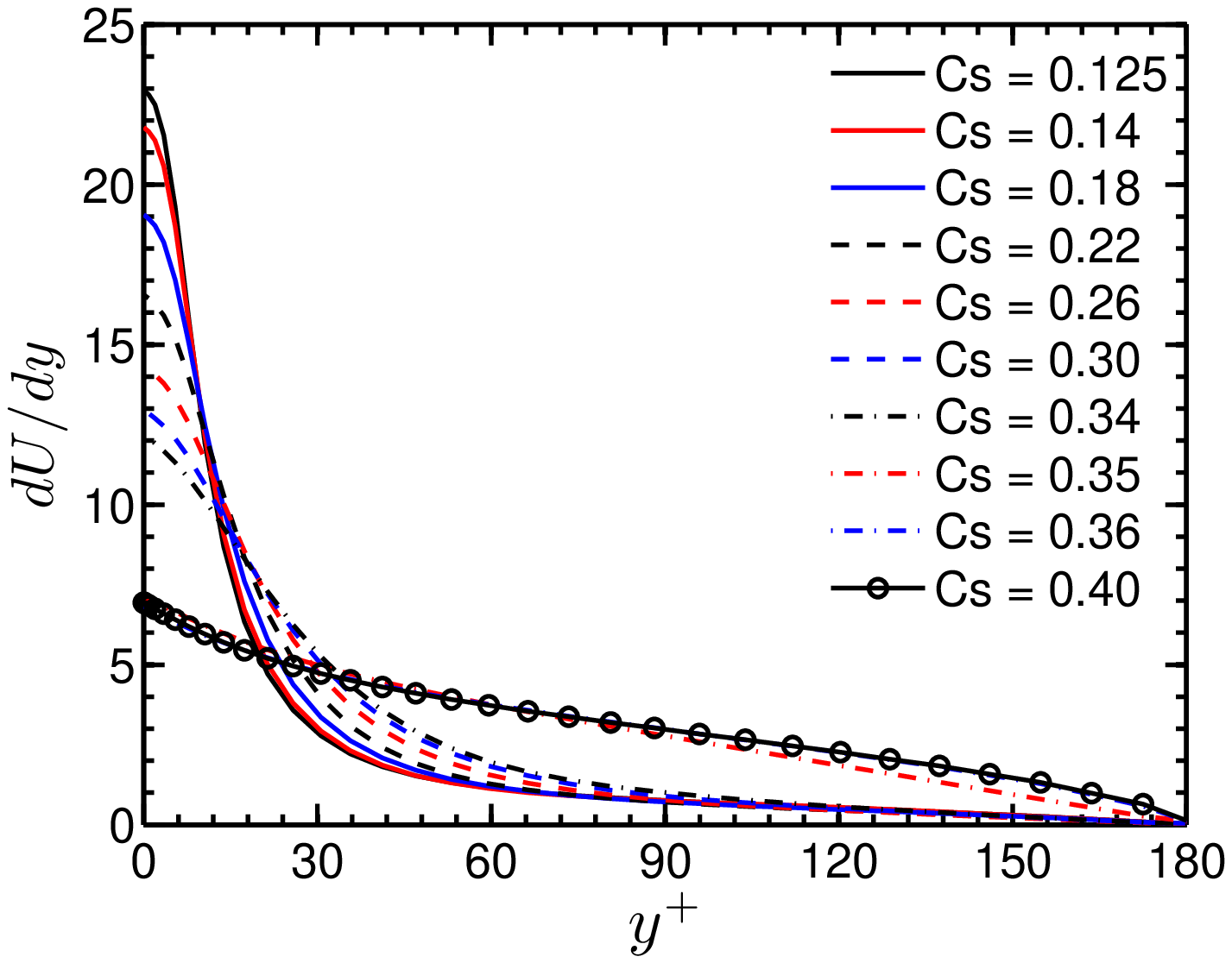}
	\caption{ $Re_b = 5600$}
	\endminipage 
	\minipage{0.35\textwidth}
	\includegraphics[width=\textwidth]{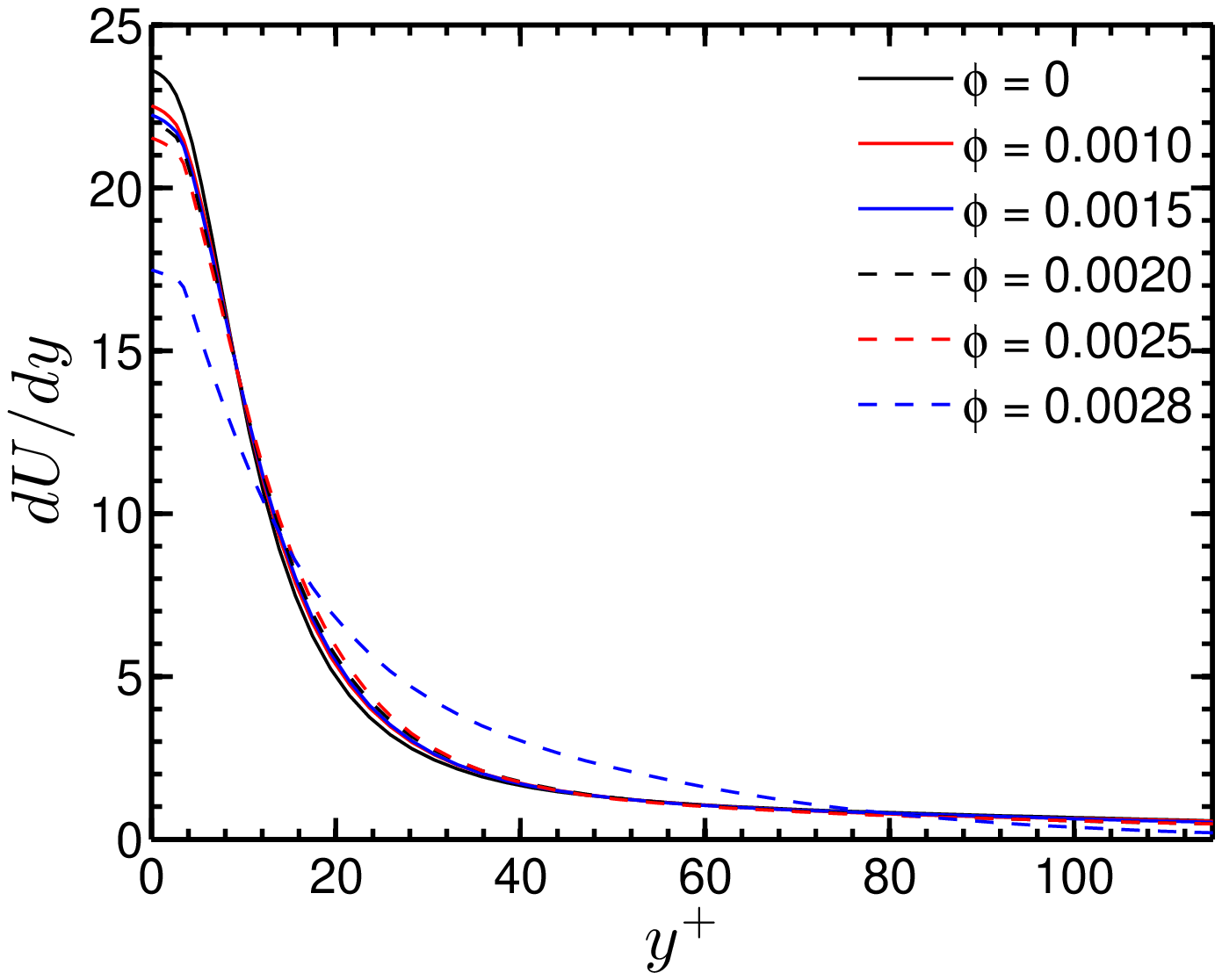}
	\caption{  $Re_b = 5600$, St = 210.93}
	\endminipage\\
	\minipage{0.35\textwidth}
	\includegraphics[width=\textwidth]{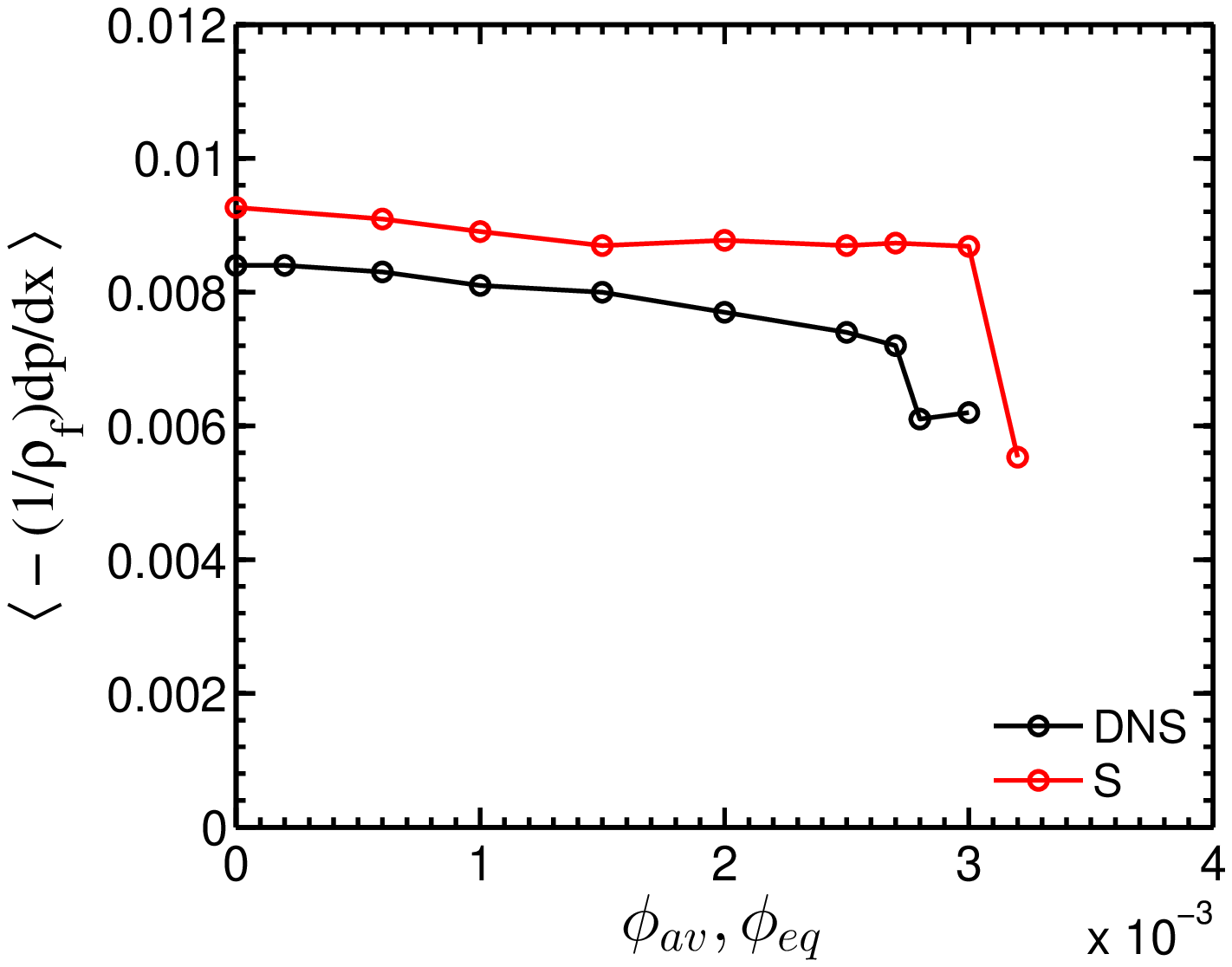}
	\caption{ $Re_b = 5600$}
	\endminipage 
	\end{subfigure} 
	\caption{The streamwise mean velocity (Figs.~(a and b)), mean velocity gradient  (Figs.~(c and d)) and pressure gradient  (Fig.~(e)) for $Re_b = 5600$. The streamwise velocity in Figs.~(a and b) are normalized with unladen frictional velocity ($u_\tau$). The mean velocity and pressure gradient in Figs.~(c-e) are normalized with fluid bulk velocity and channel width. In Figs.~(a and c), the simulations are performed with different $C_s$ for the Smagorinsky model. Figs.~(b and d) show the results from particle-laden DNS for  St = 210.93. The pressure gradient for a range of $\phi_{av}$ and $\phi_{eq}$ is compared in Fig.~(e) where legends are as, S: Smagorinsky model, and DNS: Direct numerical simulation.}
	\label{mean_profiles}
\end{figure}

The feedback term from particle-laden DNS and eddy viscosity-based dissipation due to modified $C_s$ from the Smagorinksy model is compared in Fig.~\ref{Fp_DNS_LES}. Here, the dissipation due to eddy viscosity for the unladen simulation ($\nu_t (d\widetilde{U}/dy)^2$ for the $C_s = 0.125$ case) is subtracted from the $F_{p}$ to depict the equivalent particle dissipation only. The feedback term in DNS is taken for St = 105.47 and 210.93 for $Re_b = 3300$ and 5600, respectively. The feedback term as a function of volume fraction ($\phi_{av}$) from particle-laden DNS and as a function of equivalent volume fraction ($\phi_{eq}$) is plotted from the Smagorinsky model. As the turbulence collapse happen at different $\phi_{av}$ and $\phi_{eq}$ values for particle-laden DNS and Smagorinsky coefficient (Fig.~\ref{avg_rms_S_DNS}). The $\phi_{av}$ and $\phi_{eq}$ are divided by the $\phi_{av,cr}$ and $\phi_{eq,cr}$ which are the critical loadings for DNS and Smagorinsky model, respectively, where turbulence collapse is observed. The Smagorinsky model predicts the feedback term and the trend with reasonable accuracy for both Reynolds numbers. For $Re_b = 3300$ in Fig.~\ref{Fp_DNS_LES}(a), the Smagorinsky model accurately predict the dissipation at low $\phi_{eq}$, while underpredict particle dissipation near the $\phi_{eq,cr}$. In case of moderate Reynolds number ($Re_b = 5600$), the prediction by Smagorinksy model matches with DNS, Fig.~\ref{Fp_DNS_LES}(b). In the case of the Smagorinsky model, it is to be noted that particles act as a source/sink in particle-laden cases depending on the local relative velocity. However, in the case of modified average $C_s$, it will be a dissipative effect only. Therefore, further analysis is required so that $C_s$ can be expressed as a function of Reynolds number, Stokes number, and wall-normal distance to capture the effect of particles more accurately. This will be an interesting future scope.

The mean velocity, the gradient of mean velocity, and the pressure gradient across the channel are plotted in Fig.~\ref{mean_profiles}. The plots are shown for the simulations performed with Smagorinsky model with varying $C_s$ for $Re_b = 5600$, and the particle-laden DNS for $Re_b = 5600$ and St = 210.93. The mean velocity profiles in Fig.~(a and b) show that the mean velocity decreases in the buffer region and increases in the channel center with an increase in $C_s$. However, the extent of decreases predicted by modified $C_s$ is more than that predicted by DNS, shown in Fig.~\ref{mean_profiles}(b). In the case of mean velocity gradient, a significant decrease is observed in the near-wall region for the Smagorinsky model (Fig.~\ref{mean_profiles} (c)) than the DNS case (Fig.~\ref{mean_profiles} (d)). The Smagorinksy model captures the qualitative behavior of pressure gradient, which is observed in particle-laden DNS, Fig.~\ref{mean_profiles} (e). Here,  $C_s$ is denoted in terms of equivalent volume fraction ($\phi_{eq}$). Thus, the variation of $C_s$ in the Smagorinksy model captures the effect of particles qualitatively, and further analysis will lead to new LES models in the future.

\section{CONCLUSIONS}
\label{sec:conclusion}
Direct numerical simulations are performed for particle-laden turbulent channel flows at two bulk Reynolds numbers and different Stokes numbers over a range of particle volume fractions. It is observed that the local isotropy of small and large scales decreases with an increase in particle volume loading. We report the variation of the Kolmogorov constant with an increase in volume loading at low and moderate Reynolds numbers due to an increase in anisotropy of fluid velocity fluctuations. In the near-wall region ($y^+ = 15$), a decrease in the Kolmogorov constant is observed when estimated via the second-order velocity structure function, while it remains unaltered if estimated using the compensated energy spectrum. Both analyses show no variation in the Kolmogorov constant at $y^+ = 50$. And an almost linear decrease in the Kolmogorov constant is observed in the channel center region for the considered Reynolds numbers and Stokes numbers. The Kolmogorov constant increases from the wall to the channel center for unladen wall-bounded flows. However, the present study reveals that in the case of particle-laden flows, the Kolmogorov constant at the channel center is lower than the near-wall region ($y^+ = 15$) for high volume fraction. Thus, from the present study, it can be concluded that the Kolmogorov constant for turbulent channel flows is not only a function of wall-normal location but also a function of particle volume loading. 

The present analysis highlights two important points related to modeling particle-laden turbulent flows. First, the increase in local anisotropy of fluid fluctuations with an increase in particle loading depicts that inhomogenous anisotropic models will be a better choice to capture the dynamics of particle-laden turbulent flows at high particle volume loadings. Second, the variation of the Kolmogorov constant as a function of particle volume fraction is to be considered to predict fluid phase dynamics. In the proposed modeling approach, the variation of the Smagorinsky coefficient is estimated from the variation of the Kolmogorov constant. The simulations are performed to predict the dynamics of the fluid phase without solving the particle phase equations simultaneously. The new method captures the qualitative trend of turbulence attenuation and the sudden collapse of turbulence similar to the behavior observed for particle-laden turbulent flows. However, the model quantitatively shows some deficiency in capturing the fluid phase fluctuations and equivalent particle feedback dissipation. This happens as the Smagorinsky model is based on isotropic scalar eddy viscosity formulation. Development of an anisotropic eddy viscosity model as a function of particle loading, Stokes number, etc., will be an interesting future scope. It is worth mentioning that the present study is conducted at low and moderate Reynolds numbers. Nevertheless, the demonstrated variations of the Kolmogorov constant in the case of low Reynolds number particle-laden turbulent flows will be helpful in developing better turbulence models in LES and the stochastic modeling approach.

\bibliography{refrences}

\end{document}